\def\Title{Evolutions of Gowdy, Brill and Teukolsky initial data on a smooth lattice.}
\def\Author{Leo Brewin}

\pdfoutput=1

\documentclass[a4paper,12pt]{article}
\usepackage{ifpdf}
\ifpdf
\usepackage[pdftex]{graphicx}
\usepackage[usenames,dvipsnames]{xcolor}
\else
\usepackage{graphicx}
\usepackage[usenames,dvips]{xcolor}
\fi
\usepackage{amsmath}
\usepackage{amssymb}
\usepackage[style=numeric,sorting=none,sortcites=true]{biblatex}
\usepackage{breqn}
\usepackage{hyperref}
\usepackage{caption}
\usepackage{paper}

\hypersetup{
colorlinks=true,
citecolor=blue,
linkcolor=red,
pdfauthor= \Author,
pdftitle = \Title,
}

\numberwithin{equation}{section}

\begin{document}

\title{\Title}
\author{%
\Author\\[10pt]%
School of Mathematical Sciences\\%
Monash University, 3800\\%
Australia}
\date{26-Mar-2016}
\date{5-Feb-2017}
\date{13-Feb-2017}
\date{1-Mar-2017}
\reference{Preprint}

\maketitle

\begin{abstract}
\noindent
Numerical results, based on a lattice method for computational general relativity, will be
presented for Cauchy evolution of initial data for the Brill, Teukolsky and polarised Gowdy
space-times. The simple objective of this paper is to demonstrate that the lattice method
can, at least for these space-times, match results obtained from contemporary methods. Some
of the issues addressed in this paper include the handling of axisymmetric instabilities (in
the Brill space-time) and an implementation of a Sommerfeld radiation condition for the
Brill and Teukolsky space-times. It will be shown that the lattice method performs
particularly well in regard to the passage of the waves through the outer boundary.
Questions concerning multiple black-holes, mesh refinement and long term stability will not
be discussed here but may form the basis of future work.
\end{abstract}

\section{Introduction}
\label{sec:Intro}

With the recent successful detection of gravitational waves, and the reasonable expectation
of more to follow, there will soon be a wealth of new information about the universe
allowing ever more detailed questions to be asked. But the computational methods that have
served us well for today's questions may well prove to be inadequate for the questions that
arise in the near future. So it seems that there is good reason to continue to develop new
approaches to computational general relativity. One such approach, known as smooth lattice
general relativity, will be described in this paper. As its name suggests it is based on a
lattice and it employs a metric that is locally smooth.

The smooth lattice method differs from traditional numerical methods in computational
general relativity in a number of important aspects. The space-time manifold consists of a
large collection of overlapping computational cells with local Riemann normal coordinates
used in each cell. The computational cells are a set of vertices and legs that define small
subsets of the manifold. The use of local Riemann normal coordinates in each each cell not
only reduces the complexity of the evolution equations but it also explicitly incorporates
the Einstein equivalence principle into the formalism. The lattice method provides an
elegant separation between the topological properties of the space-time (by specifying
combinatoric data such as the connections between cells, vertices etc.) and the metric
properties (by specifying data such as leg-lengths, curvature components etc. within each
cell). A key element of the lattice method is that it uses the second Bianchi identities to
evolve the Riemann curvatures. More details of the lattice method will be given later in
section (\ref{sec:SmoothLattices}).

Previous applications of the lattice method includes the
Schwarzschild~\cite{brewin:2010-02}, Oppenheimer-Snyder~\cite{brewin:2009-05} and
Kasner~\cite{brewin:2014-01} space-times. Though these were important tests of the lattice
method, they lacked some of the more challenging aspects expected in full 3-dimensional
computational general relativity, in particular the presence of gravitational waves and
their interactions with the outer boundaries on a finite computational grid. In this paper
evolutions of a smooth lattice with zero shift for the Gowdy~\cite{gowdy:1971-01},
Brill~\cite{brill:1959-01} and Teukolsky~\cite{teukolsky:1982-01} spacetimes will be
presented. The objective is not to explore any new features of these space-times but rather
to use them as examples of the smooth lattice method.

The boundaries in the Gowdy space-time will be handled using standard periodic boundary
conditions while the Brill and Teukolsky space-times will require an out outgoing radiation
condition. The Brill space-time adds the extra complexity of the numerical instabilities
that arise from the use of a lattice adapted to the axisymmetry. These issues will be
addressed in the following sections.

This class of space-times has been studied extensively by other authors.
See~\cite{berger:1993-01,hern:1998-01,garfinkle:2004-01} for the Gowdy space-time,
\cite{eppley:1979-01,Choptuik:2003-01,alcubierre:2000-04,miyama:1981-01} for Brill waves and
\cite{baumgarte:1998-01,abrahams:1998-01} for Teukolsky waves.

The structure of this paper is as follows. The notation used in this paper will be defined
in the following section. Sections (\ref{sec:SmoothLattices},\ref{sec:CauchyEvol}) provide a
broad summary of the smooth lattice method including details of the evolution equations on a
typical lattice. The specific details of the lattice, the construction of the initial data
and the evolution equations for each of the three spacetimes are given sections
(\ref{sec:GowdyIntro},\ref{sec:BrillIntro},\ref{sec:TeukolskyWaves}). This is followed by a
short discussion on the use of the Einstein toolkit~\cite{loffler:2011-01} before the
results are presented in section (\ref{sec:Results}). Most of the algebraic calculations are
deferred to the appendices (\ref{sec:TransMatrix}--\ref{sec:FullRiemannEqtns}).

\section{Notation}
\label{sec:Notation}

Throughout this paper Greek letters will denote space-time indices while spatial indices
will be denoted by just three Latin letters, $i,j$ and $k$. The remaining Latin letters will
serve as vertex labels. One small exception to these rules will be noted in
Appendix~(\ref{sec:CartanEqtns}) where Latin indices will be used (extensively) to record
frame components for differential forms.

The coordinates for a typical Riemann normal frame will be denoted by either $(t,x,y,z)$
or $x^\mu$ while globally defined coordinates will be denoted by the addition of a tilde
such as $(\Tt,\Tx,\Ty,\Tz)$ or ${\Tx}^\mu$. A tilde will also be used to denote tensor
components in the global frame, e.g., ${\tilde{T}_{xy}}$ would be the $\Tx\Ty$ component
of the tensor $T$ in the global coordinate frame. Note that the global coordinates are
not an essential part of the smooth lattice method. They appear in this paper solely to
assist in setting the initial data and also when comparing the evolved data against
the exact solution or against data obtained by other numerical means (e.g., a finite
difference code).

A key element of the smooth lattice method is that it employs many local Riemann normal
frames. This introduces a minor bookkeeping issue -- if a tensor is defined across two
frames, how should its components in each frame be recorded? Let $\Ba$ and $\Bb$ be the
Riemann normal frames associated with the pair of vertices $a$ and $b$. Consider a
vector $v$ defined over this pair of frames. Then the components, in the frame $\Bb$, of
the vector $v$ at vertex $a$ will be denoted by $v^\alpha{}_{a\Bb}$ while
$v^\alpha{}_{a\Ba}$ denotes the components, in $\Ba$, of $v$ at $a$. Similar notation
will be used for other tensors, for example $R^\alpha{}_{\beta p\Bq}$ would denote the
components of the Ricci tensor at the vertex $p$ in the frame $\Bq$.

It is customary to denote the Cauchy time parameter by the symbol $t$. However, that
symbol is reserved for the time coordinate of a typical local Riemann normal frame and
thus some other symbol is required, for example $\Tt$ with a corresponding time
derivative operator $d/d\Tt$. The proliferation of tildes that would follow from this
choice can be avoided with the following convention -- replace $d/d\Tt$ with $d/dt$ and
take the $d/dt$ to be the time derivative operator associated with the Cauchy time
parameter $\Tt$. This convention applies only to the operator $d/dt$, thus a (partial)
time derivative such as $v^\mu{}_{,t}$ should be understood as a derivative with respect
to the Riemann normal coordinate $t$.

The signature for the metric, Riemann and Ricci tensors follows that of Misner, Thorne
and Wheeler~\cite{mtw:1973-01}.

\section{Smooth lattices}
\label{sec:SmoothLattices}

A smooth lattice is a discrete entity endowed with sufficient structure to allow it to be
used as a useful approximation to a smooth geometry (which in the context of computational
general relativity is taken to be a solution of the Einstein equations). The typical
elements of a smooth lattice are combinatoric data such as vertices, legs, etc. and
geometric data such as a coordinates, the Riemann and metric tensors and any other geometric
data needed to make the approximation to the smooth geometry meaningful.


An $n$-dimensional smooth lattice can be considered as a generalisation of an
$n$-dimensional piecewise linear manifold. The later are constructed by gluing together a
collection of flat $n$-simplices in such a way as to ensure that the resulting object is an
$n$-dimensional manifold, that the points common to any pair of $n$-simplices form
sub-spaces of dimension $n-1$ or less and that the metric is continuous across the interface
between every pair of connected $n$-simplices.

In a smooth lattice the cells need not be simplices, they are required to overlap with their
neighbours and the curvature may be non-zero throughout each cell. The picture to bear in
mind is that the cells of a smooth lattice are akin to the collection of coordinate charts
that one would normally use to cover a manifold. The overlap between each pair of charts is
non-trivial and allows for coordinate transformations between neighbouring charts. So too
for the smooth lattice -- each pair of neighbouring cells overlap to the extent that a well
defined transition function can be constructed. This is an essential element of the smooth
lattice formalism -- it is used extensively when computing various source terms in the
equations that control the evolution of the lattice (see appendix (\ref{sec:TransMatrix})
for further details). Another important feature of the smooth lattice is that each cell of
the lattice need not be flat. The intention here is to better allow the smooth lattice to
approximate smooth geometries than could otherwise be achieved using piecewise flat simplices
(compare the approximation of a sphere by spherical triangles as opposed to flat triangles).
The smooth lattice should also provide smoothly varying estimates for various quantities
(for example the geodesic length of a leg) in the overlap region between a pair of cells.
The use of the adjective \emph{smooth} in the name \emph{smooth lattice} is intended to
capture the idea that all quantities on the lattice should vary smoothly (as best as
possible) across the lattice.

Denote the smooth geometry by $(g,M)$ where $g$ is the metric on the $n$-dimensional
manifold $M$. A smooth lattice representation of $(g,M)$ can be constructed in a number of
stages, in particular, choose a set of cells $M_i,\>i=1,2,3,\cdots$ that cover $M$, add the
vertices and legs and finally add the metric data to the lattice.

The cells $M_i,\>i=1,2,3,\cdots$ must be chosen so that each point in $M$ is contained in at
least one $M_i$ and each point in each $M_i$ should also be a point in $M$. Now decorate $M$
by introducing a set of vertices $V$ and a set of legs $L$ as follows. Add one or more
vertices to each cell and in each cell label one of these as the central vertex for that
cell (which will later serve as the origin of a set of coordinates local to the cell). Thus
each cell will contain one central vertex as well as other vertices (which are also the
central vertices of other neighbouring cells). The legs $L$ of the lattice are chosen as the
geodesics that connects the central vertices between pairs of neighbouring cells. Paths other
than geodesics could be used but since the geodesic is defined intrinsically by the
underlying smooth geometry it is a natural choice. There is, however, the issue of the
uniqueness of the geodesic -- if the curvature is too large or the vertices too far apart
there may not exist a unique geodesic joining the pair of vertices. This problem can be
overcome by a suitable choice of cells -- in regions where the curvature is large the cells
should be small and closely packed while in other regions, where the curvature is weak, the
cells can be well spaced out. It is well known that such a construction is always possible
(in the absence of curvature singularities).

The next step in the construction is to assign metric data to the cells. In each cell $M_i$,
expand the metric around the central vertex in terms of a local set of Riemann normal
coordinates $x^\alpha$ (see~\cite{willmore:1996-01,synge:1960-01,brewin:2009-03}), that is
\begin{align}
   \label{eqn:dsExact}
	ds^2 = \left (g_{\alpha\beta}
	              - \frac{1}{3} R_{\alpha\mu\beta\nu} x^\mu x^\nu
                 - \frac{1}{6} R_{\alpha\mu\beta\nu,\gamma} x^\mu x^\nu x^\gamma
					  + \cdots
			 \right) dx^\alpha dx^\beta
\end{align}
The coefficients $g_{\alpha\beta}$, $R_{\alpha\mu\beta\nu}$ etc. can be obtained by
projecting their corresponding quantities from the smooth metric onto a local orthonormal
basis on the central vertex.

At this stage the lattice is an exact copy of the original smooth geometry but with
additional structure (the vertices, legs, cells, coordinates etc.). The approximation is
introduced by truncating the series expansion for the metric at some finite order. The
lattice will then no longer be an exact copy of the original smooth metric and should be
considered an entity in its own right and will be denoted by $(g,M,V,L)$. The original
smooth geometry will now be denoted by $(\Tg,\TM)$.

For the space-times considered in this paper the metric in each cell will be taken as
\begin{align}
   \label{eqn:dsApprox}
	ds^2 = \left(g_{\alpha\beta}
	              - \frac{1}{3}R_{\alpha\mu\beta\nu} x^\mu x^\nu
			 \right) dx^\alpha dx^\beta
\end{align}
where $g_{\alpha\beta} = \diag (-1,1,1,1)$. This form of the metric will lead to estimates
for the geodesic lengths that differ from that given by $(\Tg,\TM)$. By inspection of the
(\ref{eqn:dsExact}) and (\ref{eqn:dsApprox}) it is should be clear that for a typical leg
$(p,q)$ in $\TM$ and $M$, the geodesic lengths, using the two metrics $\Tg$ and $g$, will
differ by a term of order $\BigO{RL^5}$ where $R$ and $L$ are estimates of the largest
curvatures and lengths in any of the cells that contain this leg.

If $(p,q)$ is a leg in the smooth lattice then the (squared) geodesic length can be
estimated (see~\cite{brewin:2009-03,synge:1960-01}) on the smooth lattice using
\begin{align}
   \Lsqpq = \gab \Dxa_{pq} \Dxb_{pq}
          - \frac{1}{3} \Racbd \xa_p \xb_p \xc_q \xd_q
          + \BigO{RL^5}
   \label{eqn:RNCLsqpq}
\end{align}
where where $\Dxa_{pq}=\xa_q-\xa_p$. Of course other sources of truncation errors will arise
as part of the numerical evolution of the lattice data so this $\BigO{RL^5}$ truncation is
the best that can expected at this level of approximation. To obtain higher order
approximations would require not only retaining more terms in the series expansion for the
metric but would also require the cells to overlap beyond nearest neighbours.

Imagine for the moment that the truncation errors on the right hand side of
(\ref{eqn:RNCLsqpq}) where discarded. This leaves one equation that links the vertex
coordinates, the leg lengths and the curvatures. It might be thought that given sufficiently
many leg-lengths that the curvatures and coordinates could be computed by solving
(\ref{eqn:RNCLsqpq}). Past experience shows that even though the equations can be solved (in
some cases) the resulting evolution of the lattice did not converge to the continuum
space-time. It was found that correct evolutions could be obtained by evolving either the
leg-lengths and curvatures or equally by evolving the coordinates and the curvatures. Both
approaches will be discussed in more detail in section (\ref{sec:DotLegs}).

\subsection{Continuous time smooth lattices}
\label{sec:ContinuousTime}
The construction of the smooth lattice as described above would naturally lead, for the case
of computational general relativity, to a structure that is discrete in both space and time.
There is, however, an alternative picture in which the lattice evolves smoothly in time
while retaining its discrete spatial structure. This allows for a fairly simple construction
of a Cauchy initial value problem on such a lattice (as described later in the following
section. For the remainder of this paper, the smooth lattice, its coordinates, leg lengths
and Riemann curvatures should be considered to evolve smoothly with time.

\section{Cauchy evolution of a smooth lattice}
\label{sec:CauchyEvol}

Suppose that the spacetime $(\Tg,\TM)$ can be foliated by a one parameter family of spatial
hypersurfaces $\Sigma(\Tt)$ (i.e., each $\Sigma(\Tt)$ is a Cauchy surface in $(\Tg,\TM)$).
Each element of this family could be represented by a lattice with 3-dimensional
computational cells denoted by $\Sigma_i$. The 4-dimensional computational cells $M_i$ of
$M$ will be taken as the space-time volume swept out by the corresponding $\Sigma_i$ for an
infinitesimal increment in the Cauchy time parameter $\Tt$. Thus a single $M_i$ is a
4-dimensional cylinder, with a 3-dimensional base $\Sigma_i$, that connects a pair of
infinitesimally close Cauchy surfaces while the set of all $M_i, i=1,2,3,\cdots$ fills out
the space-time region between that pair of Cauchy surfaces.

The dynamical variables on a smooth lattice can be chosen to include the Riemann curvatures
on the central vertex and either the (squared) leg-lengths or the Riemann normal coordinates
for each vertex in each cell. In either case, the addition of the extrinsic curvatures (at
the central vertex) allows the full set of evolution equations for the lattice to be given
in first order form.

\subsection{Lapse and shift}
\label{sec:LapseShift}

In the standard formulation of the Cauchy initial value problem for general relativity the
lapse function and shift vectors can be freely specified at each point in the space-time.
This naturally carries over to the smooth lattice by allowing the lapse function and shift
vector to be freely specified on the central vertex of each cell.

In computational general relativity it is usually the case that once the lapse function and
shift vector have been fully specified then there are no remaining coordinate freedoms. This
is not exactly true on a smooth lattice -- each cell carries its own local set of coordinates
and specifying the lapse and shift at one point in that cell is not sufficient to properly
constrain the coordinates on the remaining vertices. What remains is the freedom to orient
the coordinate axes within each cell. Thus using boosts and spatial rotations the $t$-axis
can be aligned with the world-line of the central vertex (for the case of zero shift) while
the spatial axes can be given some preferred alignment with some of the remaining vertices of
the cell.\footnote{This picture changes slightly if the coordinates are evolved, see the
comment at the end of section (\ref{sec:DotLegs}).} This is a choice that depends on the
structure of the cells and possibly on any symmetries that might exist in the space-time.

In each of the space-times considered in this paper the shift vector will be set equal to
zero (i.e., the world-lines of the vertices will be normal to the Cauchy surfaces) while the
lapse function will be given as a function on the set of central vertices.

\subsection{Evolving the legs and coordinates}
\label{sec:DotLegs}

The only legs that will be evolved in a cell are those that are directly connected to the
central vertex. There are two reasons for making this choice. First, legs that are not tied
to the central vertex are likely to incur a larger truncation error than legs closer to the
central vertex (such as those tied to that vertex). Second, there is no contribution to the
leg-length from the Riemann tensor for legs directly connected to the central vertex thus
avoiding any issues of accounting for time derivatives of such terms.

Consider a typical cell with central vertex $o$ and let $q$ be any of its vertices. A
standard result from differential geometry, known as the first variation of
arc-length~\cite{berger-marcel:2003,chavel:2006-01,hicks:1965-01}, states that for a
one-parameter family of geodesics, the arc-length $\Loq$ will evolve according to
\begin{align}
   \label{eqn:FirstVariationA}
   \dotLoq = [v_\mu (Nn^\mu)]_o^q
\end{align}
where $v^\alpha$ is the (forward pointing) unit tangent vector to the geodesic, $n^\mu$ is
the (future pointing) unit tangent vector to the vertex world-line and $N$ is the lapse
function. For a short leg, where the lapse and extrinsic curvatures are approximately
constant across the leg, this result can be estimated by~\cite{brewin:2009-04}\footnote{%
This paper contains a number of small errors that do not effect the final results.
A corrected version can be found on ArXiv:0903.5365}
\begin{align}
   \dotLoq = - N \cKab \va_{oq} \vb_{oq} \Loq + \BigO{L^2}
\end{align}
Since $N$ and $\cKab$ are defined on the vertices there is an ambiguity in attempting to
apply this equation to any leg -- each leg is defined by two vertices so which vertex should
supply the required values? As there is no clear reason to prefer one vertex over the other
it seems reasonable to take the average from both vertices, that is\footnote{This result can
also be obtained directly from (\ref{eqn:FirstVariationA}) as shown in Appendix
(\ref{sec:EvolveLenoq}).}
\begin{align}
   \label{eqn:FirstVariationB}
   \dotLoq
   =
   - \frac{1}{2}
        \left(  \left(N \cKab\right)_{q\Bq} \va_{qo\Bq} \vb_{qo\Bq}
              + \left(N \cKab\right)_{o\Bo} \va_{oq\Bo} \vb_{oq\Bo} \right) \Loq
   + \BigO{L^2}
\end{align}

A simple generalisation of this result can be obtained by noting that any 3-geodesic within
a Cauchy surface can be arbitrarily approximated by a large sequence of short 4-geodesics of
the space-time. The arc-length for each short 4-geodesic is subject to the above evolution
equation and thus, on summing over all contributions to the path and taking a suitable
limit, it follows that
\begin{align}
   \label{eqn:IntegralDotL}
\dotBarLoq
   &= - \int_o^q\>N \cKab \va_{oq} \vb_{oq} ds
\end{align}
where $s$ is the proper distance along the path and $\BarLoq = \int_o^q ds$ is the
arc-length of the 3-geodesic.

Using this equation to evolve the leg-lengths requires a re-appraisal of how the legs of the
lattice are interpreted. In the standard formulation~\cite{brewin:2014-01}, the legs of the
lattice are geodesics in space-time (and will appear as chords connecting the vertices)
whereas in this alternative interpretation the geodesics now lie entirely within a Cauchy
surface.

The evolution equation (\ref{eqn:IntegralDotL}) is suitable for simple lattices, such as the
Gowdy lattice, where information about $N$ and $\cKab$ can be deduced along the entire path.
In all other cases, such as the Brill and Teukolsky lattices, the former evolution equation
(\ref{eqn:FirstVariationB}) must be used.

As the leg-lengths evolve, so too must the Riemann normal coordinates. So it is natural to
ask: What are the appropriate evolution equations for the $\xa$? A simple calculation, as
detailed in~\cite{brewin:2014-01}, shows that for any vertex $p$ in a cell
\begin{align}
\label{eqn:DotLegsA}
\dotXap
   &= - N \Kab \xb_p
\end{align}
A short independent derivation of this equation can also be found in Appendix
(\ref{sec:EvolveRNC}). Note that in choosing to evolve the coordinates, the freedom to adapt
the coordinates to the lattice, as described in section (\ref{sec:LapseShift}), can only be
imposed either on the initial Cauchy surface or at future times by applying suitable
rotations.

\subsection{Evolving the extrinsic curvatures}
\label{sec:DotKab}

In~\cite{brewin:2014-01} the evolution equations for the extrinsic curvatures where given
for the particular case of a unit lapse. The method employed in that paper can be easily
repeated for the more general case of a non-constant lapse. The results are as
follows\footnote{These equations can also be obtained directly by projecting the Arnowitt,
Deser and Misner (ADM) 3+1 equations~\cite{mtw:1973-01}, with zero shift, onto a local
orthonormal frame.}
\begin{align}
	\label{eqn:DotKabA}
      \dotKxx &= -\dNxx + N \left(\Rtxtx + \Kxx^2 - \Kxy^2 - \Kxz^2\right)\\[5pt]
	\label{eqn:DotKabB}
      \dotKyy &= -\dNyy + N \left(\Rtyty + \Kyy^2 - \Kxy^2 - \Kyz^2\right)\\[5pt]
	\label{eqn:DotKabC}
      \dotKzz &= -\dNzz + N \left(\Rtztz + \Kzz^2 - \Kxz^2 - \Kyz^2\right)\\[5pt]
	\label{eqn:DotKabD}
      \dotKxy &= -\dNxy + N \left(\Rtxty - \Kxz\Kyz \right)\\[5pt]
	\label{eqn:DotKabE}
      \dotKxz &= -\dNxz + N \left(\Rtxtz - \Kxy\Kyz \right)\\[5pt]
	\label{eqn:DotKabF}
      \dotKyz &= -\dNyz + N \left(\Rtytz - \Kxy\Kxz \right)
\end{align}
These equations apply at the central vertex where, in the Riemann normal frame of this
vertex, $n^\alpha = \delta^\alpha_t$ and where the covariant derivatives
$N_{;\alpha\beta}$ coincides with the partial derivatives $N_{,\alpha\beta}$.

\subsection{Evolving the Riemann curvatures}
\label{sec:DotRabcd}

In 4-dimensions there are 20 algebraically independent components of the Riemann tensor
at any one point and in each cell these are taken to be
\begin{gather}
\Rxyxy,\> \Rxyxz,\> \Rxyyz,\> \Rxzxz,\> \Rxzyz,\> \Ryzyz\notag\\
\Rtxxy,\> \Rtyxy,\> \Rtzxy,\> \Rtxxz,\> \Rtyxz,\> \Rtzxz,\> \Rtyyz,\> \Rtzyz\\
\Rtxtx,\> \Rtyty,\> \Rtztz,\> \Rtxty,\> \Rtxtz,\> \Rtytz\notag
\end{gather}
Of these, the first 14 will be evolved while the remaining 6 will be set by applying
the vacuum Einstein equations (see section (\ref{sec:VacuumEqtns})).

The evolution equations for the Riemann curvatures are based upon the second
Bianchi identities. At the origin of the local frame (i.e., the central vertex)
the connection vanishes and thus these equations take the simple form
\bgroup
\def\P{\phantom{{}+{}}}
\def\M{{}-{}}
\begin{align}
\label{eqn:DotRabcdA}\DRxyxyDt &= \P \DRtyxyDx - \DRtxxyDy\\
\label{eqn:DotRabcdB}\DRxyxzDt &= \P \DRtzxyDx - \DRtxxyDz\\
\label{eqn:DotRabcdC}\DRxyyzDt &= \P \DRtzxyDy - \DRtyxyDz\\
\label{eqn:DotRabcdD}\DRxzxzDt &= \P \DRtzxzDx - \DRtxxzDz\\
\label{eqn:DotRabcdE}\DRxzyzDt &= \P \DRtzxzDy - \DRtyxzDz\\
\label{eqn:DotRabcdF}\DRyzyzDt &= \P \DRtzyzDy - \DRtyyzDz\\
\label{eqn:DotRabcdG}\DRtxxyDt &= \M \DRxyxyDy - \DRxyxzDz\\
\label{eqn:DotRabcdH}\DRtyxyDt &= \P \DRxyxyDx - \DRxyyzDz\\
\label{eqn:DotRabcdI}\DRtzxyDt &= \P \DRxyxzDx + \DRxyyzDy\\
\label{eqn:DotRabcdJ}\DRtxxzDt &= \M \DRxyxzDy - \DRxzxzDz\\
\label{eqn:DotRabcdK}\DRtyxzDt &= \P \DRxyxzDx - \DRxzyzDz\\
\label{eqn:DotRabcdL}\DRtzxzDt &= \P \DRxzxzDx + \DRxzyzDy\\
\label{eqn:DotRabcdM}\DRtyyzDt &= \P \DRxyyzDx - \DRyzyzDz\\
\label{eqn:DotRabcdN}\DRtzyzDt &= \P \DRxzyzDx + \DRyzyzDy
\end{align}
\egroup
There is, however, a small bump in the road in using these equations to evolve the
curvatures -- the only data immediately available are the point values for the curvatures in
each cell and thus some process must be applied to estimate the partial derivatives in each
cell. It is possible to use a finite difference approximation using data from neighbouring
cells but in doing so a proper account must be made of the different orientations of the
neighbouring frames. This is clearly true for the spatial derivatives where neighbouring
frames may differ by boosts and rotations. It is also true for the time derivatives due to
progression of boosts needed to keep the world-line of the origin of the local frame normal
to the Cauchy surfaces. Thus $\DRtzxyDx$, for example, will consist not only of the raw
partial derivatives (i.e., taking the raw data from neighbouring frames without regard for
coordinate transformations) but also of terms that account for the boosts and rotations
between neighbouring frames. The details are spelt out in full, for the particular class of
lattices used in this paper, in Appendix (\ref{sec:SourceTerms}) leading to expression such
as
\begin{align}
   \label{eqn:CartanRabcd}
   \DRabcdDe = \dRabcdde - \mfae \Rfbcd - \mfbe \Rafcd - \mfce \Rabfd - \mfde \Rabcf
\end{align}
in which the $\dRabcdde$ are the raw partial derivatives of $\Rabcd$ and the $\mabc$
are geometrical data built solely from the structure of the lattice (i.e., they depend
only on the leg-lengths and Riemann normal coordinates). This result is very much like
the usual definition of a covariant derivative. This does of course lead to a
significant increase in the number of terms in each equation. The full set of equations
(for a zero shift) can be found in Appendix (\ref{sec:FullRiemannEqtns}).

\subsection{The vacuum Einstein equations}
\label{sec:VacuumEqtns}

The second Bianchi identities provide no information about the time derivatives of the
Riemann components such as $\Rtxtx$. Consequently such components can not be evolved but
rather must be determined algebraically by applying the (vacuum) Einstein equations. Thus
the 6 curvatures $\Rtxtx,\Rtxty\cdots\Rtytz$ are obtained from
\begin{align}
0 &= \Rxx = -\Rtxtx + \Rxyxy + \Rxzxz\label{eqn:RiemEvolO}\\
0 &= \Ryy = -\Rtyty + \Rxyxy + \Ryzyz\label{eqn:RiemEvolP}\\
0 &= \Rzz = -\Rtztz + \Rxzxz + \Ryzyz\label{eqn:RiemEvolQ}\\
0 &= \Rxy = -\Rtxty + \Rxzyz\label{eqn:RiemEvolR}\\
0 &= \Rxz = -\Rtxtz - \Rxyyz\label{eqn:RiemEvolS}\\
0 &= \Ryz = -\Rtytz + \Rxyxz\label{eqn:RiemEvolT}
\end{align}

\subsection{Constraint equations}
\label{sec:Constraints}

The constraints consist not only of the four standard Hamiltonian and momentum constraints,
which on a lattice take the form
\begin{align}
0 &= \Rtt = \phantom{-} \Rtxtx + \Rtyty + \Rtztz\label{eqn:EinsConstA}\\
0 &= \Rtx = \phantom{-} \Rtyxy + \Rtzxz\label{eqn:EinsConstB}\\
0 &= \Rty = - \Rtxxy + \Rtzyz\label{eqn:EinsConstC}\\
0 &= \Rtz = - \Rtxxz - \Rtyyz\label{eqn:EinsConstD}
\end{align}
but also the extra constraints that arise from allowing the Riemann curvatures to be
evolved. These constraints follow from the second Bianchi identities, namely
\begin{align}
0 &= \DRxyxyDz + \DRxyyzDx - \DRxyxzDy\label{eqn:BianConstA}\\
0 &= \DRxyxzDz + \DRxzyzDx - \DRxzxzDy\label{eqn:BianConstB}\\
0 &= \DRxyyzDz + \DRyzyzDx - \DRxzyzDy\label{eqn:BianConstC}\\
0 &= \DRtyxyDz + \DRtyyzDx - \DRtyxzDy\label{eqn:BianConstE}\\
0 &= \DRtzxyDz + \DRtzyzDx - \DRtzxzDy\label{eqn:BianConstF}\\
0 &= \DRtxxyDz + \DRtxyzDx - \DRtxxzDy\label{eqn:BianConstD}
\end{align}
Note that $\Rtxyz$ is not one of the 20 chosen $\Rabcd$ but it can be computed directly using
$\Rtxyz=\Rtyxz-\Rtzxy$.

\section{Gowdy polarised cosmologies}
\label{sec:GowdyIntro}

Polarised Gowdy cosmologies on $T^3\times R$ are a class of solutions of the vacuum Einstein
equations that posses two linearly independent spatial Killing vectors. The metric, in
coordinates adapted to the symmetries, is commonly written in the form~\cite{new-kc:1998-01,alcubierre:2004-02}
\begin{align}
   \label{eqn:GowdyMetric}
   ds^2 = \Tt^{-1/2} e^{\lambda/2}\left(-d\Tt^2+d\Tz^2\right)
        + \Tt\left(e^{P}d\Tx^2+e^{-P}d\Ty^2\right)
\end{align}
where $P$ and $\lambda$ are functions of $(\Tt,\Tz)$ and where $\partial/\partial\Tx$ and
$\partial/\partial\Ty$ are the two Killing vectors. Each of the spatial coordinates
$(\Tx,\Ty,\Tz)$ are required to be periodic (to respect the $T^3$ topology). The functions
$P$ and $\lambda$ used in this paper are those given by New-Watt etal~\cite{new-kc:1998-01},
namely,
\begin{align}
   \label{eqn:GowdyExactA}
   P (\Tt,\Tz) ={}& J_0(2\pi \Tt)\cos(2\pi \Tz)\\
   \label{eqn:GowdyExactB}
   \lambda (\Tt,\Tz) ={}&
                  -2\pi \Tt J_0(2\pi t) J_1(2\pi \Tt)\cos^2(2\pi \Tz)
                  + 2(\pi \Tt)^2\left(J^2_0(2\pi \Tt)+J^2_1(2\pi \Tt)\right)\notag\\
                 &-2\pi^2\left(J^2_0(2\pi)+J^2_1(2\pi)\right)
                  -\pi J_0(2\pi) J_1(2\pi)
\end{align}
with $\Tz$ restricted to $\left[-0.5,0.5\right]$. The domain for $\Tx$ and $\Ty$ can be
chosen as any finite interval, e.g., $\left[0,1\right]$.

The metric is singular only at $\Tt=0$ and consequently initial data should be set at some
other time (e.g., at $\Tt=1$ as described below). The Gowdy initial data will be evolved away
from the $\Tt=0$ singularity.

\subsection{A Gowdy lattice}
\label{sec:GowdyLattice}

A lattice that represents the spatial part of this metric is rather easy to construct.
Start by discretising the $\Tz$ axis into a finite number of points labelled from $0$ to
$N_z$ with the point labelled $0$ identified with that labelled $N_z$ (i.e., two labels for
a single point). These points will soon be identified as the vertices of the lattice. Note
that there are no legs at this stage, these will be added later. Now use the Killing
vectors $\partial/\partial \Tx$ and $\partial/\partial \Ty$ to drag the discretised $\Tz$
axis along the $\Tx$ and $\Ty$ axis. The legs of the lattice can now be constructed as the
space-time geodesics that connect pairs of points (now taken as vertices of the lattice).
This leads to the simple lattice shown in figure (\ref{fig:GowdyLattice}) consisting of $N_z$
computational cells labelled from $0$ to $N_z$ with cell $0$ identified with cell $N_z$. This
lattice contains three classes of legs, one for each of the three coordinate axes, namely,
$\Lxx,\Lyy$ and $\Lzz$. Other data that must be carried by the lattice include the
extrinsic curvatures, $\cKab$, the Riemann curvatures, $\Rabcd$ and the lapse function $N$.

Consider a typical computational cell, as shown in figure (\ref{fig:GowdyLattice}), and ask
the question: How should the Riemann normal frame be constructed? Let $\partial_\alpha$ be
the unit basis vectors for the Riemann normal frame. Now choose the origin of the Riemann
normal frame to be (permanently) attached to the central vertex. Next, use boosts to ensure
that $\partial_t$ is normal to the Cauchy surface, then use rotations to ensure that the
vertices of $\Lzz$ lie on the $z$-axis and also for the vertices of $\Lxx$ to lie in the
$xz$-plane. Given the symmetries of the Gowdy space-time it is no hard to appreciate that
the $(t,x,y,z)$ coordinates of the seven vertices of the cell $M_p$ will be of the following
form
\begin{equation}
   \begin{aligned}
   x^\mu_{0\Bp} &= (0,0,0,0)^\mu\\
   x^\mu_{1\Bp} &= (t_1,0,0,(\Lzz)_{p})^\mu&\qquad
   x^\mu_{2\Bp} &= (t_2,0,0,-(\Lzz)_{p-1})^\mu\\
   x^\mu_{3\Bp} &= (t_3,0,(\Lyy)_p,0)^\mu&
   x^\mu_{4\Bp} &= (t_4,0,-(\Lyy)_p,0)^\mu\\
   x^\mu_{5\Bp} &= (t_5,(\Lxx)_p,0,0)^\mu&
   x^\mu_{6\Bp} &= (t_6,-(\Lxx)_p,0,0)^\mu
   \end{aligned}
\end{equation}
where the time coordinate is given by $2t = -\cKab\xa\xb$ (see~\cite{brewin:2010-03}).

Note that this construction also ensures that the Riemann normal axes are aligned with
their Gowdy counterparts (as a consequence of the Gowdy metric being diagonal).

\subsection{Initial data}
\label{sec:GowdyInitialData}

A straightforward computation on the Gowdy metric reveals that there are three non-trivial
extrinsic curvatures, $\TKxx, \TKyy$ and $\TKzz$ and five non-trivial Riemann curvatures,
$\TRxyxy,\TRxzxz,\TRyzyz,\TRtxxz$ and $\TRtyyz$. The lattice values for the extrinsic and
Riemann curvatures, $\cKab$ and $\Rabcd$, were computed by projecting their counterparts,
$\TKab$ and $\TRabcd$, onto the local Riemann normal frame. This provides not only a way to
identify the non-trivial components on the lattice but also a simple way to assign the
initial data.

The leg-lengths $\Lxx,\Lyy$ and $\Lzz$ were set as follows. The $\Lxx$ were computed as the
length of the geodesic connecting $(1,0,0,\Tz)$ to $(1,\delta \Tx,0,\Tz)$ with $\delta
\Tx=0.0001$. A similar approach was used to compute the $\Lyy$ this time using the points
$(1,0,0,\Tz)$ and $(1,0,\delta \Ty,\Tz)$ with $\delta \Ty = \delta \Tx = 0.0001$. A common
value for $\Lzz$ was chosen for all cells, namely
\begin{align}
   \Lzz = \frac{1}{N_z}\int_{-0.5}^{0.5} \sqrt{{\tilde{g}}_{zz}}\>d\Tz
\end{align}
This in turn required the $\Tz$ coordinate to be unequally spaced from cell to cell. Starting
with $\Tz_0=-0.5$ the successive $\Tz_p$ for $p=1,2,3\cdots N_z-1$ where found by treating
the equation
\begin{align}
   0 = \Lzz - \int_{\Tz_{p-1}}^{\Tz_{p}} \sqrt{{\tilde{g}}_{zz}}\>d\Tz
\end{align}
as a non-linear equation for $\Tz_{p}$ given $\Tz_{p-1}$.

\subsection{Evolution equations}
\label{sec:GowdyEqtns}

The evolution equations for $\Lxx,\Lyy$ and $\Lzz$ follow directly from equation
(\ref{eqn:IntegralDotL}) by making appropriate use of the symmetries built into the Gowdy
lattice, in particular that the legs are aligned to the coordinate axes and thus
$v^\alpha_{ox}=(0,1,0,0)$, $v^\alpha_{oy}=(0,0,1,0)$ and $v^\alpha_{oz}=(0,0,0,1)$ while
rotational symmetry ensures that the integrand in (\ref{eqn:IntegralDotL}) is constant along
the $x$ and $y$ axes. This leads to the following evolution equations for $\Lxx,\Lyy$ and
$\Lzz$ in cell $p$,
\begin{align}
   \label{eqn:DotGowdyLxx}\dotLxx &= -N\Kxx\Lxx\\[5pt]
   \label{eqn:DotGowdyLyy}\dotLyy &= -N\Kyy\Lyy\\[5pt]
   \label{eqn:DotGowdyLzz}\dotLzz &= -\int_{p}^{p+1}\>N\Kzz\>ds
\end{align}
and where $s$ is the arc-length along the leg connecting successive cells (i.e., along the
$\Tz$-axis of the lattice) and where the limits $(p,p+1)$ are understood to denote the
corresponding vertices.

The evolution equations for the extrinsic and Riemann curvatures can be constructed in at
least two ways. In the first approach the evolution equations for the $\TKab$ and $\TRabcd$
can be projected onto the the local Riemann normal frame. The second approach is to impose
the known symmetries on the the complete set of equations given in Appendix
(\ref{sec:FullRiemannEqtns}). Both approaches lead to the following set of equations
for the extrinsic curvatures,
\begin{align}
   \label{eqn:DotGowdyKxx}\dotKxx &= -N_{,xx} + N\left(\Kxx^2 + \Rxyxy +\Rxzxz\right)\\[5pt]
   \label{eqn:DotGowdyKyy}\dotKyy &= -N_{,yy} + N\left(\Kyy^2 + \Rxyxy +\Ryzyz\right)\\[5pt]
   \label{eqn:DotGowdyKzz}\dotKzz &= -N_{,zz} + N\left(\Kzz^2 + \Rxzxz +\Ryzyz\right)
\end{align}
and for the Riemann curvatures,
\begin{align}
%
%
   \label{eqn:DotGowdyRxyxy}
   \dotRxyxy ={}&
        N(\Ryzyz+2\Rxyxy) \Kxx
       +N(\Rxzxz+2\Rxyxy) \Kyy
      \notag\\
      &-N\mxzx \Rtyyz
       -N\myzy \Rtxxz
      \\[5pt]
   \label{eqn:DotGowdyRxzxz}
   \dotRxzxz ={}&
        N(\Ryzyz+2\Rxzxz) \Kxx
       +N(\Rxyxy+2\Rxzxz) \Kzz
      \notag\\
      &-N\mxzx \Rtxxz
      -2\Rtxxz \dNz
      -N\dRtxxzdz
      \\[5pt]
   \label{eqn:DotGowdyRyzyz}
   \dotRyzyz ={}&
        N(\Rxzxz+2\Ryzyz) \Kyy
       +N(\Rxyxy+2\Ryzyz) \Kzz
      \notag\\
      &-N\myzy \Rtyyz
       -2\Rtyyz \dNz
       -N\dRtyyzdz
      \\[5pt]
   \label{eqn:DotGowdyRtxxz}
   \dotRtxxz ={}&
        N(\Kyy+2\Kzz) \Rtxxz
       +N(\Rxyxy-\Rxzxz) \myzy
      \notag\\
      &-(\Rxyxy+2\Rxzxz) \dNz
       -N\dRxzxzdz
      \\[5pt]
   \label{eqn:DotGowdyRtyyz}
   \dotRtyyz ={}&
        N(\Kxx+2\Kzz)\Rtyyz
       +N(\Rxyxy-\Ryzyz) \mxzx
      \notag\\
      &-(\Rxyxy+2\Ryzyz) \dNz
       -N\dRyzyzdz
\end{align}
where
\begin{gather}
   N_{,z}  = \frac{\partial N}{\partial{ s}}\qquad
   N_{,zz} = \frac{\partial^2 N}{\partial{ s}^2}\\[5pt]
   N_{,xx} = \frac{1}{\Lxx}\frac{\partial\Lxx}{\partial s}
                           \frac{\partial N}{\partial{ s}}\qquad
   N_{,yy} = \frac{1}{\Lyy}\frac{\partial\Lyy}{\partial s}
                           \frac{\partial N}{\partial{ s}}\\[5pt]
   \label{eqn:GowdyDRabcd}
   \dRtxxzdz = \frac{\partial \Rtxxz}{\partial{ s}}\qquad
   \dRtyyzdz = \frac{\partial \Rtyyz}{\partial{ s}}\\[5pt]
   \label{eqn:GowdyMabc}
   \mxzx =\frac{1}{\Lxx}\frac{\partial\Lxx}{\partial s}\qquad
   \myzy =\frac{1}{\Lyy}\frac{\partial\Lyy}{\partial s}
\end{gather}

\subsection{The lapse function}
\label{sec:GowdyLapse}

The lapse function can be freely chosen across the lattice either by way of an explicit
function (e.g. $N=1$) or by evolving the lapse along with other lattice data. This second
choice will taken in this paper where three different methods for evolving the lapse will be
used, namely
\begin{align}
   \dotN &= -2N\TrK &&\text{1+$\log$}\\[5pt]
   \dotN &= -N^2\TrK &&\text{Harmonic}\\[5pt]
   \dotN &= -N^2\Kzz &&\text{Exact}
\end{align}
where $\TrK = \Kxx + \Kyy + \Kzz$. The $1+\log$ and harmonic lapse equations are standard
gauge choices and need no explanation while the third equation, as its name suggests, is
designed to track the exact solution. This exact lapse equation can be obtained as
follows. First note that for the exact solution $N^2 = \Tgzz$. Then use $d\Tgzz/dt=-2N\TKzz$
to obtain $dN/dt = -\TKzz$ whereupon the result follows by noting that $\TKzz = \Tgzz\Kzz =
N^2 \Kzz$.

Many other choices are of course possible but those just given stand out as they allow for a
direct comparison with either the exact solution
(\ref{eqn:GowdyMetric}--\ref{eqn:GowdyExactB}) or with the results from the Cactus code.

Initial values for the lapse will be discussed later in section (\ref{sec:GowdyResults}).

\subsection{Constraints}
\label{sec:GowdyConstraints}

The only constraints that survive under the symmetries inherent in the Gowdy space-time
are (\ref{eqn:EinsConstA},\ref{eqn:EinsConstD},\ref{eqn:BianConstA}) and can be written as
\begin{align}
   \label{eqn:GowdyC1}
      0 = C_1 ={}& \Rxyxy+\Rxzxz+\Ryzyz\\
   \label{eqn:GowdyC2}
      0 = C_2 ={}& \Rtxxz+\Rtyyz\\
   \label{eqn:GowdyC3}
      0 = C_3 ={}&
      \dRxyxydz
      +\Kxx\Rtyyz
      +\Kyy\Rtxxz\notag\\
      &+(\Rxyxy-\Ryzyz)\mxzx
       +(\Rxyxy-\Rxzxz)\myzy
\end{align}
where $\dRxyxydz$, $\mxzx$ and $\myzy$ are given by
(\ref{eqn:GowdyDRabcd},\ref{eqn:GowdyMabc}). Note also that trivial factors have been
cleared from the first two equations. This set of constraints were not imposed during the
evolution but were instead used as a quality control on the evolved data (see section
(\ref{sec:GowdyResults})).

\subsection{Numerical dissipation}
\label{sec:GowdyDissip}

It was found that for some choices of the lapse function, most notably the $1+\log$ choice,
the addition of some numerical dissipation could significantly prolong the evolution.

The particular form of numerical dissipation used here is based upon the familiar
Kreiss-Oliger approach in which an additional term is added to the right hand side of
selected evolution equations, in our case, the evolution equations for the extrinsic and
Riemann curvatures. In each case the modified evolution equation in cell $p$ was of the form
\begin{align}
   \label{eqn:GowdyKODissip}
   \frac{dY}{dt} = \left(\frac{dY}{dt}\right)_{\epsilon=0}
                 - \frac{2\epsilon}{(\Lzz)_{p}+(\Lzz)_{p+1}}
                   (&Y_{p+3}-6Y_{p+2}+15Y_{p+1}-20Y_{p}\notag\\
                    &+Y_{p-3}-6Y_{p-2}+15Y_{p-1})
\end{align}
where $\epsilon$ is a small number (in the results described below $\epsilon=0.8$). The
first term on the right hand side is the right hand side of the evolution equations
(\ref{eqn:DotGowdyKxx}-\ref{eqn:DotGowdyRtyyz}) while the second term is a naive
approximation to $\epsilon\Lzz^5d^6Y/ds^6$. The important point is that the dissipation
scales as $\BigO{\Lzz^5}$ and thus will vanish in the limit as $\Lzz\rightarrow0$.

\section{Brill waves}
\label{sec:BrillIntro}

Brill waves~\cite{brill:1959-01} are time and axisymmetric solutions of the vacuum Einstein
equations generated by initial data of the form
\begin{align}
   \label{eqn:BrillMetric}
   ds^2 = \psi^4\left(e^{2q}\left(d\Trho^2 + d\Tz^2\right) + \Trho^2 d\Tphi^2\right)
\end{align}
in which $(\Trho,\Tphi,\Tz)$ are cylindrical polar coordinates and where $\psi(\Trho,\Tz)$
and $q(\Trho,\Tz)$ are a class of functions subject to the conditions of asymptotic
flatness, the vacuum Einstein equations and reflection symmetry across both $\Tz=0$ and
$\Trho=0$. The reflection symmetry across $\Trho=0$ follows from the condition that the data
be well behaved at $\Trho=0$. However, the condition that the data be reflection symmetric
across $\Tz=0$ has no physical basis and is introduced only to reduce the bulk of the
numerics (i.e., the data can be evolved in the quarter plane ($\Trho>0,\Tz>0$) rather than
the half plane ($\Trho>0,\vert\Tz\vert<\infty$)).

Brill showed that the initial data will have a finite ADM mass when the functions $q$ and
$\psi$ behave as $q=\BigO{\Tr^{-2}}$ and $\psi=1 + \BigO{\Tr^{-1}}$ as
$\Tr\rightarrow\infty$ where $\Tr^2=\Trho^2+\Tz^2$. He also showed that for the initial data
to be well behaved near the $\Trho=0$ coordinate singularity, $q$ must behave like
$q=\BigO{\Trho^2}$ as $\Trho\rightarrow0$ which can also be expressed as
\begin{align}
   0 = \lim_{\Trho\rightarrow0} q\>, \quad
   0 = \lim_{\Trho\rightarrow0} \left(\frac{\partial q}{\partial\Trho}\right)
\end{align}
while the reflection symmetric conditions on $q$ and $\psi$ requires
\begin{gather}
   \label{eqn:BrillBCa}
    0 = \lim_{\Trho\rightarrow0} \left(\frac{\partial q}{\partial\Trho}\right)\>,\quad
    0 = \lim_{\Tz\rightarrow0}   \left(\frac{\partial q}{\partial\Tz}\right)\\[5pt]
   \label{eqn:BrillBCb}
    0 = \lim_{\Trho\rightarrow0} \left(\frac{\partial\psi}{\partial\Trho}\right)\>,\quad
    0 = \lim_{\Tz\rightarrow0}   \left(\frac{\partial\psi}{\partial\Tz}\right)
\end{gather}
The condition that $\psi=1 + \BigO{\Tr^{-1}}$ as $\Tr\rightarrow\infty$ was implemented
using a standard mixed outer boundary condition,
\begin{align}
   \label{eqn:BrillBCc}
   \frac{\partial\psi}{\partial\Tr}
   =
   \frac{1-\psi}{\Tr}\quad\text{as }\Tr\rightarrow\infty
\end{align}

Finally, the vacuum Einstein equations requires $\psi$ to be a solution of the
Hamiltonian constraint which in this case takes the form
\begin{align}
   \label{eqn:BrillH}
   \nabla^2 \psi = - \frac{\psi}{4}\left( \frac{\partial^2 q}{\partial\Trho^2}
                                         +\frac{\partial^2 q}{\partial\Tz^2}\right)
\end{align}
where $\nabla^2$ is the (flat space) Laplacian in the cylindrical coordinates
$(\Trho,\Tphi,\Tz)$. The three momentum constraints provide no new information as they are
identically satisfied for any choice of $q$ and $\psi$.

\subsection{Eppley Initial data}
\label{sec:BrillInitialData}
The function $q(\Trho,\Tz)$ was chosen as per Eppley~\cite{eppley:1977-01}, namely
\begin{align}
   q(\Trho,\Tz) = \frac{a \rho^2}{1+(\Trho^2+\Tz^2)^{n/2}}
\end{align}
with $n=5$ (any $n\ge4$ would be sufficient to satisfy $q=\BigO{\Trho^{-2}}$). The parameter
$a$ governs the wave amplitude with $a=0.01$ in the results presented below. Even though
this is a weak amplitude it is sufficient to test the lattice method.

The Hamiltonian constraint (\ref{eqn:BrillH}), subject to the boundary conditions
(\ref{eqn:BrillBCb}--\ref{eqn:BrillBCc}), was solved for $\psi$ using standard second order
centred finite differences (including on the boundaries). The grid comprised
$2048\times2048$ equally spaced points covering the rectangle bounded by $\Trho=\Tz=0$ and
$\Trho=z=20$. The finite difference equations were solved (with a maximum residual of
approximately $10^{-13}$) using a full multigrid code. The full Brill 3-metric was then
constructed using the reflection symmetry across $z=0$ and the rotational symmetry around the
$z$-axis.

Since the Brill initial data is axisymmetric it is sufficient to use a 2-dimensional lattice
on which to record the initial data for the lattice. An example of such a lattice is shown in
figure (\ref{fig:BrillLattice}). Each cell contains legs that are (at $\Tt=0$) aligned to
the Brill $(\Trho,\Tz)$ axes as well as a set of diagonal legs. A full 3-dimensional lattice
could be constructed by rotating this 2-dimensional lattice around the symmetry axis (as
indicated in figure (\ref{fig:BrillLattice})). In our computer code the right portion of
lattice covered the domain bounded by $\Trho=\Tz=0$, $\Tz=\pm5$ and $\Trho=5$ while the left
portion was obtained by reflection symmetry across $\Trho=0$. This places the symmetry axis
mid-way from left to right across the lattice (this is the blue axis shown in figure
(\ref{fig:BrillLattice})).

Each cell of the lattice contains 9 vertices $o,a,b,\cdots,h$ plus one additional vertex $p$
connected just to the central vertex $o$. The purpose of the extra vertex $p$ is that the
collection of all such vertices defines the image of the 2-dimensional lattice under the
action of the rotational symmetry. Figure (\ref{fig:BrillLattice}) shows two such additional
lattices in which each yellow leg has vertices of the form $(o,p)$.

In each cell the local Riemann normal coordinates $(t,x,y,z)$ were chosen as follows
\begin{align}
   x^\alpha_{p\Bo} &= (0,0,y_p,0)\\
   x^\alpha_{d\Bo} &= (0,x_d,0,z_d)&
   x^\alpha_{c\Bo} &= (0,0,0,z_c)&
   x^\alpha_{b\Bo} &= (0,x_b,0,z_b)\\
   x^\alpha_{e\Bo} &= (0,x_e,0,z_e)&
   x^\alpha_{o\Bo} &= (0,0,0,0)&
   x^\alpha_{a\Bo} &= (0,x_a,0,z_a)\\
   x^\alpha_{f\Bo} &= (0,x_f,0,z_f)&
   x^\alpha_{g\Bo} &= (0,x_g,0,z_g)&
   x^\alpha_{h\Bo} &= (0,x_h,0,z_h)
\end{align}
for some set of numbers $x_a,z_a,\cdots y_p$ and where the labels $o,a,b,\cdots,h$
follow the pattern shown in figure (\ref{fig:CartanConn}).

The leg-lengths and Riemann normal coordinates were set by first distributing the $N_x\times
N_z$ vertices as equally spaced points in the $(\Trho,\Tz)$ domain, $(-5,-5)$ to $(5,5)$, and
then integrating the geodesic equations as a two-point boundary value problem for each leg in
each cell.

The remaining initial data on the lattice consists of the non-zero components of the Riemann
and extrinsic curvatures along with either the leg-lengths or the vertex
coordinates\footnote{The choice depends on which evolution scheme is used -- evolving the
leg-lengths or evolving the coordinates.}. Given the symmetries of the Brill metric it is
not hard to see that the there are only 4 non-trivial extrinsic curvatures, $\Kxx$, $\Kyy$,
$\Kzz$ and $\Kxz$ and 8 non-trivial extrinsic curvatures, $\Rxyxy$, $\Ryzyz$, $\Rxzxz$,
$\Rxyyz$, $\Rtxxz$, $\Rtzxz$, $\Rtyxy$ and $\Rtyyz$. Each of these 12 curvatures were given
initial values by projecting their counterparts from the Brill metric (extended to 3+1 form
using a unit lapse and setting $d\psi/dt=dq/dt=0$ at $\Tt=0$) onto the local orthonormal
frame.

\subsection{Evolution equations}
\label{sec:BrillEqtns}

The initial data just described has only 12 non-trivial components for the Riemann and
extrinsic curvatures. It is easy to see that that this situation is preserved by the
evolution equations. For example, equation (\ref{eqn:DotKabD}) shows that $d\cKxy/dt = 0$
for this particular set of initial data. Thus all of the symmetries in the initial data
will be preserved throughout the evolution (e.g., $\cKxy$ will remain zero for all time).
This leads to the following set of evolution equations for the 4 extrinsic curvatures
\bgroup
\def\P{\phantom{{}+{}}}
\def\M{{}-{}}
\begin{align}
   \label{eqn:DotBrillKxx}\dotKxx &= \P \Rxyxy + \Rxzxz + \Kxx^2 - \Kxz^2 \\[5pt]
   \label{eqn:DotBrillKyy}\dotKyy &= \P \Rxyxy + \Ryzyz + \Kyy^2 \\[5pt]
   \label{eqn:DotBrillKzz}\dotKzz &= \P \Rxzxz + \Ryzyz + \Kzz^2 - \Kxz^2 \\[5pt]
   \label{eqn:DotBrillKxz}\dotKxz &= \M \Rxyyz
\end{align}
\egroup
while the evolution equations for the 8 Riemann curvatures are
\begin{align}
   \label{eqn:DotBrillRxyxy}
   \dotRxyxy ={}
   &
     (\Ryzyz+2\Rxyxy) \Kxx
   + (\Rxzxz+2\Rxyxy) \Kyy
   - \Kxz \Rxyyz
   \notag\\
   &
   - \mxyy \Rtyxy
   - \mxzx \Rtyyz
   + \dRtyxydx
   \\[5pt]
   \label{eqn:DotBrillRyzyz}
   \dotRyzyz ={}
   &
     (\Rxzxz+2\Ryzyz) \Kyy
   + (\Rxyxy+2\Ryzyz) \Kzz
   - \Kxz \Rxyyz
   \notag\\
   &
   - \mxzz \Rtyxy
   - \mxyy \Rtzxz
   - \dRtyyzdz
   \\[5pt]
   \label{eqn:DotBrillRxzxz}
   \dotRxzxz ={}
   &
     (\Ryzyz+2\Rxzxz) \Kxx
   + (\Rxyxy+2\Rxzxz) \Kzz
   + 2\Kxz \Rxyyz
   \notag\\
   &
   - \mxzx \Rtxxz
   - \mxzz \Rtzxz
   + \dRtzxzdx
   - \dRtxxzdz
   \\[5pt]
   \label{eqn:DotBrillRxyyz}
   \dotRxyyz ={}
   &
     (\Kzz + 2\Kyy) \Rxyyz
   - (\Ryzyz+2\Rxyxy) \Kxz
   + \mxzz \Rtyyz
   \notag\\
   &
   - \dRtyxydz
   \\[5pt]
   \label{eqn:DotBrillRtxxz}
   \dotRtxxz ={}
   &
     (\Kyy + 2\Kzz) \Rtxxz
   - 2\Kxz \Rtzxz
   - \mxyy \Rxyyz
   \notag\\
   &
   - \dRxzxzdz
   \\[5pt]
   \label{eqn:DotBrillRtzxz}
   \dotRtzxz ={}
   &
     (\Kyy + 2\Kxx) \Rtzxz
   + (\Ryzyz - \Rxzxz) \mxyy
   - 2\Kxz \Rtxxz
   \notag\\
   &
   + \dRxzxzdx
   \\[5pt]
   \label{eqn:DotBrillRtyxy}
   \dotRtyxy ={}
   &
     (\Kzz + 2\Kxx) \Rtyxy
   + (\Ryzyz - \Rxyxy) \mxzz
   - \Kxz \Rtyyz
   \notag\\
   &
   - 2\mxzx\Rxyyz
   + \dRxyxydx
   - \dRxyyzdz
   \\[5pt]
   \label{eqn:DotBrillRtyyz}
   \dotRtyyz ={}
   &
     (\Kxx + 2\Kzz) \Rtyyz
   + (\Rxyxy - \Ryzyz) \mxzx
   - \Kxz \Rtyxy
   \notag\\
   &
   - 2\mxzz\Rxyyz
   + \dRxyyzdx
   - \dRyzyzdz
\end{align}
where $\mxyy$, $\mxzx$ and $\mxzz$ are solutions of
\begin{align}
   \label{eqn:mxyy}
      v^y_{qs\Ba} + v^y_{uw\Be}
      &=
      \mxyy \left(v^x_{ea\Bo} v^y_{tp\Bo}-v^x_{tp\Bo} v^y_{ea\Bo}\right)\\
   \label{eqn:mxzx}
      v^x_{hb\Ba} + v^x_{bd\Bc} + v^x_{df\Be} + v^x_{fh\Bg}
      &=
      \mxzx \left(v^z_{ea\Bo} v^x_{gc\Bo}-v^z_{gc\Bo} v^x_{ea\Bo}\right)\\
   \label{eqn:mxzz}
      v^z_{hb\Ba} + v^z_{bd\Bc} + v^z_{df\Be} + v^z_{fh\Bg}
      &=
      \mxzz \left(v^z_{ea\Bo} v^x_{gc\Bo}-v^z_{gc\Bo} v^x_{ea\Bo}\right)
\end{align}
where $v^\alpha_{ab\Bc} = x^\alpha_{b\Bc}-x^\alpha_{a\Bc}$. The equations for $\mxyy$,
$\mxzx$ and $\mxzz$ were obtained by a simple application of equation
(\ref{eqn:FinalMabcEqtns}) to the $xz$-plane (leading to equations (\ref{eqn:mxzx}) and
(\ref{eqn:mxzz})) and the $yz$-plane (leading to equation (\ref{eqn:mxyy})).

The final set of evolution equations required are those for the leg-lengths or the vertex
coordinates. In contrast to the Gowdy lattice it was decided to evolve the vertex
coordinates. There are two reasons for doing so. First, the above evolution equations for
the $\Rabcd$ refer directly to the vertex coordinates and second, solving the coupled set of
non-linear equation (\ref{eqn:RNCLsqpq}) for the the vertex coordinates involves not only
extra work but was observed to lead to asymmetric evolutions (i.e., the evolved data failed
to be reflection symmetric across the symmetry axis). This loss of symmetry was attributed
to the algorithm~\cite{brewin:2014-01} used to solve these equations\footnote{The algorithm
in~\cite{brewin:2014-01} computes the coordinates one by one visiting the vertices in a
clock wise order. But for two cells on either side of the symmetry axis, one cell should be
processed clockwise and the other anti-clockwise.}.

\subsection{Numerical dissipation}
\label{sec:BrillKreissOliger}

Other authors~\cite{Choptuik:2003-01,garfinkle:2001-01} have noted that the singular
behaviour of the evolution equations on the symmetry axis can cause numerical instabilities
to develop along the symmetry axis. This problem can be avoided by either using a fully
3-dimensional formulation (which is computationally expensive) or mitigated by introducing
numerical dissipation. Similar instability problems were expected on the 2-dimensional
axisymmetric lattice. By direct experiment it was found that good damping of the numerical
instabilities could be obtained by applying a Kreiss-Oliger dissipation to the evolution
equations. The standard practice is to weight the dissipation term by powers of the
discretisation scale (i.e., powers of $L$) to ensure that the dissipation terms do not
dominate the truncation errors inherent in the numerical integrator. For a 4th-order
Runge-Kutta integrator (as used here) this would require a dissipation term of order
$\BigO{L^6}$ which would be the case for a 6th-order derivative term (as used in the Gowdy
lattice (\ref{eqn:GowdyKODissip})). However, on this simple Brill lattice, where cells
interact only by nearest neighbours, the best that can be done is to use a 2nd-derivative
dissipation term. The choice used in the results given below was
\begin{align}
   \label{eqn:BrillKODissip}
   \frac{dY}{dt} = \left(\frac{dY}{dt}\right)_{\epsilon=0}
                 + \epsilon \left(Y_a+Y_c+Y_e+Y_g-4Y_o\right)
\end{align}
where $\epsilon$ is a small number and the first term on the right hand side is time
derivative without dissipation while the second term is a crude estimate of
$\BigO{L^2}\nabla^2Y$ on the cell (the subscripts correspond to the vertices displayed in
figure (\ref{fig:CartanConn})). The dissipation was applied only to the Riemann curvatures as
no significant gains were noted when the dissipation was also applied to the extrinsic
curvatures. In the results presented below $\epsilon=1.0$ (this was the smallest value of
$\epsilon$ that allowed the evolution to remain stable to at least $t=10$).

\subsection{Inner boundary conditions}
\label{sec:BrillInnerBCs}

Figure (\ref{fig:BrillLattice}) show three copies of the 2-dimensional lattice sharing the
common symmetry axis. Away from the symmetry axis the three copies of the lattice provide
sufficient data to estimate $y$ derivatives of data on the lattice. However, this
construction clearly fails at the symmetry axis. One consequence of this can be seen in
equation (\ref{eqn:mxyy}) which, when expressed in terms of the coordinates and leg-lengths,
leads to $\mxyy\approx - (1/\Lyy)(d\Lyy/dx)$ where $x$ is the proper distance measured along
the $x$-axis. This shows that $\mxyy$ is singular on the symmetry axis (where $\Lyy=0$). The
upshot is that any $y$ derivative, on this choice of lattice, will by singular on the
symmetry axis (e.g., all of the $y$ derivatives in equations
(\ref{eqn:DotRabcdA}--\ref{eqn:DotRabcdN})).

One approach to dealing with this problem is to return to equations
(\ref{eqn:DotRabcdA}--\ref{eqn:DotRabcdN}) and make direct use of the rotational symmetry to
express all of the $y$ derivatives in terms of the (manifestly non-singular) $x$ derivatives
on the symmetry axis. As an example, let $V_{\alpha\beta}$ be the components of a tensor $V$
on the lattice. Now consider a copy of the lattice rotated by $\pi/2$ about the symmetry
axis. Denote the components of $V$ on the second lattice by $V'_{\alpha\beta}$. Then
$V'_{\alpha\beta} = V_{\alpha\beta}$ by rotational symmetry. However, on the symmetry axis
the coordinates for both lattices are related by $x'=y$, $y'=-x$ and $z'=z$ thus the usual
tensor transformation law would give $V'_{xy}=-V_{yx}$. But $V'_{xy}=V_{xy}$ and thus
$V_{xy}=-V_{yx}$ on the symmetry axis. Now suppose $V_{\alpha\beta} = W_{\alpha,\beta}$ for
some tensor $W$. It follows that $W_{x,y}=-W_{y,x}$ on the symmetry axis. This idea can be
applied to any tensor on the lattice in particular to the derivatives of $\Rabcd$.

It is also possible to gain information about the curvature components by considering a
rotation of $\pi$ rather than $\pi/2$. Following the steps described above, the result is
that any component of a tensor with an odd number of $x$ indices will be anti-symmetric
across the symmetry axis while the remaining components will be symmetric. This shows
immediately that $\Kxz$, $\Rxyyz$, $\Rtyxy$ and $\Rtzxz$ must vanish on the symmetry axis.

The upshot is that the evolution equations (\ref{eqn:DotRabcdA}--\ref{eqn:DotRabcdN}) can be
reduced, on the symmetry axis, to just 5 non-zero equations
\begin{align}
   \label{eqn:BrillSymmAxisA}
   \dotRxyxy ={}
   &
     2(\Rxzxz+2\Rxyxy)\Kxx
   - 2\mxzx\Rtxxz
   + \dRtyxydx
   \\[5pt]
   \label{eqn:BrillSymmAxisB}
   \dotRxzxz ={}
   &
     3\Kxx\Rxzxz
   + (\Rxyxy+2\Rxzxz)\Kzz
   - \mxzx\Rtxxz
   \notag\\
   &
   + \dRtzxzdx
   - \dRtxxzdz
   \\[5pt]
   \label{eqn:BrillSymmAxisC}
   \dotRyzyz ={}
   &
     3\Kxx\Rxzxz
   + (\Rxyxy+2\Rxzxz)\Kzz
   - \mxzx\Rtxxz
   - \dRtyyzdz
   \\[5pt]
   \label{eqn:BrillSymmAxisD}
   \dotRtxxz ={}
   &
    (\Kxx+2\Kzz)\Rtxxz
   + \mxzx\Rxyxy
   - \mxzx\Rxzxz
   - \dRxzxzdz
   \\[5pt]
   \label{eqn:BrillSymmAxisE}
   \dotRtyyz ={}
   &
    (\Kxx+2\Kzz)\Rtxxz
   + \mxzx\Rxyxy
   - \mxzx\Rxzxz
   \notag\\
   &
   + \dRxyyzdx
   - \dRyzyzdz
\end{align}
Though these equations are non-singular there remains a numerical problem with cells near
the symmetry axis -- their proximity to the symmetry axis can lead to instabilities in the
evolution.

A better approach, described in more detail below, is to excise a strip of cells containing
the symmetry axis (as shown in figure (\ref{fig:BrillLattice})) and to interpolate from
outside the strip to recover the time derivatives of the Riemann curvatures within the strip.
This, along with numerical dissipation, proved to be crucial in obtaining stable evolutions.

The interpolation near the symmetry axis was implemented as follows. The cells of the
2-dimensional lattice where indexed by rows and columns aligned to the $\Tx$ and $\Tz$ axes.
Each cell was given an index pair such as $(i,j)$ with $i$ denoting the number of columns
from the $\Tx=0$ axis (i.e., the symmetry axis) and $j$ the number of rows from the $\Tz=0$
axis. The interpolation used data from the cells $i=3,4,5,6,7$, for a given $j$, to supply
data for the cells with $i=-2,-1,0,1,2$, for the same $j$. In each case the interpolation was
tailored to respect the known symmetry of the data across the symmetry axis. Thus for
$d\Rxyxy/dt$, which is symmetric across $\Tx=0$, a polynomial of the form $y(x) = a_0 + a_2
x^2 + \cdots a_8 x^8$ was used. For anti-symmetric data the polynomial was of the form $y(x)
= a_1 x + a_3 x^3 + \cdots a_9 x^9$. The five coefficients $a_0,a_2,\cdots a_8$ and
$a_1,a_3,\cdots a_9$ were determined using trivial variations of standard methods for
polynomial interpolation. The choice of interpolation indices $i=3,4,5,6,7$, which correspond
to the light blue strip in figure (\ref{fig:BrillLattice}), was found by trial and error as
it gave stable evolutions (in conjunction with the numerical dissipation) without being overly
expensive.

There is a simple variation on this interpolation scheme in which the data from the symmetry
axis (i.e., equations (\ref{eqn:BrillSymmAxisA}--\ref{eqn:BrillSymmAxisE})) is included in
the data used to build the polynomial. Thus data on the cells $i=0,3,4,5,6,7$ would be used
to build data for cells $i=-2,-1,1,2$. The evolutions that resulted form this construction
were highly unstable and crashed at approximately $t=4.7$.

\subsection{Outer boundary conditions}
\label{sec:BrillOuterBCs}

The outer boundary of the lattice is defined to be a skin of cells one cell deep on the outer
edges of the lattice (as indicated by the orange region in figure (\ref{fig:BrillLattice})).
In each of the boundary cells the Riemann and extrinsic curvatures were evolved by way of an
outgoing radiation boundary condition of the form
\begin{align}
   \label{eqn:BrillOuterBC}
   \frac{\partial f}{\partial t}
   =
   - \frac{f}{\Tr}
   - \frac{\Tr}{\Tx^i n_i}\frac{\partial f}{\partial n}
\end{align}
where $f$ is one of the Riemann and extrinsic curvatures and $n$ is the outward pointing unit
normal to the cell (at the central vertex). The $\Tx^i$ are constants set equal to the
Brill the coordinates $(\Trho,\Tz)$ of the central vertex at $t=0$. Finally,
$\Tr=(\Trho^2+\Tz^2)^{1/2}$. The leg-lengths and Riemann normal coordinates in each cell were
not evolved but rather copied across from the nearest inward neighbouring cell.

This is an extremely simplistic set of boundary conditions (particularly so for the
leg-lengths and coordinates). It was chosen simply to get a numerical scheme up and running.
The surprise it that it works very well (as discussed below in section
(\ref{sec:BrillResults})).

\subsection{Constraints}
\label{sec:BrillConstraints}

Only five of the ten constraints (\ref{eqn:EinsConstA}--\ref{eqn:BianConstD}) survive once
the axisymmetry of the Brill space-time is imposed. The surviving equations are
(%
\ref{eqn:EinsConstA},%
\ref{eqn:EinsConstB},%
\ref{eqn:EinsConstD},%
\ref{eqn:BianConstA},%
\ref{eqn:BianConstC}%
) and can be written in the form
\begin{align}
   \label{eqn:BrillConstC1}
   0 = C_1 ={} & \Rxyxy+\Rxzxz+\Ryzyz \\
   %
   \label{eqn:BrillConstC2}
   0 = C_2 ={} & \Rtyxy+\Rtzxz\\
   %
   \label{eqn:BrillConstC3}
   0 = C_3 ={} & \Rtxxz+\Rtyyz\\
   %
   \label{eqn:BrillConstC4}
   0 = C_4 ={}
   &
   (\Rxyxy-\Ryzyz) \mxzx
   - \mxyy \Rxyyz
   -2 \mxzz \Rxyyz
   \notag\\
   &
   +\Kxx \Rtyyz
   +\Kyy \Rtxxz
   +\Kxz \Rtyxy
   +\dRxyxydz
   +\dRxyyzdx
   \\
   %
   \label{eqn:BrillConstC5}
   0 = C_5 ={}
   &
   (\Ryzyz-\Rxyxy) \mxzz
   +(\Ryzyz-\Rxzxz) \mxyy
   -2 \mxzx\Rxyyz
   \notag\\
   &
   +\Kyy \Rtzxz
   +\Kzz \Rtyxy
   +\Kxz \Rtyyz
   -\dRxyyzdz
   -\dRyzyzdx
\end{align}
where some simple numerical factors have been factored out.

\section{Teukolsky linearised waves}
\label{sec:TeukolskyWaves}

The results for the Gowdy and Brill spacetimes are promising but a proper test of the smooth
lattice method requires that it be applied to truly 3-dimensional data, i.e., initial data
devoid of any symmetries such as the Teukolsky linearised waves~\cite{teukolsky:1982-01}
described by the metric
\begin{align}
   \label{eqn:TeukPolarMetric}
   \begin{split}
      ds^2
      =& -d\Tt^2 +d\Tr^2 + \Tr^2d\Omega^2\\
       &  + \left(2-3\sin^2\Ttheta\right)A(\Tt,\Tr)dr^2\\[5pt]
       &  - \left( A(\Tt,\Tr) - 3(\sin^2\Ttheta)C(\Tt,\Tr) \right)r^2 d\Ttheta^2\\[5pt]
       &  - \left( A(\Tt,\Tr) + 3(\sin^2\Ttheta)\left(C(\Tt,\Tr)-A(\Tt,\Tr)\right)
            \right)\Tr^2\sin^2\Ttheta d\Tphi^2\\[5pt]
       &  - 6r\left(\sin\Ttheta\cos\Ttheta\right)B(\Tt,\Tr) d\Tr d\Ttheta
   \end{split}
\end{align}
where
\begin{align}
   \label{eqn:funA}
      A(\Tt,\Tr) &= \frac{3}{\Tr^5} \left(  \Tr^2F^{(2)}
                                          -3\Tr F^{(1)}
                                          +3F\right)\\[5pt]
   \label{eqn:funB}
      B(\Tt,\Tr) &= \frac{-1}{\Tr^5}\left( -\Tr^3F^{(3)}
                                          +3\Tr^2F^{(2)}
                                          -6\Tr F^{(1)}
                                          +6F\right)\\[5pt]
   \label{eqn:funC}
      C(\Tt,\Tr) &= \frac{1}{4\Tr^5}\left(   \Tr^4F^{(4)}
                                           -2\Tr^3F^{(3)}
                                           +9\Tr^2F^{(2)}
                                          -21\Tr F^{(1)}
                                          +21F\right)\\[5pt]
   \label{eqn:funF}
      F^{(n)}&= \frac{1}{2} \left( \frac{d^nQ(\Tt+\Tr)}{d\Tr^n}
                                  -\frac{d^nQ(\Tt-\Tr)}{d\Tr^n}\right)
\end{align}
and where $Q(x)$ is an arbitrary function of $x$. Note that this form of the metric differs
slightly from that given by Teukolsky. Here the function $F$ has been expressed as an
explicit combination of ingoing and outgoing waves (thus ensuring time symmetric initial
data). Note also that the derivatives of $F$ are taken with respect to $\Tr$ rather than $x$
as used by Teukolsky. Consequently, the signs of the odd-derivatives of $F$ in the
expressions for $A$, $B$ and $C$ have been flipped.

Following Baumgarte and Shapiro~\cite{baumgarte:1998-01}, the function $Q(x)$ was chosen to
be
\begin{align}
   Q(x) = a x e^{-x^2}\qquad\text{with } a > 0
   \label{eqn:funQ}
\end{align}
as this produces initial data describing a compact wave centred on the origin with a wave
amplitude controlled by the parameter $a$.

Note that the metric (\ref{eqn:TeukPolarMetric}) is not an exact solution of the vacuum
Einstein equations but rather a solution of the linearised equations in the sense that
$G_{ab}(g) = \BigO{a^2}$.

This form of the metric requires some care when setting the initial data near $\Tr=0$ (where
the coordinates are singular). A better choice is to express the metric in standard Cartesian
coordinates. At the moment of time symmetry, $\Tt=0$, the Cartesian components, $h_{ij}$, of
the 3-metric are given by
\bgroup
\def\P{\phantom{{}+{}}}
\def\M{{}-{}}
\begin{align}
   h_{\Tx\Tx}&=1 - 24 a \left( 1+(\Tr^2-4) \Ty^2-\Tx^2 \Tz^2 \right)
              e^{-\Tr^2}\label{eqn:hxx}\\[5pt]
   h_{\Ty\Ty}&=1 - 24 a \left( 1+(\Tr^2-4) \Tx^2-\Ty^2 \Tz^2 \right)
              e^{-\Tr^2}\label{eqn:hyy}\\[5pt]
   h_{\Tz\Tz}&=1 + 24 a \left( (\Tx^2 + \Ty^2 - 2)^2 - 2 \right)
              e^{-\Tr^2}\label{eqn:hzz}\\[5pt]
   h_{\Tx\Ty}&=\P24 a \Tx \Ty \left( \Tr^2 + \Tz^2 - 4 \right)
              e^{-\Tr^2}\label{eqn:hxy}\\[5pt]
   h_{\Tx\Tz}&=\M24 a \Tz \Tx \left( \Tx^2 + \Ty^2 - 2 \right)
              e^{-\Tr^2}\label{eqn:hxz}\\[5pt]
   h_{\Ty\Tz}&=\M24 a \Ty \Tz \left( \Tx^2 + \Ty^2 - 2 \right)
              e^{-\Tr^2}\label{eqn:hyz}
\end{align}
\egroup
where $\Tr=(\Tx^2+\Ty^2+\Tz^2)^{1/2}$.

The 3-dimensional lattice was built by a simple generalisation of the 2-dimensional lattice
used for the Brill waves. The grid was built from a set of $N_x\times N_y\times N_z$ equally
spaced points in a the 3-dimensional volume bounded by $\vert x\vert=\vert y\vert=\vert
z\vert=5$. The points were then identified as the vertices of the lattice while on each of
the $\Tx\Ty$, $\Tx\Tz$ and $\Ty\Tz$ planes, legs were added in exactly the same pattern as
for the 2-dimensional Brill lattice, recall figure (\ref{fig:BrillLattice}). Consequently
many of the ideas discussed in regard to the Brill lattice carry over to the this lattice.
Initial data for the coordinates and leg-lengths were assigned by integrating the geodesic
equations as two-point boundary problems for each leg of the lattice (this was time
consuming but only needed to be done once). The outer boundary conditions were exactly as
per equation (\ref{eqn:BrillOuterBC}) but on this occasion applied to all six faces of the
lattice. Geodesic slicing was used (i.e., zero shift and unit lapse) and as there are no
symmetries, the full set of evolution equations (\ref{eqn:DotKabA}--\ref{eqn:DotKabF}) and
(\ref{eqn:DotRabcdA}--\ref{eqn:DotRabcdN}) were used (see also Appendix
(\ref{sec:FullRiemannEqtns})). The implementation of the numerical dissipation is in this
case slightly different to that for the 2-dimensional lattice. The appropriate version of
(\ref{eqn:BrillKODissip}) for the 3-dimensional lattice is
\begin{align}
   \frac{dY}{dt} = \left(\frac{dY}{dt}\right)_{\epsilon=0}
                 + \epsilon \left(-6Y_o + \sum_i Y_i\right)
\end{align}
where the sum on the right hand side includes contributions from the 6 immediate neighbouring
cells. The term in the second set of brackets in this expression is an approximation to
$\BigO{L^2}\nabla^2 Y$ and thus will converge to zero on successively refined lattices.

Since the Teukolsky space-time carries no symmetries it follows that none of the
constraints (\ref{eqn:EinsConstA}--\ref{eqn:BianConstD}) will be trivially satisfied
throughout the evolution. Including results for all 10 constraints is somewhat of an
overkill so results will be presented (in section (\ref{sec:TeukolskyResults})) for just
the Hamiltonian constraint, namely,
\begin{align}
   \label{eqn:TeukConstC1}
   0 = C_1 = \Rxyxy+\Rxzxz+\Ryzyz
\end{align}

\section{Cactus}
\label{sec:CactusCode}

The combination of the open source code Cactus~\cite{goodale:2002-01} and the Einstein
Toolkit~\cite{loffler:2011-01} (collectively referred to here as the Cactus code) provide a
well understood framework for computational general relativity. The Cactus code was used
largely \emph{out of the box} but with some simple extensions for setting the initial data
for the Brill and Teukolsky space-times. A new thorn was written for the Brill space-time to
set the initial data from the discretised metric provided by the same multigrid code used to
set the lattice initial data. For the Teukolsky metric the {\tt EinsteinInitialData/Exact}
thorn was extended to include the exact 3-metric given in equations
(\ref{eqn:hxx}--\ref{eqn:hyz}). These changes were made to ensure that the lattice and
Cactus evolutions were based on exactly the same initial data.

The Cactus initial data were built over the same domain as used in the corresponding lattice
initial data. The initial data were integrated using the standard BSSN and ADM thorns. The
BSSN thorn used a 4th order Runge-Kutta integrator and artificial dissipation was applied to
all dynamical variables with a dissipation parameter equal to $0.1$. The ADM integrations
used a two-step iterated Crank-Nicholson scheme without artificial dissipation. The time
step in each case was chosen to ensure a Courant factor of $1/8$.

The Cactus code does not provide values for the components of either the 3 or 4
dimensional Riemann tensor. However the spatial components, such as $\TRxyxy$, can be
reconstructed from the 3 dimensional components of the Ricci tensor and metric using a
combination of the Gauss-Codazzi equations
\begin{align}
   \bot \TRabcd = {}^3 \TRabcd + \TKac\TKbd - \TKad\TKbc
\end{align}
and the equation
\begin{align}
   {}^3\TRabcd =   \TRac\Thbd-\TRad\Thbc
                 + \Thac\TRbd-\Thad\TRbc
                 - \frac{\TR}{2} \left(\Thac\Thbd-\Thad\Thbc\right)
\end{align}
where $\Thab$ is the 3-metric, $\TRab$ is the 3-Ricci tensor and $\TR = \uThab \TRab$.

Since the Cactus and lattice data are expressed in different frames some post-processing of
the data is required before the two sets of data can be compared. There are two aspects to
this, first, mapping points between the respective spaces (e.g., given a point in the Cactus
coordinates what is the corresponding point in the lattice?) and second, comparing the data
at those shared points. Recall that when constructing the initial data for the Brill and
Teukolsky lattices, the vertices of the lattice were taken as the uniformly distributed grid
points in the Brill and Teukolsky coordinates. This correspondence is preserved throughout
the evolution by the zero shift condition. This is not the case for the Gowdy space-time
where the initial data was constructed on an unequally spaced grid (see section
(\ref{sec:GowdyLattice})) while in contrast the Cactus code uses an equally spaced grid. In
this case the conversion of tensor components, such as $\TRabcd$, from the Cactus data into
a form suitable for comparison with the lattice data entails two steps, first, the tensor
is projected onto a local orthonormal frame, second, the radial $\Tz$ coordinate is
converted to a radial proper distance $\Ts$. Since the Gowdy metric is diagonal the
projection onto the coordinate aligned orthonormal frame is trivial, for example $\Rxyxy =
\Thxx \Thyy \TRxyxy$, while the proper distance between successive grid points can be
computed by
\begin{align}
   \Delta \Ts_{ii+1} = \int_i^{i+1} \sqrt{\Thzz}\>d\Tz
\end{align}
where the limits $(i,i+1)$ are understood to represent the corresponding grid points.
The integral was estimated by a cubic polynomial based on the grid points
$(i-1,i,i+1,i+2)$.

\section{Results}
\label{sec:Results}

The evolution equations for the Brill and Teukolsky lattices were integrated using a 4th
order Runge-Kutta routine with a fixed time step $\delta t$ chosen to satisfy a Courant
condition of the form $\delta t < C\delta L$ where $\delta L$ is the shortest leg-length on
the lattice and where $C$ is a Courant factor with $0<C<1$. The same integration scheme was
used for the Gowdy lattice apart from one small change where the Courant condition was based
upon $N\delta t < C{\rm min}(\Lzz)$ where $N$ is the largest lapse on the lattice. This
Courant condition uses the shortest $\Lzz$ for the simple reason that the evolution
equations (\ref{eqn:DotGowdyLxx},\ref{eqn:DotGowdyLyy}) for $\Lxx$ and $\Lyy$ admit a
re-scaling of $\Lxx$ and $\Lyy$ and thus their values can not influence $\delta t$.

A trial and error method was first used to find any time step that yielded a stable
evolution (despite the cost). This allowed a more informed judgement to made by careful
examination of the history of the leg-lengths. Thus for the Gowdy lattices the time step was
chosen as $\delta t = 0.0512/N_z$ corresponding to a Courant factor of $1/20$, while for the
Brill and Teukolsky lattices the time step, with $C=1/8$, was set by $\delta t =
1.25/(N_z-1)$.

\subsection{Gowdy}
\label{sec:GowdyResults}

There are two obvious tests that can be applied to the lattice data, first, a comparison
against the exact data and, second, a comparison against numerical results generated by the
Cactus code. Other tests that can be applied include basic convergence tests as well as
observing the behaviour of the constraints.

The initial data for the lapse was chosen according to the comparison being made. The
comparisons with the Cactus data were based on a unit lapse, $N=1$, while the comparisons
with the exact solution used initial values taken from the exact solution, $N=e^\lambda/4$
at $\Tt=1$.

The dissipation parameter $\epsilon$ (see equation (\ref{eqn:GowdyKODissip})) was set equal
to $0.8$ (which was found by trial and error as the smallest value that ensured good
stability for the $1+\log$ lapse). The integral in equation (\ref{eqn:DotGowdyLzz}) was
estimated using a 4th order interpolation built from 5 cells centred on this leg.

Selected results can be seen in figures (\ref{fig:GowdyA}--\ref{fig:GowdyE}) and show that
the lattice method works well with excellent agreement against the exact and numerical
solutions. Note that since the lattice expands by factors of order 100, the $\Lzz$ have been
uniformly scaled to squeeze the lattice into the range $[-0.5,0.5]$. Figure
(\ref{fig:GowdyA}) shows a comparison of the original and scaled data. Figures
(\ref{fig:GowdyD},\ref{fig:GowdyE}) show the behaviour of selected constraints as well as
basic convergence tests.

\subsection{Brill}
\label{sec:BrillResults}

The results for the Brill initial data are shown in figures
(\ref{fig:BrillProfile05}--\ref{fig:BrillDissip}). In all cases the dissipation parameter
$\epsilon$ for the lattice was set equal to $1.0$ (except as noted in figure
(\ref{fig:BrillDissip})). The Cactus BSSN data was computed on a full 3-dimensional grid and
thus there is no reason to expect any instabilities on the symmetry axis. This allows a much
small dissipation parameter, $\epsilon=0.1$, to be used for the BSSN evolutions. The Cactus
ADM thorn does not appear to support any form of Kreiss-Oliger numerical dissipation.

The expected behaviour for the Brill wave is that the curvature will be propagated away from
the symmetry axis with the wave hitting the edges of the outer boundary by about $t=5$
followed by the four corners by about $t=7$ and will completely cross the boundary by about
$t=10$. As the wave moves across the grid it should leave zero curvature in its wake (though
the extrinsic curvatures need not return to zero).

The results for all three methods at $t=5$ are shown in figure (\ref{fig:BrillProfile05})
where it is clear that though there is some good agreement in the propagation of the main the
wave there are also some notable differences. The ADM method shows a series of parallel
waves propagating in from the outer boundary towards the symmetry axis (such waves will later
be referred to as boundary waves, these waves are particular evident in movies from $t=0$ to
$t=10$) while the BSSN data shows a non-propagating bump close to the origin. In contrast the
lattice data shows a smooth behaviour in the wave with no apparent boundary waves nor any
sign of a bump. By $t=10$ (see figure (\ref{fig:BrillProfile10})) the ADM data shows not only
the boundary waves but also reflected waves from the outer boundary. Similar reflected waves
can also be seen in the BSSN results though with a significantly smaller amplitude. The bump
in the BSSN data has remained in place and has grown in amplitude. The lattice data shows no
signs of reflection but there is a very small bump that correlates with the wings of the BSSN
bump.


It is reasonable to ask why the three methods should give such different results in the
region behind the main wave. The smooth profile in the lattice data might be due to the
large dissipation parameter compared to that used in the ADM and BSSN data. The boundary
waves in the ADM data are clearly associated with the boundary conditions while the cause of
the bump in the BSSN data is not so easy to identify from these plots. A more detailed
analysis will be given later when discussing the Teukolsky data where similar behaviour
was observed.

The effects of changing dissipation parameter on the evolution of the lattice data is shown
in figure (\ref{fig:BrillDissip}). This shows clearly how crucial the numerical dissipation
is in controlling the instabilities. The figure also shows that despite the significant
dissipation ($\epsilon=1.0$) required to suppress the axis instability, the broad features of
the main wave are largely unaffected.

Figure (\ref{fig:BrillConstC1C4}) shows the behaviour of the constraints $C_1$
(\ref{eqn:BrillConstC1}) and $C_4$ (\ref{eqn:BrillConstC4}) over the period $t=0$ to $t=10$.
The remaining three constraints are not shown as they show much the same behaviour. Each plot
contains four curves corresponding to different lattices scales, $N_z=101$ (red), $N_z=201$
(blue), $N_z=401$ (green) and $N_z=801$ (black). These show that the constraints appear to
decrease as $N_z$ is increased. It also appears that the constraints settle to a non-zero
value as $t$ increases. This could be due to truncation errors inherent in the solution of
the Hamiltonian equation (\ref{eqn:BrillH}) coupled with the interpolation to the lattice
(though this claim was not tested). The two bumps in the left figure, one just after $t=5$
and one close to $t=10$ are most likely due to reflections from the outer boundary (this too
was not tested).

\subsection{Teukolsky}
\label{sec:TeukolskyResults}

The Teukolsky data is specified on a full 3-dimensional grid/lattice and is thus not
susceptible to the axis instability seen in the Brill data. This allows for a much smaller
dissipation parameter to be used for the lattice, ADM and BSSN codes, in this case
$\epsilon=0.1$.

The results for the Teukolsky initial data are shown figures
(\ref{fig:TeukProfile05}--\ref{fig:TeukBCTest}) and bear some similarities with the results
for the Brill initial data. However, in this case the boundary and reflected waves appear to
be much less noticeable while the bump in the BSSN data is still present and is more
pronounced than in the Brill wave data.

The plots in figure (\ref{fig:TeukBump}) show that the bump in the BSSN data is a numerical
artefact. The figure shows that as the spatial resolution is decreased (i.e., increasing
$N_z$) the amplitude of the bump, at $t=5$, decreases. The figure also shows that the
amplitude of the bump grows with time. No attempt was made to determine the source of the
bump.

In order to better understand the influence of the outer boundary condition on the evolution
it was decided to run the lattice, ADM and BSSN codes on two different sets of initial data,
each with the same spatial resolution but with one grid twice the size of the other (i.e.,
one grid had boundaries at $\pm5$ and the other at $\pm10$). The influence of the outer
boundary condition on the evolution was then be measured by comparing the evolution on the
common region. The results are shown in figure (\ref{fig:TeukBCTest}). The right panel shows
the evolution of $\Rxyxy$ on the lattice on both grids with $N_z=101$ for the red curve and
$N_z=201$ for the blue curve. Notice how the red curve lies entirely on top of the blue
curve even as the wave passes through the $\pm5$ boundary. The left panel shows the
difference in $\Rxyxy$ between the two grids for the lattice data (red curve) and for the
BSSN data (green curve, using $N_z=100$ and $N_z=200$). This shows clearly that the boundary
waves for both methods are present well before the main wave hits the boundary. It also
shows that the amplitude for the BSSN data is much larger than for the lattice data. Note
also that the boundary waves do not propagate very far into the grid (in stark contrast to
the ADM Brill waves). By $t=10$ the main wave has left the smaller grid and the data in the
left panel describes a mix of waves dominated by the reflected waves. This figure also shows
that the BSSN data contains a long wavelength mode while the waves in the lattice data are
much smaller in amplitude and are dominated by high frequency modes (which are rapidly
suppressed by the numerical dissipation).

The evolution of the Hamiltonian constraint (\ref{eqn:TeukConstC1}) is shown in figure
(\ref{fig:TeukConstC1}). The linear growth in the constraint for the BSSN data is due solely
to the growth of the BSSN bump at the origin. The sharp rise in the constraint for the
lattice data for $N_z=201$ is due to the onset of a small instability in the lattice near
the origin. This can also be seen in the small bump in the lower right plot of figure
(\ref{fig:TeukBCTest}). This instability can be suppressed by increasing the dissipation
parameter but at the expense of compromising the quality of the evolution. The source of
this instability is thought to be due to the residual extrinsic curvatures driving the
lattice vertices in different directions leading to distorted computational cells that break
the near-planar assumptions built into the derivation of equations
(\ref{eqn:FinalMabcEqtns}). This is an important issue for the viability of the lattice
method and will be explored in more detail in subsequent work.

\section{Discussion}
\label{sec:Discuss}

The passage of the waves through the outer boundaries appear to be better handled by the
lattice method than both the ADM and BSSN methods. This is particularly true for the Brill
waves but less so for the Teukolsky waves. It is reasonable to ask if this is a generic
feature of the lattice method and if so, then which features of the lattice method gives rise
to this result? An argument can be made that this behaviour may well be germane to the
lattice method. The basis of the argument is the simple observation that in any small region
of space-time covered by Riemann normal coordinates the first order coupled evolution
equations for the Riemann curvatures (\ref{eqn:DotRabcdA}--\ref{eqn:DotRabcdN}) can be
de-coupled to second order equations in which the principle part is the wave
operator\footnote{This is shown in detail in section 4.3 and 4.4 of~\cite{brewin:2010-03}
but note that the author failed to explicitly state that all computations were for the
principle part of the equations.}. That is, for each Riemann component such as $\Rxyxy$,
\begin{align}
   R_{xyxy,tt} = R_{xyxy,xx} + R_{xyxy,yy} + R_{xyxy,zz} + \BigO{R^2}
\end{align}
where the term $\BigO{R^2}$ is a collection of terms quadratic in the $\Rabcd$. The natural
outgoing boundary condition for this wave equation is the Sommerfeld condition as per
equation (\ref{eqn:BrillOuterBC}). Thus it is not surprising that the lattice method works as
well as it does. This result is a direct consequence of the use of Riemann normal
coordinates. In a generic set of coordinates the principle part would not be the wave
operator.

As encouraging as the results may appear to be there remain many questions about the method.
How does it behave for long term integrations? What are its stability properties? How can it
be extended to higher order methods? How can mesh refinement be implemented? How well does
it work on purely tetrahedral meshes? How well does it work for non-unit lapse functions?
How can black holes be incorporated into a lattice (punctures or trapped surfaces?) and how
would these holes move through the lattice? How can energy flux, ADM mass and other
asymptotic quantities be computed on a lattice?

These are all important question and must answered before the lattice method can be
considered for serious work in computational general relativity. These questions will be
addressed in later papers.

\appendix

\section{The transition matrices}
\label{sec:TransMatrix}

The transition matrices play a central role in the computation of the derivatives such as
$R_{xyxy,z}$. They are used to import data from neighbouring cells so that the vertices of a
chosen cell are populated with data expressed in the frame of that cell. A finite difference
estimate can then be made for the required partial derivatives.

The purpose of this appendix is to extend the approach given in~\cite{brewin:2014-01}. In
that paper particular attention was paid to the form of the transition matrix for a cubic
lattice. It was argued that, with sufficient refinement of the lattice, the transition
matrices should vary smoothly across the lattice and should converge to the identity matrix
in the continuum limit\footnote{Both of these conditions apply to cubic lattices but need
not apply for other lattices.}.

The particular feature of the cubic lattice that makes it attractive for our purposes is
that it is easily sub-divided in a manner that preserves its original structure. This allows
a whole family of cubic lattices to be constructed, with arbitrarily small cells, and
thus it is easy to investigate the continuum limit of the lattice.

For a vertex $p$ with neighbour $q$ the transition matrix\footnote{There is one such matrix
for each pair $(p,q)$. In this paper the transition matrix will be denoted by $M$ rather
than $M(p,q)$ as used in~\cite{brewin:2014-01}.} $M$ allows data such as $v^\alpha_{q\Bq}$
to be imported from $\Bq$ to $\Bp$ via
\begin{align}
   \label{eqn:TransMabA}
   v^\alpha_{q\Bp} = \Mab v^\beta_{q\Bq}
\end{align}

When constructing a frame within a cell there is considerable freedom in locating the origin
and orientation of the coordinate axes. A simple and natural choice is to locate the origin
on the central vertex and to align the coordinate axes with various sub-spaces of the cell
(e.g., align the $x$-axis to the leg $(0,1)$, the $y$-axis to the plane spanned by the legs
$(0,1)$ and $(0,2)$ etc.).

Without further information about the relationship of one cell to another little can be said
about the corresponding transition matrices. However, for the cubic lattice it is not
hard to see that the frames for a typical pair of cells can be chosen so that the transition
matrix will be of the form
\begin{align}
   \label{eqn:TransMabB}
   \Mab = \dab + \mab + \BigO{L^2}
\end{align}
where $\mab = \BigO{L}$ are determined from the data in the pair of cells (i.e., the
coordinates and leg-lengths). This form of $M$ ensures that it converges to the identity
matrix in the continuum limit (e.g., by successive refinements of the cubic lattice).
Note that the $\mab$ must be subject to a constraint since the resulting transition matrix
must preserve scalar products. That is, for any pair of vectors $u$ and $v$,
\begin{align}
   v_{\alpha q\Bp} u^\alpha_{q\Bp} = v_{\alpha q\Bq} u^\alpha_{q\Bq}
\end{align}
which leads immediately to
\begin{align}
   \label{eqn:AntisymmMab}
   0 = \cmab + \cmba
\end{align}
This shows that the $\cmab$ define a skew-symmetric $4\times4$ matrix determined by just six
independent entries (corresponding to three boosts and three rotations).

The $\mab$ were computed in~\cite{brewin:2014-01} by applying (\ref{eqn:TransMabA}) to a
specially chosen set of vectors. A different approach will be taken in this paper, one that
will be seen to be more in the spirt of Cartan's method of local frames (see Appendix
\ref{sec:CartanEqtns}).

First recall that the lattice is assumed to be a discrete approximation to some possibly
unknown smooth geometry. Thus it is reasonable to requite that the $\mab$ should also be
smooth functions across the lattice. This allows the $\cmab$ to be expanded as a Taylor
series based on the vertex $p$. That is
\begin{align}
   \cmab = \cmabc x^\gamma_{q\Bp} + \BigO{L^2}
\end{align}
for some set of coefficients $\cmabc$.

Now consider a closed path such as that defined by the four vertices $o,a,b,c$ in figure
(\ref{fig:CartanConn}). Clearly
\begin{align}
   \label{eqn:VectorLoop}
   0 = v^\alpha_{oa\Bo} + v^\alpha_{ab\Bo} + v^\alpha_{bc\Bo} + v^\alpha_{co\Bo}
\end{align}
where $v^\alpha_{pq\Br}$ are defined by $v^\alpha_{pq\Br} =
x^\alpha_{q\Br}-x^\alpha_{p\Br}$ and $x^\alpha_{q\Br}$ are the coordinates of vertex $q$ in
the frame $\Br$. However, the vector joining vertices $a$ to $b$ can also be expressed in
terms of the frame $\Ba$. Likewise, the vector joining $b$ to $c$ can be expressed in terms
of the frame $\Bc$. Using the transformation law given by (\ref{eqn:TransMabA}) leads to
\begin{align}
   v^\alpha_{ab\Bo} &= v^\alpha_{ab\Ba} + \mabc v^\beta_{ab\Ba} v^\gamma_{oa\Bo}\\
   v^\alpha_{bc\Bo} &= v^\alpha_{bc\Bc} + \mabc v^\beta_{bc\Bc} v^\gamma_{oc\Bo}
\end{align}
Substituting this pair of equations into (\ref{eqn:VectorLoop}) leads to
\begin{align}
   \label{eqn:SimpleMabcEqtns}
   v^\alpha_{oa\Bo} + v^\alpha_{ab\Ba} + v^\alpha_{bc\Bc} + v^\alpha_{co\Bo}
   =
   \mabc\left(-v^\beta_{ab\Ba} v^\gamma_{oa\Bo}
              -v^\beta_{bc\Bc} v^\gamma_{oc\Bo}\right)
\end{align}
This construction can be applied to each of the 6 coordinate planes leading to 24 equations
for the 24 unknowns $\mabc$. In the cases of a lattice that evolves continuously in time it
is possible (see Appendix \ref{sec:TimeCompsMabc}) to solve these equations for 15 of the
$\mabc$ in terms of the extrinsic curvatures $\Kij$ and the lapse function $N$. This leaves
just 9 equations (based on the spatial coordinate planes) for the 9 remaining $\mabc$.

Though it is possible to use the above equations (\ref{eqn:SimpleMabcEqtns}) to directly
compute the $\mabc$ doing so might introduce a systematic bias due to the asymmetric
arrangement of the legs relative to the central vertex. An improved set of equations can be
obtained simply by adding together the equations that would arise from each of the four
tiles of figure (\ref{fig:CartanConn}) attached to the central vertex $\Bo$. This leads to
the following set of equations
\begin{align}
   \label{eqn:ImprovedMabcEqtns}
   v^\alpha_{hb\Ba} + v^\alpha_{bd\Bc} + v^\alpha_{df\Be} + v^\alpha_{fh\Bg}
   =
   \mabc\left(-v^\beta_{hb\Ba} v^\gamma_{oa\Bo}
              -v^\beta_{bd\Bc} v^\gamma_{oc\Bo}
              -v^\beta_{df\Be} v^\gamma_{oe\Bo}
              -v^\beta_{fh\Bg} v^\gamma_{og\Bo}\right)
\end{align}
Now since each $v^\alpha_{pq\Br} = \BigO{L}$ it follows that the right hand side of
(\ref{eqn:ImprovedMabcEqtns}) is $\BigO{L^2}$ and thus
\begin{align}
   v^\alpha_{bd\Bc} + v^\alpha_{fh\Bg} &= \BigO{L^2}\\
   v^\alpha_{hb\Ba} + v^\alpha_{df\Be} &= \BigO{L^2}
\end{align}
which allows the terms $v^\alpha_{fh\Bg}$ and $v^\alpha_{df\Be}$ on the right hand side of
\ref{eqn:ImprovedMabcEqtns} to be replaced by their counterparts leading to
\begin{align}
   v^\alpha_{hb\Ba} + v^\alpha_{bd\Bc} + v^\alpha_{df\Be} + v^\alpha_{fh\Bg}
   =
   \mabc\left(-v^\beta_{hb\Ba} v^\gamma_{ea\Bo}
              -v^\beta_{bd\Bc} v^\gamma_{gc\Bo}\right)
\end{align}
Finally note that
\begin{align}
   v^\alpha_{hb\Ba} &= \phantom {-} v^\alpha_{gc\Bo} + \BigO{L}\\
   v^\alpha_{bd\Bc} &= - v^\alpha_{ea\Bo} + \BigO{L}
\end{align}
and therefore
\begin{align}
   \label{eqn:FinalMabcEqtns}
   v^\alpha_{hb\Ba} + v^\alpha_{bd\Bc} + v^\alpha_{df\Be} + v^\alpha_{fh\Bg}
   =
   - \mabc\left( v^\beta_{gc\Bo} v^\gamma_{ea\Bo}
                -v^\beta_{ea\Bo} v^\gamma_{gc\Bo}\right)
\end{align}
These are the equations that were used in the computer code to compute the $\mabc$.

\section{Cartan structure equations}
\label{sec:CartanEqtns}

Equations (\ref{eqn:AntisymmMab}) and (\ref{eqn:FinalMabcEqtns}) bear a striking similarity
to the Cartan structure equations\footnote{Latin indices will be used in this appendix to
denote frame components (rather than spatial indices). This follows standard notation for
differential forms.}
\begin{align}
   0 &= \omega_{ij} + \omega_{ji}\label{eqn:FirstCartan}\\
   d\omega^i &= - \omega^i{}_j \wedge \omega^j\label{eqn:SecondCartan}
\end{align}
in which $\omega^i$ are the basis 1-forms, $\omega^i{}_j$ are the connection 1-forms and
where the metric is given by $g = g_{ij} \omega^i \omega^j$ with $g_{ij} = \diag (-1,1,1,1)$.

The purpose of this appendix is to show how equations (\ref{eqn:AntisymmMab}) and
(\ref{eqn:FinalMabcEqtns}) can be obtained from the Cartan structure equations
(\ref{eqn:FirstCartan}) and (\ref{eqn:SecondCartan}).

To start the ball rolling, note that equations (\ref{eqn:AntisymmMab}) and
(\ref{eqn:FirstCartan}) agree upon choosing $m^i{}_j = \omega^i{}_j$. Showing that the
remaining pair of equations (\ref{eqn:FinalMabcEqtns}) and (\ref{eqn:SecondCartan}) agree
requires a bit more work. Start by integrating (\ref{eqn:SecondCartan}) over the tile $R$
defined by the vertices $b,d,f,h$ in figure (\ref{fig:CartanConn})
\begin{align}
   \int_R d\omega^i = - \int_R \omega^i{}_{jk} \omega^k \wedge \omega^j
\end{align}
where $\omega^i{}_j$ has been expanded as $\omega^i{}_{jk} \omega^k$.
This equation can be re-written using Stoke's theorem as
\begin{align}
   \label{eqn:IntegratedCartan}
   \int_{\partial R} \omega^i = - \int_R \omega^i{}_{jk} \omega^k \wedge \omega^j
\end{align}
The path integral on the left can be split into four pieces, one the four edges of the
tile. On each edge
set $\omega^i = dx^i$ where $x^i$ are the local Riemann normal coordinates appropriate to
the edge (e.g., along the edge $(b,d)$ use the coordinates of frame $\Bc$). Thus
\begin{align}
   \int_{\partial R} \omega^i
   = \sum_{(p,q)\in \partial R} \int_p^q dx^i
   = v^i_{hb\Ba} + v^i_{bd\Bc} + v^i_{df\Be} + v^i_{fh\Bg}
\end{align}
where $v^i_{pq\Br} = x^i_{q\Br}-x^i_{p\Br}$. The area integral on the
right hand side of (\ref{eqn:IntegratedCartan}) can be estimated to leading order in
the length scale $L$ by
approximating $\omega^i{}_{jk}$ by its value at the vertex $o$. Thus
\begin{align}
   \int_R \omega^i{}_{jk} \omega^k \wedge \omega^j
        = \omega^i{}_{jk\Bo} \int_R \omega^k \wedge \omega^j + \BigO{L^3}
\end{align}
and noting that the integrand on the right is just the area 2-form for the tile leads to
the estimate
\begin{align}
   \int_R \omega^i{}_{jk} \omega^k \wedge \omega^j
        = \omega^i{}_{jk\Bo} \left(  v^k_{ea\Bo} v^j_{gc\Bo}
                                   - v^k_{gc\Bo} v^j_{ea\Bo}\right) + \BigO{L^3}
\end{align}
The integrated form of the Cartan equation (\ref{eqn:IntegratedCartan}) can now be
re-written as
\begin{align}
   v^i_{hb\Ba} + v^i_{bd\Bc} + v^i_{df\Be} + v^i_{fh\Bg}
   =
   - \omega^i{}_{jk\Bo}  \left(  v^j_{gc\Bo} v^k_{ea\Bo}
                               - v^j_{ea\Bo} v^k_{gc\Bo} \right) + \BigO{L^3}
\end{align}
which agrees (apart from the Greek/Latin indices), to leading order in $L$, with
(\ref{eqn:FinalMabcEqtns}) provided $m^i{}_{jk} = \omega^i{}_{jk\Bo}$.

\section{Source terms}
\label{sec:SourceTerms}

A lattice would normally consist of a finite number of local frames, one for each central
vertex. But there is nothing to stop the construction of a local frame at every point in the
lattice. The new frames could be introduced by any rule but for a smooth lattice it is
reasonable to require that the frames vary smoothly across the lattice. This will certainly
be the case when the transition matrices are of the form
\begin{align}
   \Mab (x) = \dab + \mabc x^\gamma
\end{align}
The addition of these extra frames makes it easier to discuss differentiation on the lattice.

Consider a cell $p$ and some point $q$ within that cell. Let $v^\alpha$ be the components of
a typical vector at $q$ expressed in the local frame of $q$, that is $v^\alpha_q =
v^\alpha_{q\Bq}$. The components of the vector in the frame $\Bp$ would then be given by
$M^\alpha{}_{\beta q\Bp} v^\alpha$. This allows the derivatives of $v^\alpha$ at $p$ and in
$\Bp$ to be computed as follows
\begin{align}
   v^\alpha_{,\gamma}
   =
   v^\alpha_{,\gamma p\Bp}
   &= \left(\Mab v^\beta\right)_{,\gamma p}\\
   &= \Mabcp v^\beta_p + \Mabp v^\beta_{,\gamma p}\\
   \label{eqn:DerivVa}
   &= \mabc v^\beta_p + v^\alpha_{,\gamma p}
\end{align}

At this point there is a slight problem with the notation. The last term on the right hand
side above is a derivative of $v^\alpha$ formed from the raw point values of the $v^\alpha$.
That derivative takes no account of the transition matrices and thus is not the partial
derivative (indeed the partial derivative is the term on the left hand side). To emphasise
this distinction the following notation will be used. Define a new derivative operator $\D$
by\footnote{But note that mixed $\D$ derivatives need not commute.}
\begin{align}
   v^\alpha_{\D\gamma}
   =
   v^\alpha_{,\gamma p}
\end{align}
Then the equation (\ref{eqn:DerivVa}) can be written as
\begin{align}
   v^\alpha_{,\gamma} = v^\alpha_{\D\gamma} + \mabc v^\beta
\end{align}
where it is understood that all terms are evaluated at $p$ and in $\Bp$.
By following a similar line of reasoning it is not hard to see that, for example,
\begin{align}
   v_{\alpha,\gamma} &= v_{\alpha:\gamma} - m^\beta{}_{\alpha\gamma} v_{\beta}\\
   R_{\alpha\beta,\gamma} &= R_{\alpha\beta:\gamma}
                           - m^\rho{}_{\alpha\gamma} R_{\rho\beta}
                           - m^\rho{}_{\beta\gamma} R_{\alpha\rho}
\end{align}
As a consistency check it is rather easy to see that applying this notation to
$0=g_{\alpha\beta;\gamma} = g_{\alpha\beta,\gamma}$ leads directly to equation
(\ref{eqn:AntisymmMab}). To see that this is so first note that $g_{\alpha\beta q\Bq} =
\diag(-1,1,1,1)$ at every vertex $q$ and thus the derivatives $g_{\alpha\beta\D\gamma}$ are
zero everywhere. This leads immediately to equation (\ref{eqn:AntisymmMab}).

It should be noted that the hessian of lapse $\dNij$ could be computed entirely from data
within a single frame or by sharing data, such as $\dNi$, between neighbouring frames. In
the later case some care must be taken when computing terms like $N_{\D x\D y}$ since the colon
derivatives need not commute.\footnote{For example, $N_{,x,y} =
N_{\D x\D y}-\myxy\dNy-\mzxy\dNz$ while $N_{,y,x} = N_{\D y\D x}-\mxyx\dNx-\mzyx\dNz$ and as
$N_{,x,y}=N_{,y,x}$ it follows that $N_{\D x\D y}-N_{\D y\D x} =
\myxy\dNy+\mzxy\dNz-\mxyx\dNx-\mzyx\dNz$ which in general will not be zero.}

\section{The time components of $\mabc$}
\label{sec:TimeCompsMabc}

In a lattice that is discrete in both space and time there would be 24 distinct $\mabc$ in
each computational cell. However, in the case of a continuous time lattice with a zero shift
vector at each central vertex, 15 of the 24 $\mabc$ can be expressed in terms of the lapse
function $N$ and the extrinsic curvature $\cKij$, namely
\begin{gather}
   \mijt = 0\\
   \mtij = \mtji = - \cKij\\
   \mtit = (\log N)_{,i}
\end{gather}

The key to this computation will be the application of (\ref{eqn:SimpleMabcEqtns}) to two
carefully chosen tiles, in particular a time-like tile (generated by the evolution of a
spatial leg) and a spatial tile (where all of the vertices lie in one Cauchy surface).

\subsection*{Showing that $\mtij = \mtji$}

Consider a spatial tile in which all of the vertices of the tile lie within one Cauchy
surface, Thus the $t$ component of the various $v^\alpha$ in (\ref{eqn:SimpleMabcEqtns}) are
zero. This leads immediately to
\begin{align}
   0 = (\mtij-\mtji)v^i_{oa\Bo} v^j_{oc\Bo}
\end{align}
where the implied sum over $j$ includes only the spatial terms (since $v^t=0$).
Since this equation must be true for all choices of $v^i_{oa\Bo} v^j_{oc\Bo}$ it follows that
\begin{align}
   \label{eqn:mtijsymm}
   \mtij = \mtji
\end{align}

\subsection*{Showing that $\mtit = (\log N)_{,i}$}

Consider now the time-like tile generated by the leg $(oa)$ as it evolves between a pair of
nearby Cauchy surfaces (as indicated by vertices $(o,a,b,c)$ in figure
(\ref{fig:CartanConn})). The two time-like edges $(oc)$ and $(ab)$ are tangent to the
world-lines normal to the Cauchy surface while the space-like edges $(oa)$ and $(bc)$ are the
two instances of the leg $(ab)$, one at time $t$ the other at $t+\delta t$. Since the
shift vector is assumed to vanish at each central vertex, it follows that
\begin{align}
   v^\alpha_{oc\Bo} &= (N\delta t,0,0,0)^\alpha_{oc\Bo}\\
   v^\alpha_{ab\Ba} &= (N\delta t,0,0,0)^\alpha_{ab\Ba}
\end{align}
Likewise, for the spatial edges the $v^\alpha$ will have a zero $t$ component and thus
will be of the form
\begin{align}
   v^\alpha_{oa\Bo} &= (0,v^x,v^y,v^z)^\alpha_{oa\Bo}\\
   v^\alpha_{cb\Bc} &= (0,v^x,v^y,v^z)^\alpha_{cb\Bc}
\end{align}
for some choice of $v^i_{oa\Bo}$ and $v^i_{cb\Bc}$. With this choice for the $v^\alpha$ and
noting that $N_{ab\Ba}=N_{oc\Bo}+\BigO{L}$, the $t$ component of equation
(\ref{eqn:SimpleMabcEqtns}) is given by
\begin{align}
   (N_{ab\Ba} - N_{oc\Bo}) \delta t
      = (\mtit - \mtti) v^i_{oa\Bo} N_{oc\Bo} \delta t
      + \BigO{L^2\delta t}
\end{align}
Noting that $\mtti=0$ and estimating the left hand side by $N_{,io\Bo} v^i_{oa\Bo}\delta t$
leads to
\begin{align}
   N_{,io\Bo} v^i_{oa\Bo} = N_{oc\Bo}\mtit v^i_{oa\Bo}
\end{align}
and since the $v^i_{oa\Bo}$ are arbitrary, it follows that
\begin{align}
   \mtit = (\log N)_{,i}
\end{align}
in which it is understood that all terms are evaluated at $o$ in the frame $\Bo$.

\subsection*{Showing that $\mtij = -\cKij$}

This computation follows on directly from the previous computation. This time our
attention is on the spatial terms of equation (\ref{eqn:SimpleMabcEqtns}), namely
\begin{align}
   \begin{aligned}
      v^i_{oa\Bo} - v^i_{cb\Bc}
         = (\mijt-\mitj) v^j_{oa\Bo} N_{oc\Bo} \delta t
         + \BigO{L^2\delta t}
   \end{aligned}
\end{align}
Now recall that $v^\alpha_{pq\Br}$ is defined by $v^\alpha_{pq\Br} =
x^\alpha_{q\Br}-x^\alpha_{p\Br}$ and as $x^\alpha_{o\Bo}=x^\alpha_{c\Bc}=0$ it follows that
\begin{align}
   \begin{aligned}
      x^i_{a\Bo} - x^i_{b\Bc}
         = (\mijt-\mitj) x^j_{a\Bo} N_{oc\Bo} \delta t
         + \BigO{L^2\delta t}
   \end{aligned}
\end{align}
and on taking a limit as $\delta t\rightarrow0$ leads immediately to the
evolution equations
\begin{align}
   \label{eqn:DotX}
   -\left(\frac{dx^i}{dt}\right)_{a\Bo}
      = (\mijt-\mitj) x^j_{a\Bo} N_o
      + \BigO{L^2}
\end{align}
for the coordinates $x^i_{a\Bo}(t)$. Now take $d/dt$ of $g_{ij} x^i_{a\Bo} x^j_{a\Bo}$ and
use equation (\ref{eqn:DotLegsA}) to obtain
\begin{align}
   g_{ij} x^i_{a\Bo}\left(\frac{dx^j}{dt}\right)_{a\Bo} &= -N K_{ij} x^i_{a\Bo} x^j_{a\Bo}
\end{align}
which when combined with the above result leads to
\begin{align}
   \cKij x^i_{a\Bo} x^j_{a\Bo}
   = \cmijt x^i_{a\Bo} x^j_{a\Bo}
   - \cmitj x^i_{a\Bo} x^j_{a\Bo}
\end{align}
and as the first term on right vanishes due to $\cmijt = - \cmjit$ the above
can be further simplified to
\begin{align}
   0 = (\cKij + \cmitj) x^i_{a\Bo} x^j_{a\Bo}
\end{align}
But from (\ref{eqn:mtijsymm}), $\cmitj=-\cmtij=-\cmtji=\cmjti$, and as the $x^i_{a\Bo}$ are
arbitrary (since the vertex $a$ can be chosen anywhere in the cell) the previous equation can
only be true provided
\begin{align}
   \cmitj = - \cKij
\end{align}
or equally
\begin{align}
   \mtij = - \cKij
\end{align}

\subsection*{Showing that $\mijt = 0$}

The next task is to show that $\mijt=0$. This is rather easy to do. Having just shown that
$\mtij = -\cKij$ means that equation (\ref{eqn:DotX}) can also be written as
\begin{align}
   -\left(\frac{dx^i}{dt}\right)_{a\Bo}
      = (\mijt+\Kij) x^j_{a\Bo} N_o
      + \BigO{L^2}
\end{align}
which when compared with (\ref{eqn:DotLegsA}) shows that
\begin{align}
   0 = \mijt x^j_{a\Bo} N_o
\end{align}
for any choice of $x^j_{a\Bo} N_o$. This in turn requires $\mijt=0$.

\section{Evolution of $x^i$}
\label{sec:EvolveRNC}

Our aim here is to obtain evolution equations for the spatial coordinates $x^i(t)$ of
each vertex in a computational cell.

To begin, consider
two points $p$ and $q$ chosen arbitrarily in a typical cell. Equation (\ref{eqn:DotX}) can
be applied to this pair of points leading to
\begin{align}
   \label{eqn:DotXp} -\left(\frac{dx^i}{dt}\right)_{p\Bo} = (\mijt-\mitj) x^j_{p\Bo} N_o\\
   \label{eqn:DotXq} -\left(\frac{dx^i}{dt}\right)_{q\Bo} = (\mijt-\mitj) x^j_{q\Bo} N_o
\end{align}
Now combine this pair by contracting (\ref{eqn:DotXp}) with $x^j_{q\Bo}$ and
(\ref{eqn:DotXq}) with $x^j_{p\Bo}$ while noting that $\cmijt=-\cmjit$ to obtain
\begin{align}
   - g_{ij} x^j_{q\Bo} \frac{dx^i_{p\Bo}}{dt}
   - g_{ij} x^j_{p\Bo} \frac{dx^i_{q\Bo}}{dt}
   = \cKij (x^i_{q\Bo}x^j_{p\Bo}+x^j_{q\Bo}x^i_{p\Bo}) N_o
\end{align}
After shuffling terms across the equals sign this can also be re-written as
\begin{align}
   \left(N_o\cKij x^i_{p\Bo} + g_{ij}\frac{dx^i_{p\Bo}}{dt}\right) x^j_{q\Bo}
   =
   - \left(N_o\cKij x^i_{q\Bo} + g_{ij}\frac{dx^i_{q\Bo}}{dt}\right) x^j_{p\Bo}
   \label{eqn:NKijXpXq}
\end{align}
This equation must be true for all choices of $(p,q)$. As the bracketed term on the left
hand side depends only on $p$, that term must match the only $p$ dependent term on the
right hand side, namely the $x^j_{p\Bo}$. Thus it follows that
\begin{align}
   N_o\Kij x^j_{p\Bo} + \frac{dx^i_{p\Bo}}{dt}
      =   \alpha x^i_{p\Bo}\label{eqn:NKijalpha1}\\[5pt]
   N_o\Kij x^j_{q\Bo} + \frac{dx^i_{q\Bo}}{dt}
      = - \alpha x^i_{q\Bo}\label{eqn:NKijalpha2}
\end{align}
for some scalar $\alpha$. But upon setting $p=q$ in (\ref{eqn:NKijXpXq}) it follows that
\begin{align}
   g_{ik} x^k_{p\Bo} \left(N_o\Kij x^j_{p\Bo} + \frac{dx^i_{p\Bo}}{dt}\right) = 0
\end{align}
which when applied to (\ref{eqn:NKijalpha1}) leads to
\begin{align}
   0 = \alpha g_{ij} x^i_{p\Bo} x^j_{p\Bo} = \alpha \Lsqop
\end{align}
and thus $\alpha=0$. This leads immediately to
\begin{align}
   \frac{dx^i_{p\Bo}}{dt} = - N_o\Kij x^j_{p\Bo}
\end{align}
with a similar result for the point $q$. Since the point $p$ is arbitrary it follow that
this result holds for any point in the computational cell.

\section{Evolution of $\Loq$}
\label{sec:EvolveLenoq}

Equation (\ref{eqn:FirstVariationB}) can be obtained from (\ref{eqn:FirstVariationA}) as
follows. Let $(o,q)$ be a typical leg connected to the central vertex of some cell. Our
first step is to express the various vectors at $o$ and $q$ in terms of the local frames
$\Bo$ and $\Bq$. Since the shift vector is assumed to be zero across the lattice it is
follows that the unit normals take the simple form
\begin{align}
   \label{eqn:NormalO}n^\alpha_{o\Bo} &= (1,0,0,0)\\
   \label{eqn:NormalQ}n^\alpha_{q\Bq} &= (1,0,0,0)
\end{align}
while
\begin{align}
   \label{eqn:VectorMO} \va_{oq\Bo} \Loq &= \xa_{q\Bo}\\
   \label{eqn:VectorMQ} \va_{qo\Bq} \Loq &= \xa_{o\Bq}
\end{align}
which follows directly from the definition of Riemann normal coordinates $x^\alpha$. Recall
that $\xa_{a\Bb}$ are the Riemann normal coordinates of the vertex $a$ in the frame
$\Bb$. Note also that the forward pointing unit tangent vectors $\va_{o\Bo}$ and
$\va_{q\Bq}$ are given by
\begin{align}
   \label{eqn:TangentO} \va_{o\Bo} &= \phantom{-} \va_{oq\Bo}\\
   \label{eqn:TangentQ} \va_{q\Bq} &= -\va_{qo\Bq}
\end{align}
Now substitute the above equations (\ref{eqn:NormalO}--\ref{eqn:TangentQ}) into
(\ref{eqn:FirstVariationA}) to obtain
\begin{align}
   \Loq\dotLoq &= \Loq [v_\mu (Nn^\mu)]_o^q\\
               &= \Loq \left(v_\mu (Nn^\mu)\right)_q - \Loq \left(v_\mu (Nn^\mu)\right)_o\\
               &= -N_q x_{\mu o\Bq} n^\mu_{q\Bq} - N_o x_{\mu q\Bo} n^\mu_{o\Bo}\\
               &= N_q t_{o\Bq} + N_o t_{q\Bo}
\end{align}
where $t$ is the Riemann normal time coordinate. However, as shown
in~\cite{brewin:2010-03},
\begin{align}
   -2 t_{o\Bq} &= \left(\cKab\right)_{q\Bq} \xa_{o\Bq} \xb_{o\Bq} + \BigO{L^3}\\
   -2 t_{q\Bo} &= \left(\cKab\right)_{o\Bo} \xa_{q\Bo} \xb_{q\Bo} + \BigO{L^3}
\end{align}
which using (\ref{eqn:VectorMO}--\ref{eqn:VectorMQ}) can also be written as
\begin{align}
   -2 t_{o\Bq} &=
      \left(\cKab\right)_{q\Bq} \va_{qo\Bq} \vb_{qo\Bq} \Lsqoq
         + \BigO{L^3}\\
   -2 t_{q\Bo} &=
      \left(\cKab\right)_{o\Bo} \va_{oq\Bo} \vb_{oq\Bo} \Lsqoq
         + \BigO{L^3}
\end{align}
and thus
\begin{align}
   2\Loq\dotLoq = - \left(N \cKab\right)_{q\Bq} \va_{qo\Bq} \vb_{qo\Bq} \Lsqoq
                  - \left(N \cKab\right)_{o\Bo} \va_{oq\Bo} \vb_{oq\Bo} \Lsqoq
                  + \BigO{L^3}
\end{align}
which leads immediately to equation (\ref{eqn:FirstVariationB}).

\section{Complete evolution equations}
\label{sec:FullRiemannEqtns}

The following are the complete set of evolution equations for the 14 Riemann curvatures for
the particular case of a zero shift vector. These were obtained by applying the process
outlined in appendix (\ref{sec:SourceTerms}) to the second Bianchi identities
(\ref{eqn:DotRabcdA}--\ref{eqn:DotRabcdN}).


\begin{dmath}
\dotRxyxy=
N(
  \Kyz \Rxyxz
- \Kxz \Rxyyz
- 2 \Kxy \Rtxty
- \mxyx \Rtxxy
- \mxyy \Rtyxy
- \mxzx \Rtyyz
- \myzy \Rtxxz
+ ( \Rtyty+\Rxyxy ) \Kxx
+ ( \Rtxtx+\Rxyxy ) \Kyy
+ ( \Rtyxz-2\Rtzxy ) \mxzy
+ ( \Rtyxz+\Rtzxy ) \myzx
- \dRtxxydy
+ \dRtyxydx
)
- 2 \dNy \Rtxxy
+ 2 \dNx \Rtyxy
\end{dmath}
\begin{dmath}
\dotRxyxz=
N(
  \Kzz \Rxyxz
- \Kxz \Rtxty
- \mxyz \Rtyxy
- \myzz \Rtxxz
+ ( \Rtytz+\Rxyxz ) \Kxx
+ ( \Rxyyz-\Rtxtz ) \Kxy
+ ( \Rtxtx+\Rxyxy ) \Kyz
- ( \Rtxxy+\Rtzyz ) \mxzx
+ ( \Rtzxz-\Rtyxy ) \myzx
+ ( \Rtyxz-2\Rtzxy ) \mxzz
- \dRtxxydz
+ \dRtzxydx
)
+ ( \Rtyxz+\Rtzxy ) \dNx
- \dNy \Rtxxz
- \dNz \Rtxxy
\end{dmath}
\begin{dmath}
\dotRxyyz=
N(
  \Kzz \Rxyyz
+ \Kyz \Rtxty
+ \mxyz \Rtxxy
+ \mxzz \Rtyyz
+ ( \Rxyyz-\Rtxtz ) \Kyy
+ ( \Rtytz+\Rxyxz ) \Kxy
- ( \Rtyty+\Rxyxy ) \Kxz
- ( \Rtxxy+\Rtzyz ) \mxzy
+ ( \Rtzxz-\Rtyxy ) \myzy
- ( \Rtyxz+\Rtzxy ) \myzz
- \dRtyxydz
+ \dRtzxydy
)
- ( \Rtyxz-2\Rtzxy ) \dNy
+ \dNx \Rtyyz
- \dNz \Rtyxy
\end{dmath}
\begin{dmath}
\dotRxzxz=
N(
  \Kxy \Rxzyz
+ \Kyz \Rxyxz
- 2 \Kxz \Rtxtz
+ \mxyx \Rtzyz
- \mxzx \Rtxxz
- \mxzz \Rtzxz
+ \myzz \Rtxxy
+ ( \Rtztz+\Rxzxz ) \Kxx
+ ( \Rtxtx+\Rxzxz ) \Kzz
+ ( \Rtzxy-2\Rtyxz ) \mxyz
- ( \Rtyxz+\Rtzxy ) \myzx
- \dRtxxzdz
+ \dRtzxzdx
)
- 2 \dNz \Rtxxz
+ 2 \dNx \Rtzxz
\end{dmath}
\begin{dmath}
\dotRxzyz=
N(
  \Kyy \Rxzyz
- \Kyz \Rtxtz
+ \mxyy \Rtzyz
- \mxzy \Rtxxz
+  ( \Rtxty+\Rxzyz ) \Kzz
+  ( \Rtztz+\Rxzxz ) \Kxy
-  ( \Rtytz+\Rxyxz ) \Kxz
+  ( \Rtxxz-\Rtyyz ) \mxyz
-  ( \Rtyxz+\Rtzxy ) \myzy
+  ( \Rtyxy-\Rtzxz ) \myzz
+ \dRtzxzdy
- \dRtyxzdz
)
+  ( \Rtzxy-2\Rtyxz ) \dNz
+ \dNx \Rtzyz
+ \dNy \Rtzxz
\end{dmath}
\begin{dmath}
\dotRyzyz=
N(
  \Kxy \Rxzyz
- \Kxz \Rxyyz
- 2 \Kyz \Rtytz
- \mxyy \Rtzxz
- \mxzz \Rtyxy
- \myzy \Rtyyz
- \myzz \Rtzyz
+ ( \Rtztz+\Ryzyz ) \Kyy
+ ( \Rtyty+\Ryzyz ) \Kzz
- ( \Rtzxy-2\Rtyxz ) \mxyz
- ( \Rtyxz-2\Rtzxy ) \mxzy
- \dRtyyzdz
+ \dRtzyzdy
)
- 2 \dNz \Rtyyz
+ 2 \dNy \Rtzyz
\end{dmath}
\begin{dmath}
\dotRtxxy=
N(
  \Kyz \Rtxxz
+ \Kzz \Rtxxy
+ 2 \Kyy \Rtxxy
- 2 \Kxy \Rtyxy
- \mxyz \Rxyyz
+ \mxzz \Rxzyz
+ 2 \mxzy \Rxyyz
- 2 \myzy \Rxyxz
- ( \Rtyxz+\Rtzxy ) \Kxz
+ ( \Rxyxy-\Rxzxz ) \myzz
- \dRxyxydy
- \dRxyxzdz
)
- ( \Rtxtx+\Rxyxy ) \dNy
+ \dNx \Rtxty
- \dNz \Rxyxz
\end{dmath}
\begin{dmath}
\dotRtyxy=
N(
  \Kzz \Rtyxy
- \Kxz \Rtyyz
+ 2 \Kxx \Rtyxy
- 2 \Kxy \Rtxxy
+ \mxyz \Rxyxz
- \myzz \Rxzyz
- 2 \mxzx \Rxyyz
+ 2 \myzx \Rxyxz
+ ( \Rtyxz-2\Rtzxy ) \Kyz
- ( \Rxyxy-\Ryzyz ) \mxzz
+ \dRxyxydx
- \dRxyyzdz
)
+ ( \Rtyty+\Rxyxy ) \dNx
- \dNy \Rtxty
- \dNz \Rxyyz
\end{dmath}
\begin{dmath}
\dotRtzxy=
N(
- \Kxz \Rtxxy
- \Kyz \Rtyxy
+ \mxyx \Rxyyz
- \mxyy \Rxyxz
- \mxzx \Rxzyz
+ \myzy \Rxzyz
+ ( \Rtyxz+\Rtzxy ) \Kxx
- ( \Rtxxz-\Rtyyz ) \Kxy
- ( \Rtyxz-2\Rtzxy ) \Kyy
+ ( \Rxyxy-\Ryzyz ) \mxzy
- ( \Rxyxy-\Rxzxz ) \myzx
+ \dRxyxzdx
+ \dRxyyzdy
)
+ ( \Rtytz+\Rxyxz ) \dNx
- ( \Rtxtz-\Rxyyz ) \dNy
\end{dmath}
\begin{dmath}
\dotRtxxz=
N(
    \Kyy \Rtxxz
+ 2 \Kzz \Rtxxz
- 2 \Kxz \Rtzxz
+ \Kyz \Rtxxy
- \mxyy \Rxyyz
+ \mxzy \Rxzyz
- 2 \mxyz \Rxzyz
+ 2 \myzz \Rxyxz
- ( \Rtyxz+\Rtzxy ) \Kxy
+ ( \Rxyxy-\Rxzxz ) \myzy
- \dRxyxzdy
- \dRxzxzdz
)
- ( \Rtxtx+\Rxzxz ) \dNz
+ \dNx \Rtxtz
- \dNy \Rxyxz
\end{dmath}
\begin{dmath}
\dotRtyxz=
N(
- \Kxy \Rtxxz
- \Kyz \Rtzxz
+ \mxyx \Rxyyz
- \mxzx \Rxzyz
- \mxzz \Rxyxz
+ \myzz \Rxyyz
+ ( \Rtyxz+\Rtzxy ) \Kxx
- ( \Rtzxy-2\Rtyxz ) \Kzz
- ( \Rtxxy+\Rtzyz ) \Kxz
+ ( \Rxzxz-\Ryzyz ) \mxyz
- ( \Rxyxy-\Rxzxz ) \myzx
+ \dRxyxzdx
- \dRxzyzdz
)
+ ( \Rtytz+\Rxyxz ) \dNx
- ( \Rtxty+\Rxzyz ) \dNz
\end{dmath}
\begin{dmath}
\dotRtzxz=
N(
  \Kyy \Rtzxz
+ \Kxy \Rtzyz
+ 2 \Kxx \Rtzxz
- 2 \Kxz \Rtxxz
- \myzy \Rxyyz
+ \mxzy \Rxyxz
+ 2 \mxyx \Rxzyz
- 2 \myzx \Rxyxz
+ ( \Rtzxy-2\Rtyxz ) \Kyz
- ( \Rxzxz-\Ryzyz ) \mxyy
+ \dRxzxzdx
+ \dRxzyzdy
)
+ ( \Rtztz+\Rxzxz ) \dNx
- \dNz \Rtxtz
+ \dNy \Rxzyz
\end{dmath}
\begin{dmath}
\dotRtyyz=
N(
  \Kxx \Rtyyz
- \Kxz \Rtyxy
+ 2 \Kzz \Rtyyz
- 2 \Kyz \Rtzyz
- \mxyx \Rxyxz
+ \myzx \Rxzyz
+ 2 \mxyz \Rxzyz
- 2 \mxzz \Rxyyz
- ( \Rtyxz-2\Rtzxy ) \Kxy
+ ( \Rxyxy-\Ryzyz ) \mxzx
+ \dRxyyzdx
- \dRyzyzdz
)
- ( \Rtyty+\Ryzyz ) \dNz
+ \dNx \Rxyyz
+ \dNy \Rtytz
\end{dmath}
\begin{dmath}
\dotRtzyz=
N(
  \Kxx \Rtzyz
+ \Kxy \Rtzxz
+ 2 \Kyy \Rtzyz
- 2 \Kyz \Rtyyz
+ \mxzx \Rxyxz
- \myzx \Rxyyz
- 2 \mxyy \Rxzyz
+ 2 \mxzy \Rxyyz
+ ( \Rtzxy-2\Rtyxz ) \Kxz
- ( \Rxzxz-\Ryzyz ) \mxyx
+ \dRxzyzdx
+ \dRyzyzdy
)
+ ( \Rtztz+\Ryzyz ) \dNy
+ \dNx \Rxzyz
- \dNz \Rtytz
\end{dmath}

\clearpage

\printbibliography  

%

\clearpage

\def\mywd{0.49}

\def\Figure#1#2{%
\hbox to\textwidth{%
\hfill%
\includegraphics[width=#2\textwidth]{#1}%
\hfill}}
\def\FigPair#1#2#3#4{%
\hbox to\textwidth{%
\hfill%
\includegraphics[width=#2\textwidth]{#1}%
\includegraphics[width=#4\textwidth]{#3}%
\hfill}}
\def\FigTriple#1#2#3#4#5#6{%
\hbox to\textwidth{%
\hfill%
\includegraphics[width=#2\textwidth]{#1}\hfill%
\includegraphics[width=#4\textwidth]{#3}\hfill%
\includegraphics[width=#6\textwidth]{#5}%
\hfill}}
\def\FigTripleAlt#1#2#3#4#5#6#7#8#9{%
\hbox to\textwidth{%
\hfill%
\lower#3\hbox{\includegraphics[width=#2\textwidth]{#1}}\hfill%
\lower#6\hbox{\includegraphics[width=#5\textwidth]{#4}}\hfill%
\lower#9\hbox{\includegraphics[width=#8\textwidth]{#7}}%
\hfill}}
\def\FigQuad#1#2#3#4#5#6#7#8{%
\FigPair{#1}{#2}{#3}{#4}%
\FigPair{#5}{#6}{#7}{#8}}
\def\FigSix#1#2#3#4#5#6#7#8{%
\hbox to\textwidth{%
\includegraphics[width=#1\textwidth]{#3}\hfill%
\includegraphics[width=#1\textwidth]{#4}\hfill%
\includegraphics[width=#1\textwidth]{#5}%
}%
\vskip#2%
\hbox to\textwidth{%
\includegraphics[width=#1\textwidth]{#6}\hfill%
\includegraphics[width=#1\textwidth]{#7}\hfill%
\includegraphics[width=#1\textwidth]{#8}%
}}

\begin{figure}[!ht]
\FigPair%
{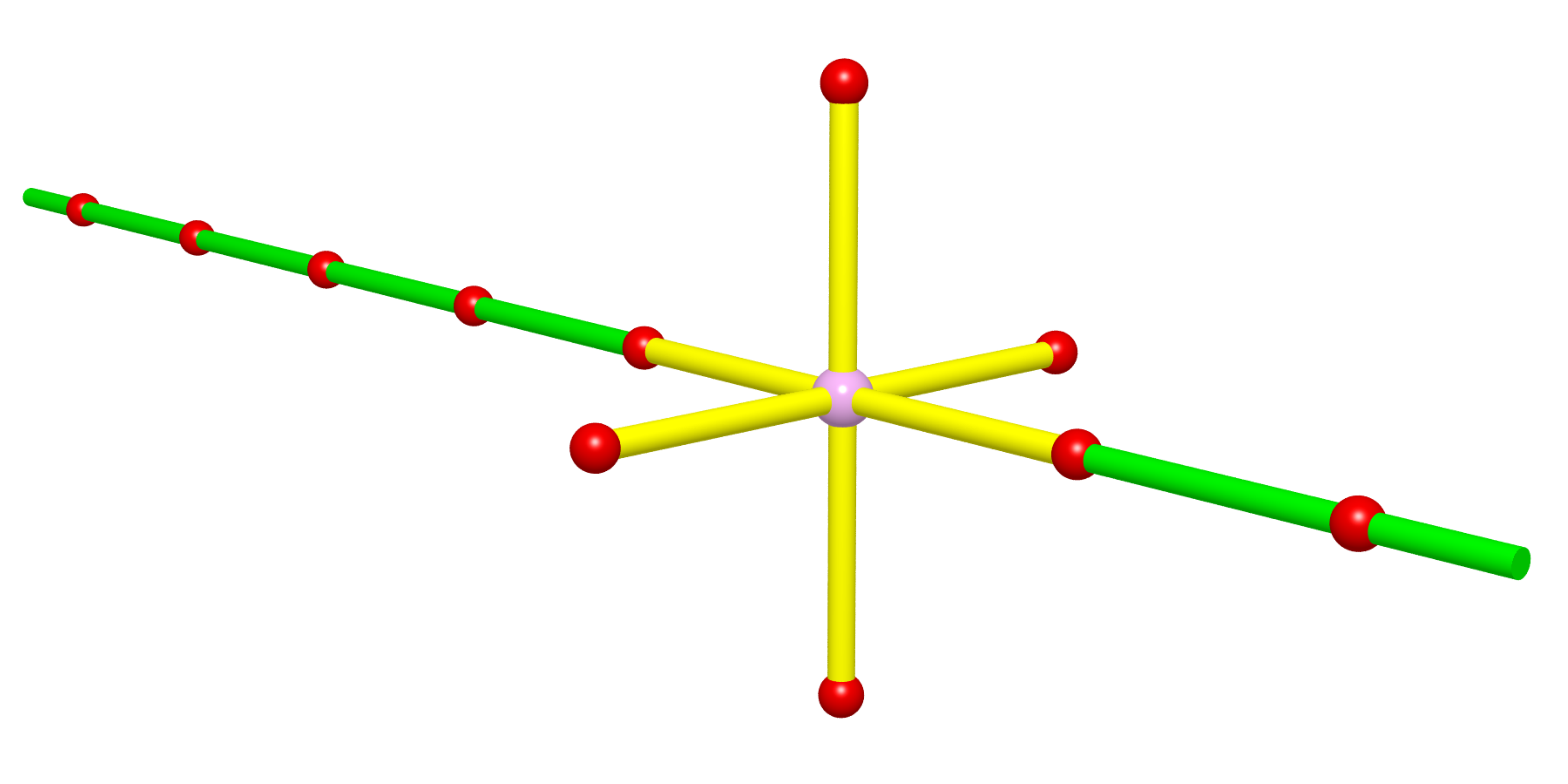}{\mywd}%
{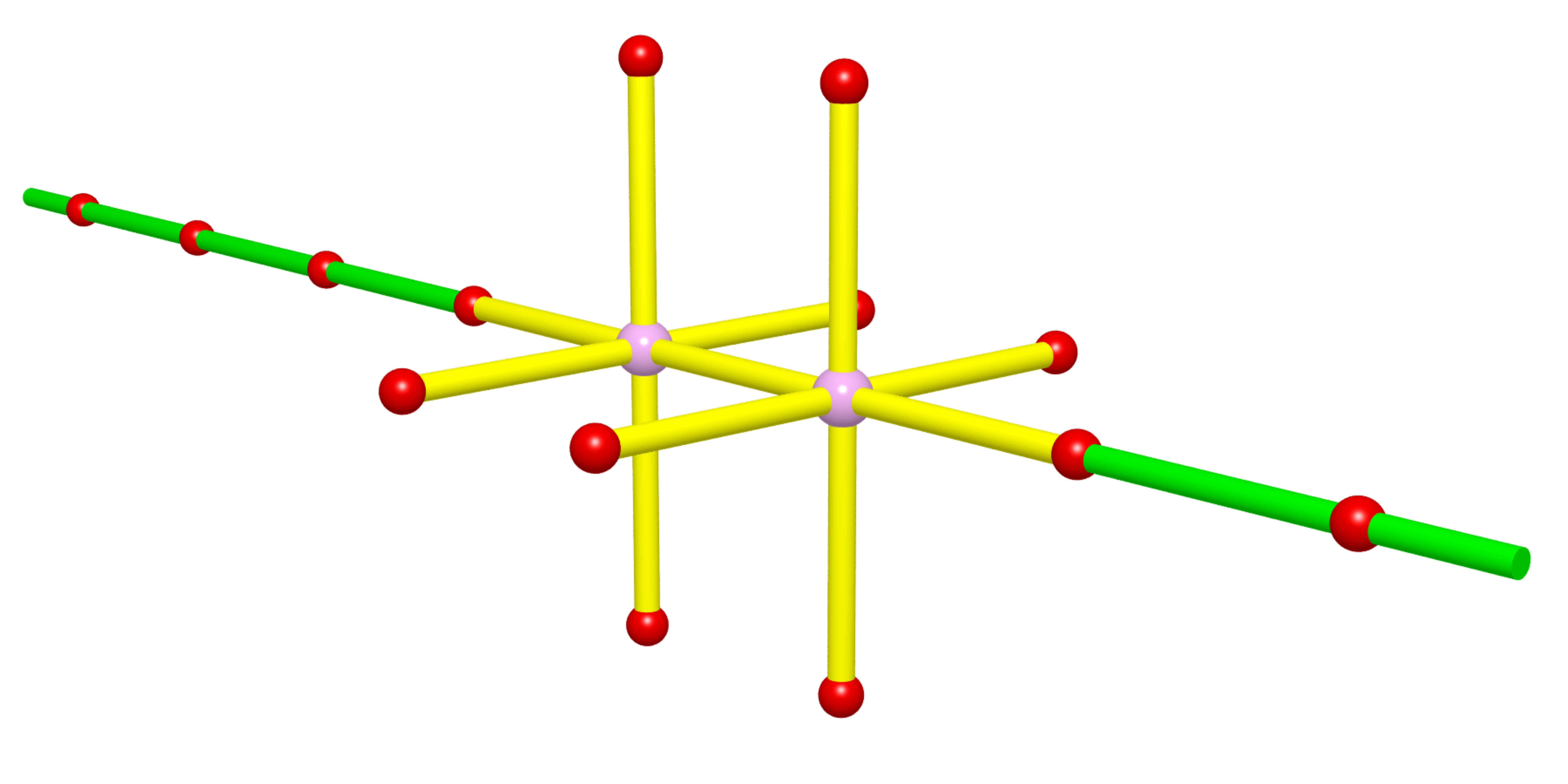}{\mywd}
\caption{Two examples of a subset of the Gowdy 1-dimensional lattice. The
left figure shows a single cell in the while the right figure shows a pair of
neighbouring cells. The purple vertices are the central vertices of their
respective cells. Note that the vertical legs pass through the central vertex
and begin and end on the red vertices. This also applies to the corresponding
horizontal legs. In contrast, the radial legs begin and end on the central
vertices.}
\label{fig:GowdyLattice}
\end{figure}

\begin{figure}[!ht]
\FigTripleAlt%
{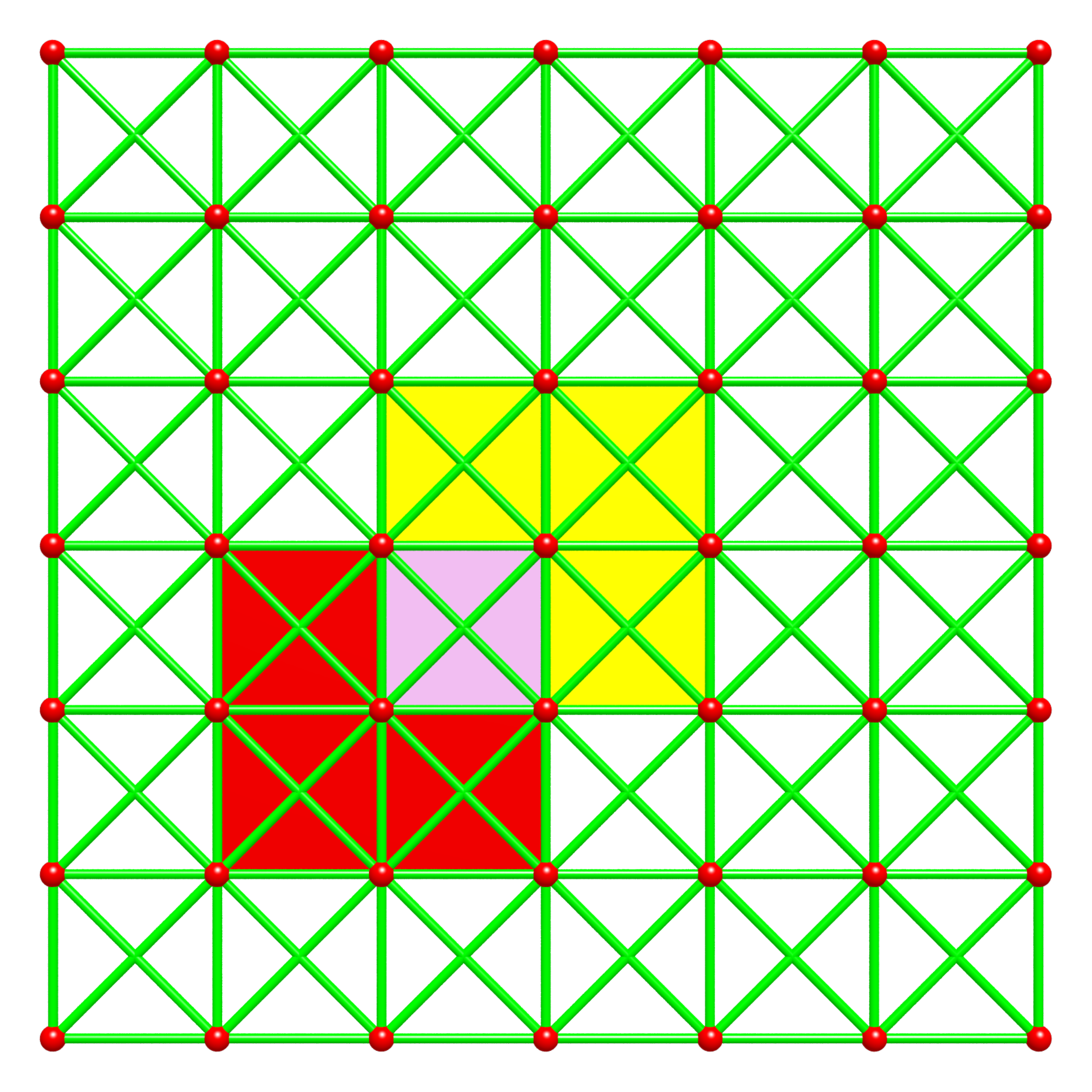}{0.28}{0.00cm}%
{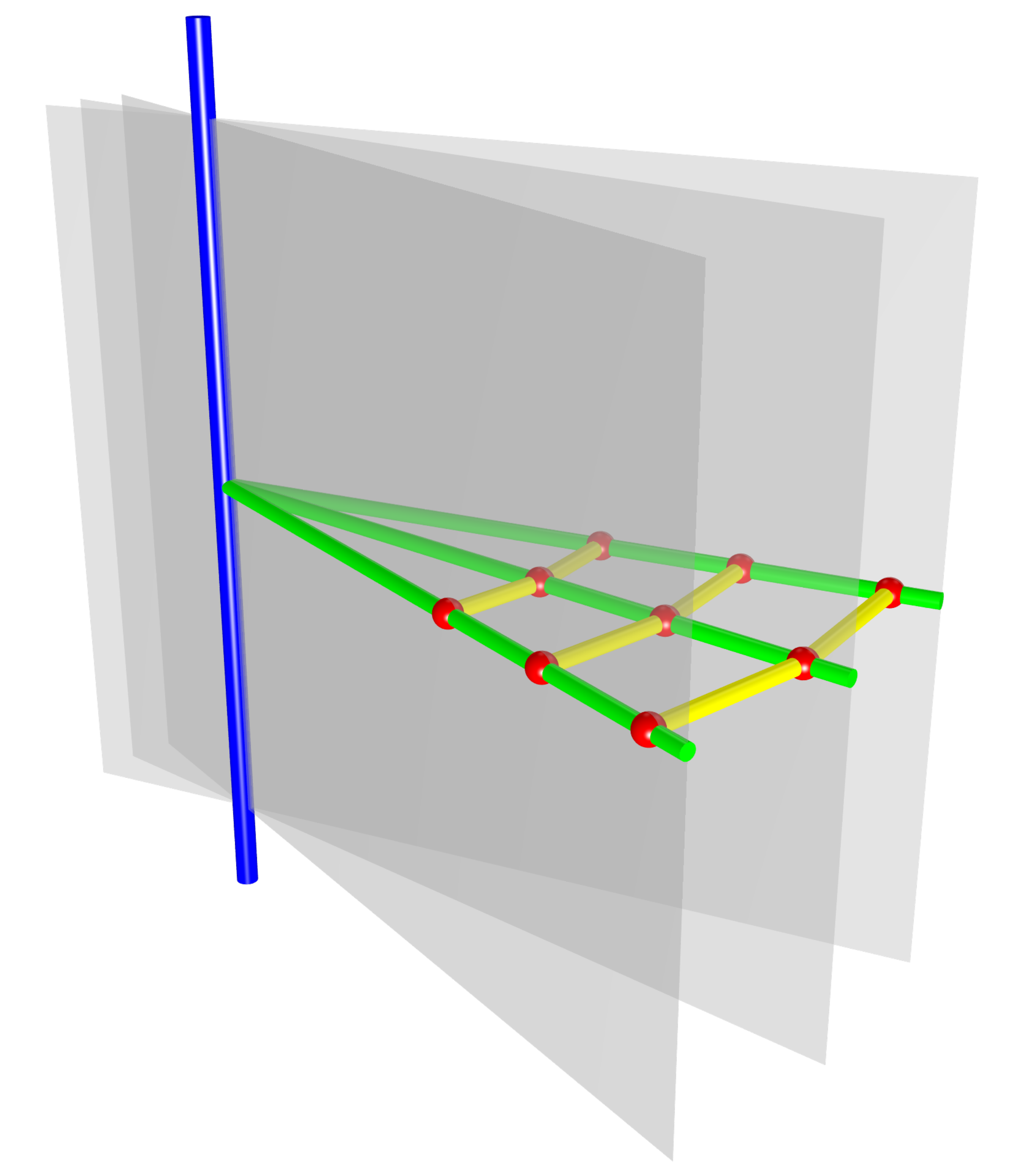}{0.28}{0.35cm}%
{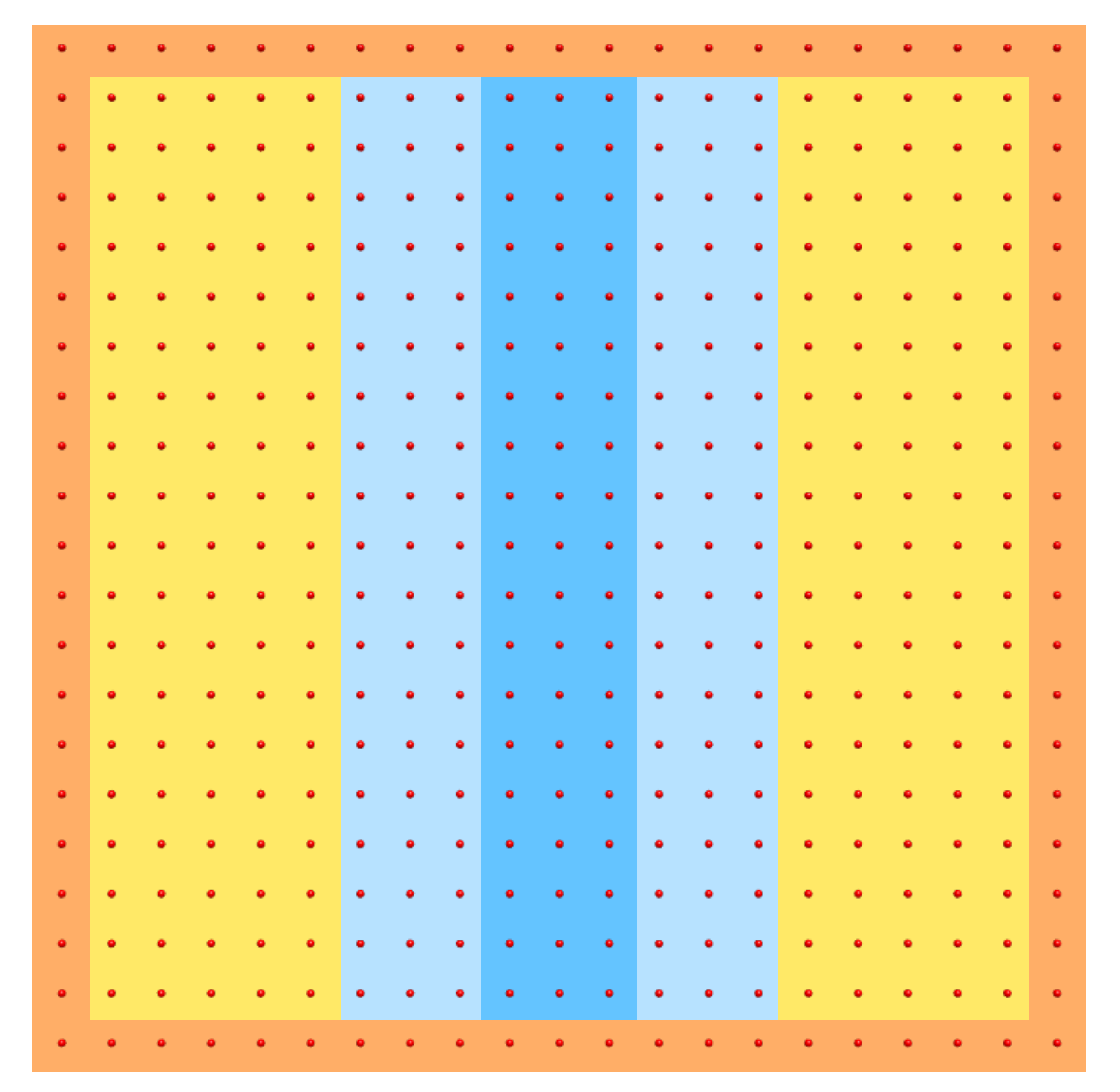}{0.28}{0.00cm}
\caption{Details of the Brill 2-dimensional lattice. The left figure shows a
subset of the lattice including two overlapping cells. Each cell is a
$2\times2$ set of vertices and legs. An axisymmetric lattice is obtained by
assembling copies of the 2-dimensional lattice in the manner shown in the
middle figure. The yellow legs in the middle figure are needed to define the
separation between the copies. The right figure shows the various subsets of
the lattice used to evolve the data and to apply various boundary conditions.
Data in the outer boundary (the orange region) were evolved using a radiation
boundary condition while the data on and near the symmetry axis (the dark
blue region) were evolved by interpolating the time derivatives from the
nearby cells (the light blue region). The remaining data (in the yellow
region) were evolved using the lattice evolution equations.}
\label{fig:BrillLattice}
\end{figure}

\begin{figure}[!ht]
\Figure%
{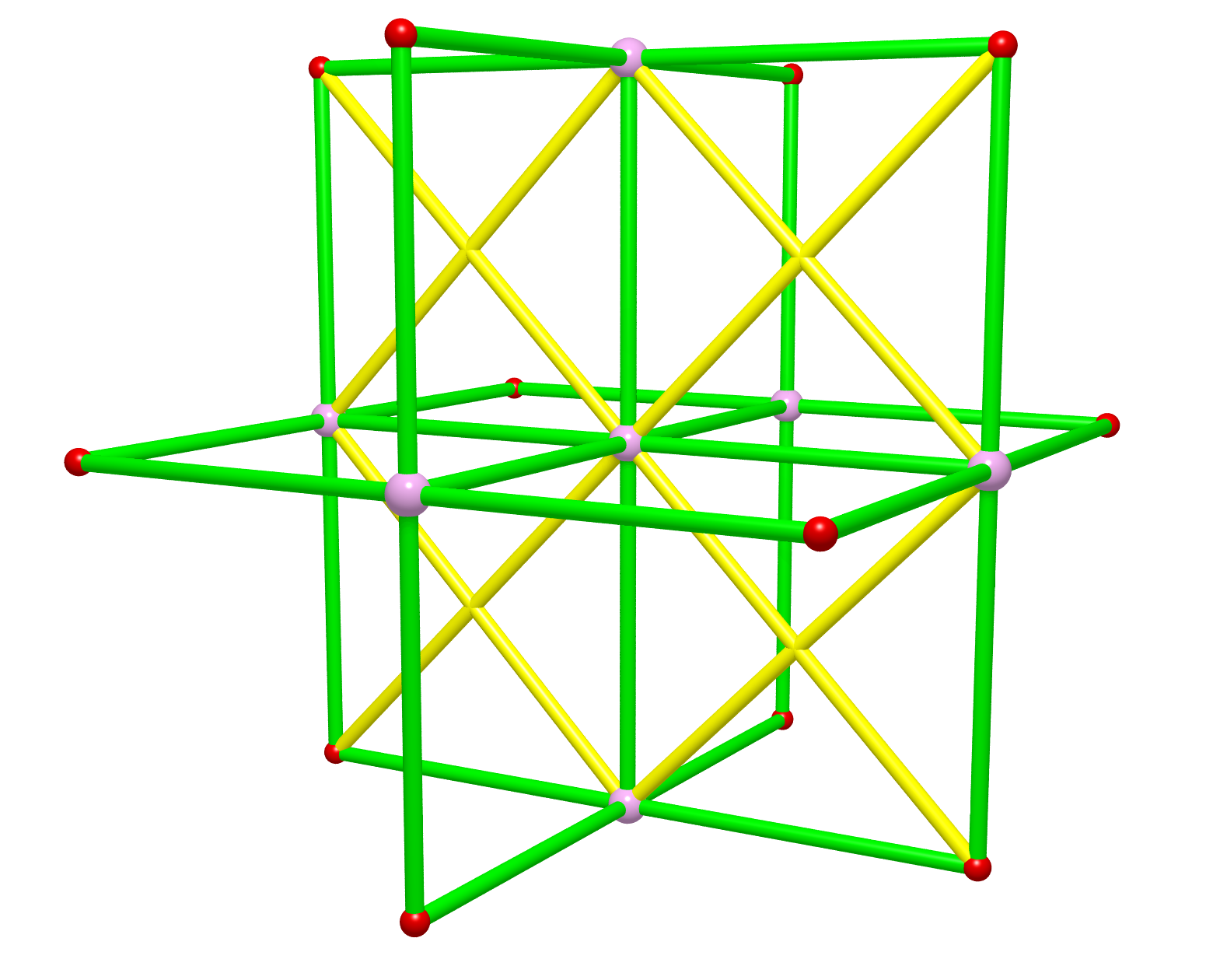}{\mywd}%
\caption{A typical computational cell for the Teukolsky lattice. This figure
shows, for simplicity, only one of three sets of yellow diagonal legs. A
proper figure would show yellow diagonal legs on each of the three coordinate
planes (bounded by the green rectangles). Note also that though this cell
looks regular (roughly equal leg-lengths and apparently orthogonal legs)
this is again just to simplify the figure. In general the leg-lengths and
their mutual angles will vary (slightly) across the cell.}
\label{fig:TeukolskyLattice}
\end{figure}

\begin{figure}[!ht]
\FigPair%
{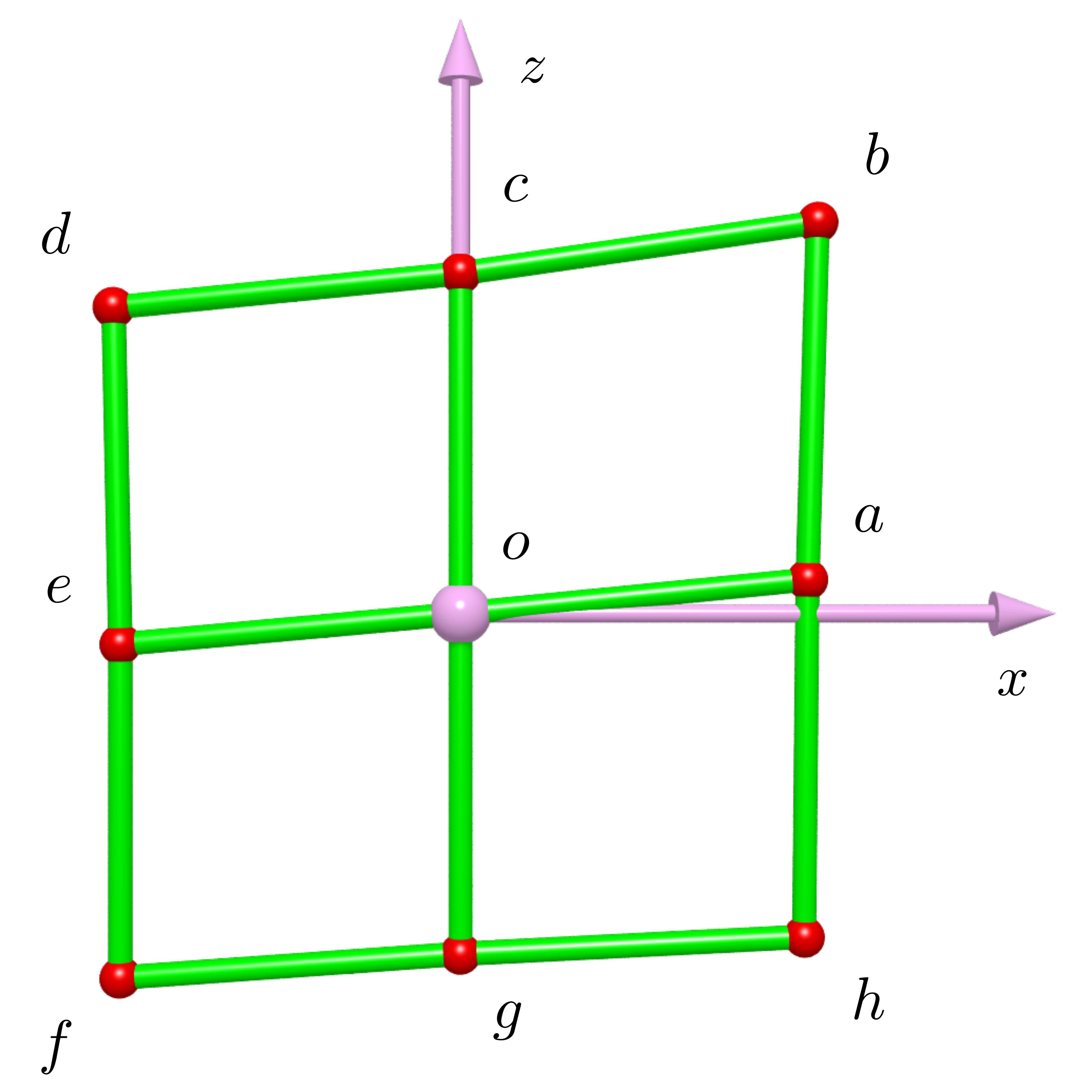}{\mywd}%
{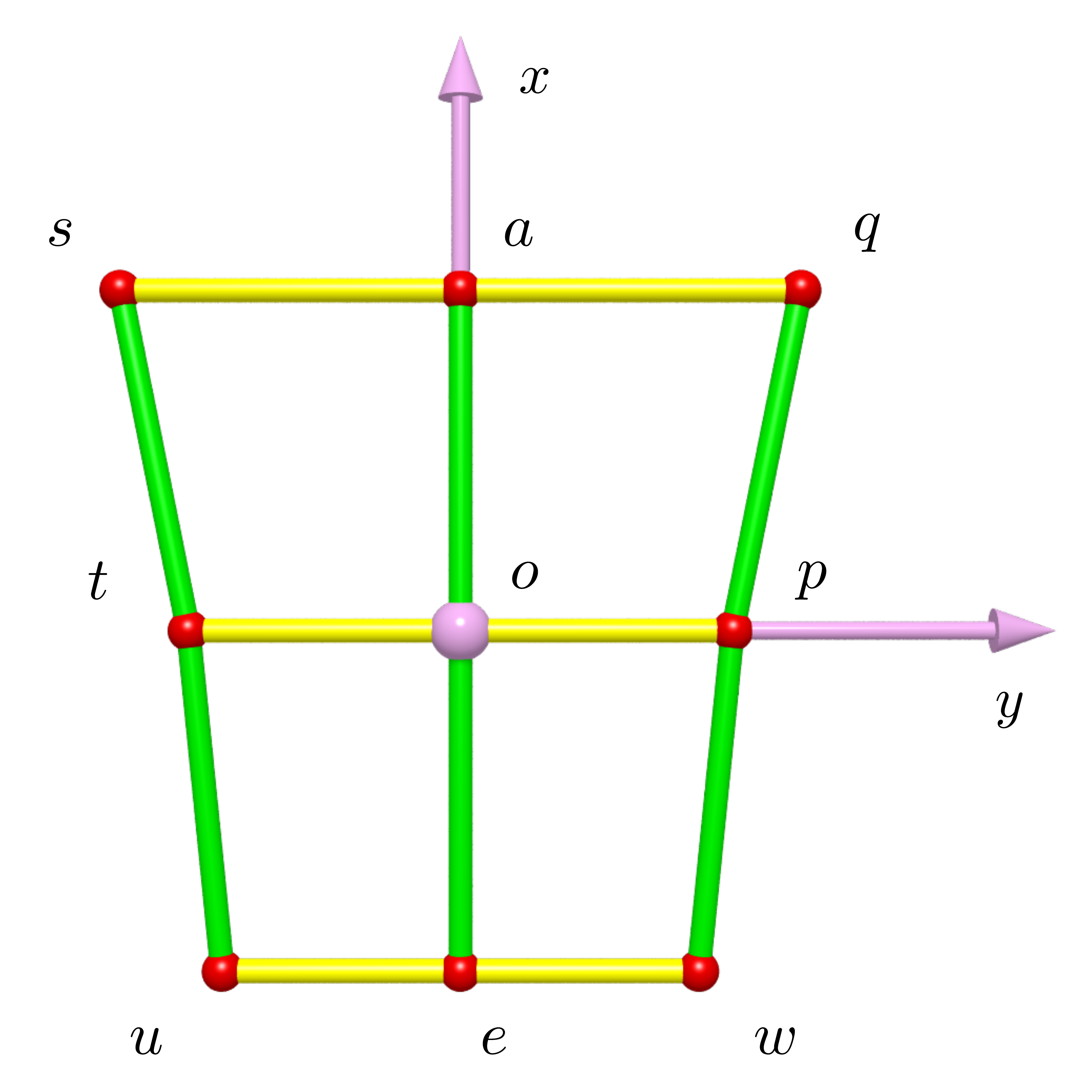}{\mywd}%
\caption{A typical set of vertices and legs used in computing the transition
matrices, $m^a{}_{bc}$. The coordinate axes in these figures are applicable
only to the 2-dimensional Brill lattice and should be ignored when reading the
discussion in Appendix (\ref{sec:TransMatrix}) particularly in the
calculations leading to equation (\ref{eqn:FinalMabcEqtns}).}
\label{fig:CartanConn}
\end{figure}

\begin{figure}[!ht]
\Figure%
{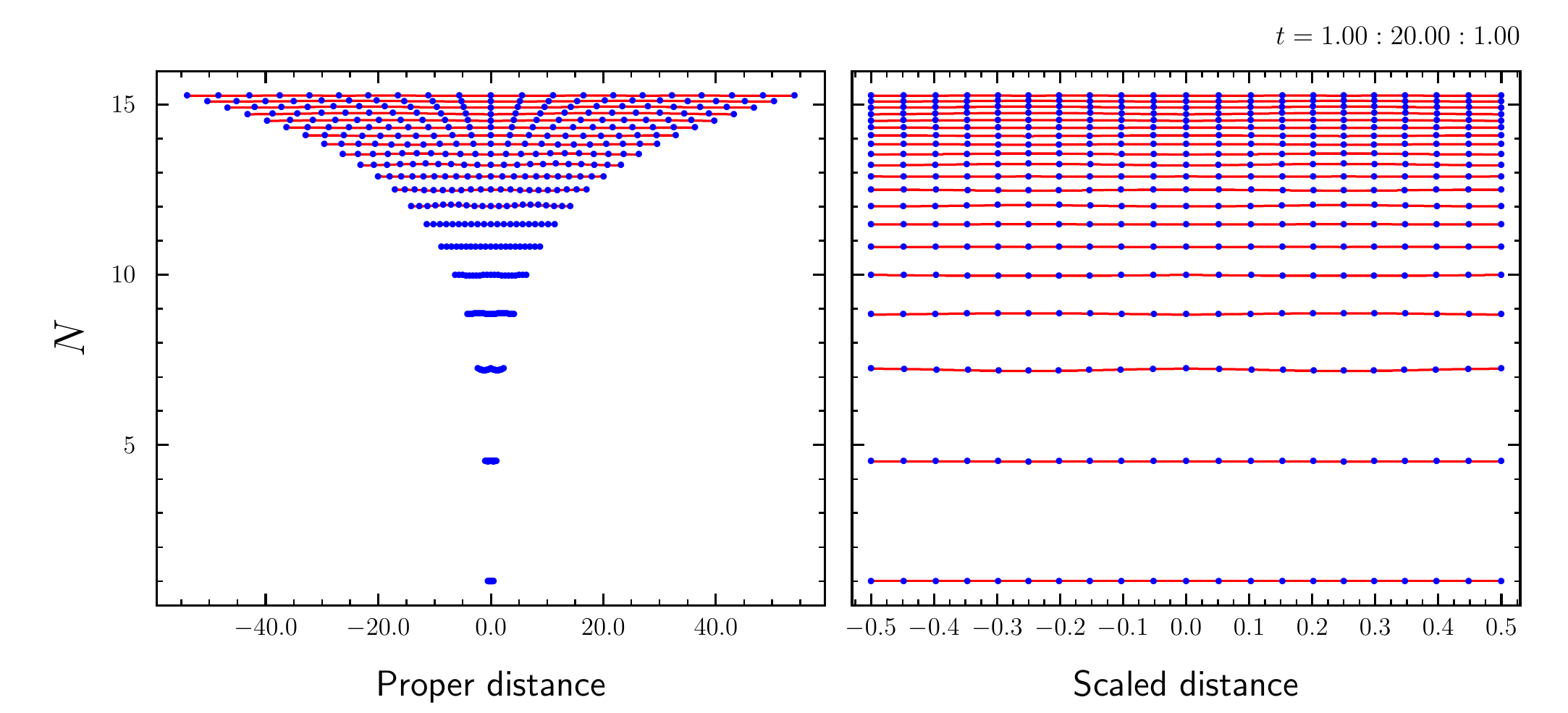}{0.98}%
\caption{This figure shows the rapid expansion (into the future of the $t=0$
singularity) of the lattice in the 1+log slicing. The left plot shows the
lapse (from $t=1$ to $t=20$ in steps of 1) as a function of the un-scaled
proper distance while the right plot shows the same data but using a re-scaled
$z$-axis. The red curves display the lattice data (for $N_z=1024$) while the
blue dots are from the Cactus data (with $N_z=400$ though only every fourth
point is shown). The agreement between the lattice and Cactus data is very
good.}
\label{fig:GowdyA}
\end{figure}

\newpage

\begin{figure}[!ht]
\FigPair%
{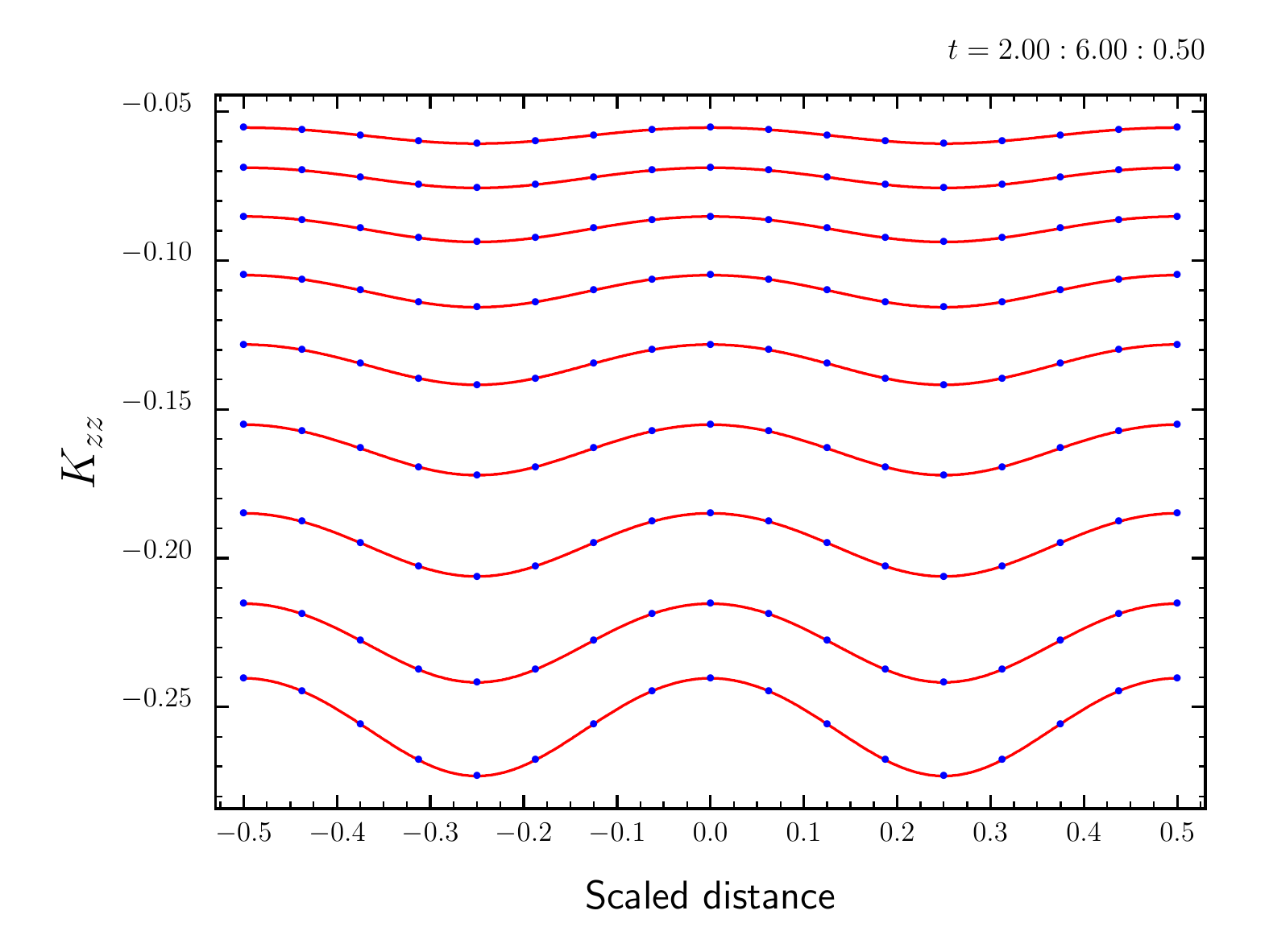}{\mywd}%
{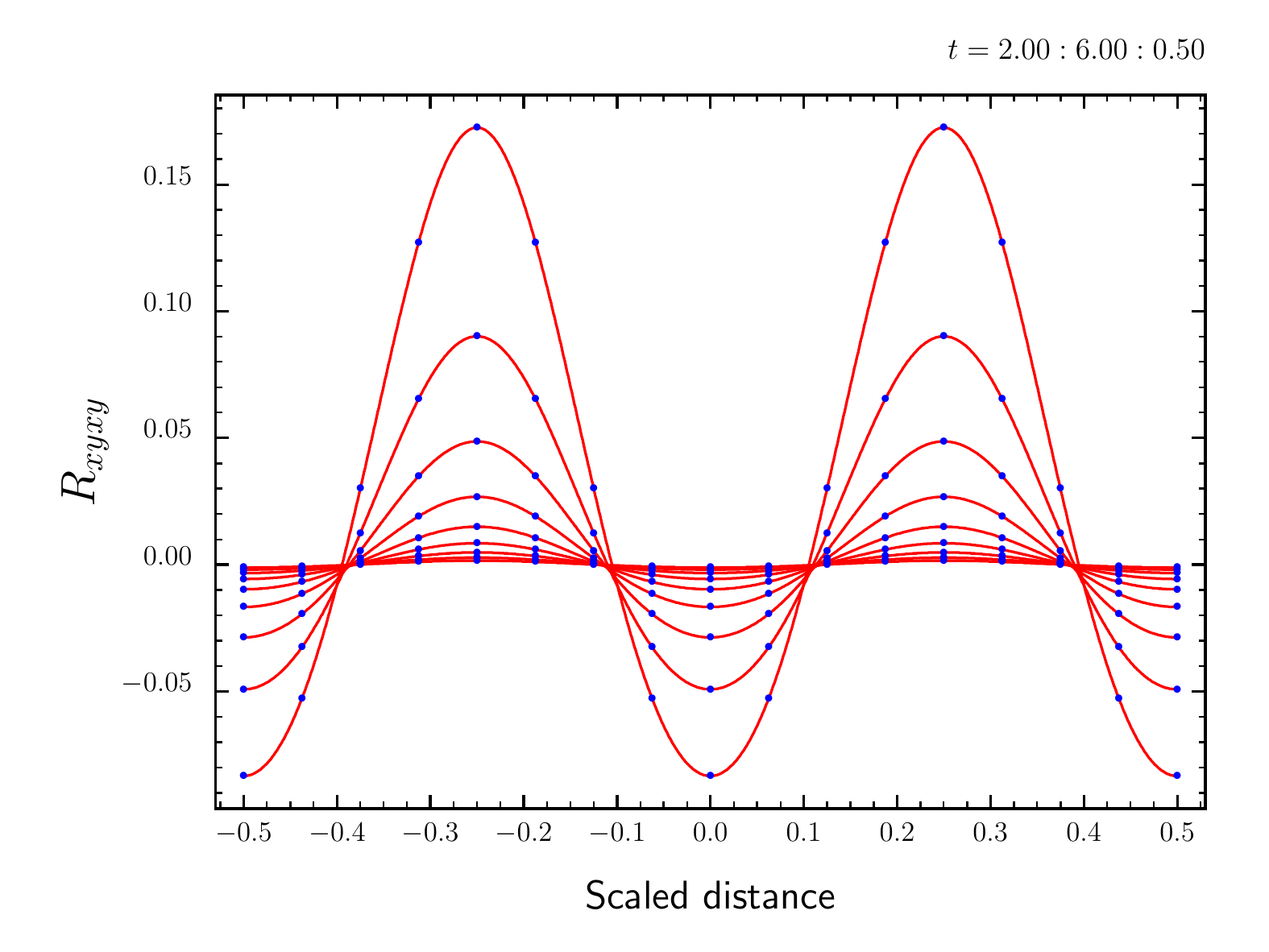}{\mywd}%
\caption{A comparison of the lattice data for the exact slicing against the
New-Watt etal~\cite{new-kc:1998-01} data. The continuous line denotes the
lattice data (using $N_z=1024$) while the New-Watt data (with $N_z=32$) are
denoted by points. It is clear that the lattice data agrees very well with
the New-Watt data. There are 9 curves in each figure representing data from
$t=2$ to $t=6$ in steps of $0.5$.}
\label{fig:GowdyB}
\end{figure}

\begin{figure}[!ht]
\FigPair%
{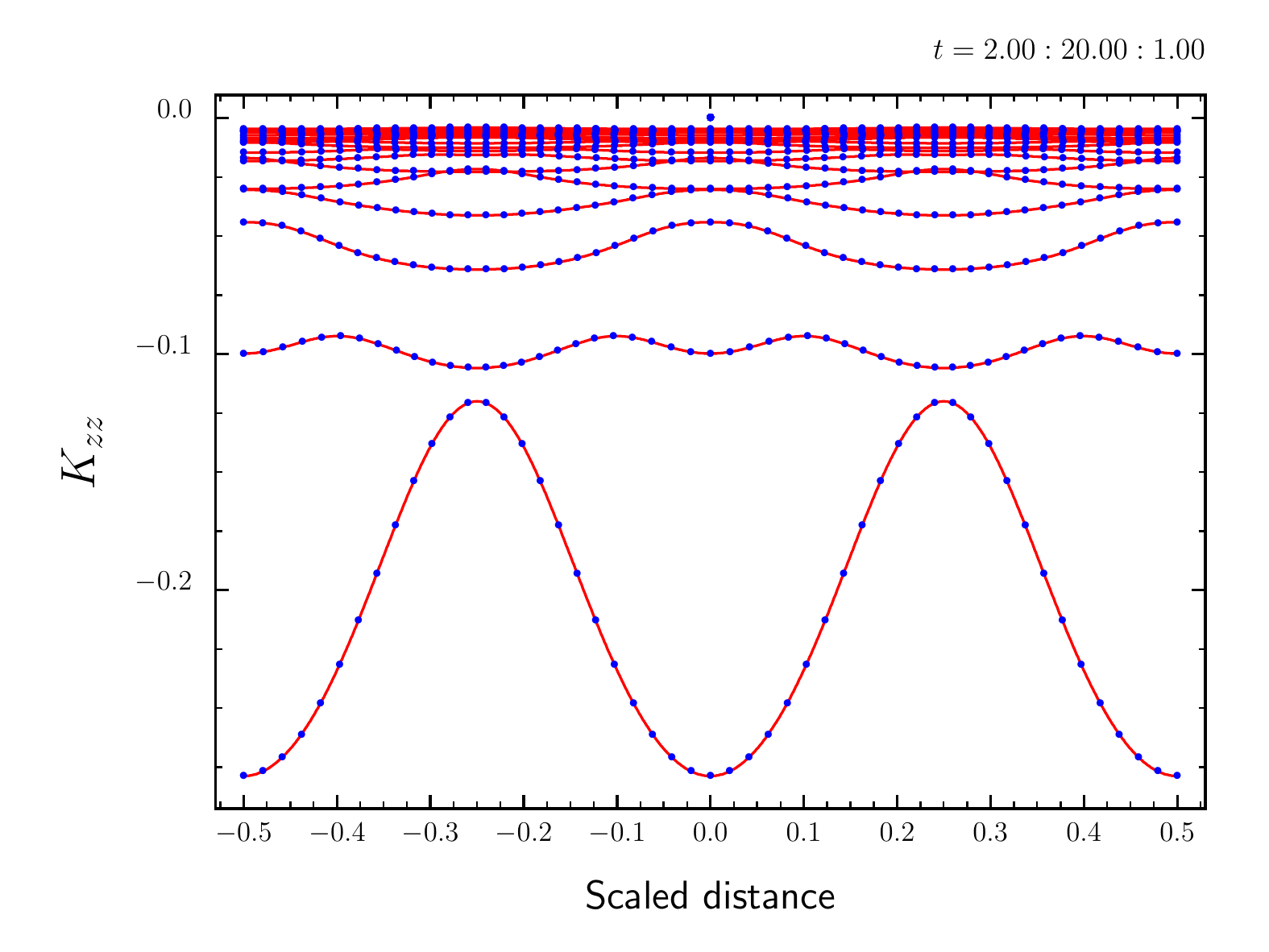}{\mywd}%
{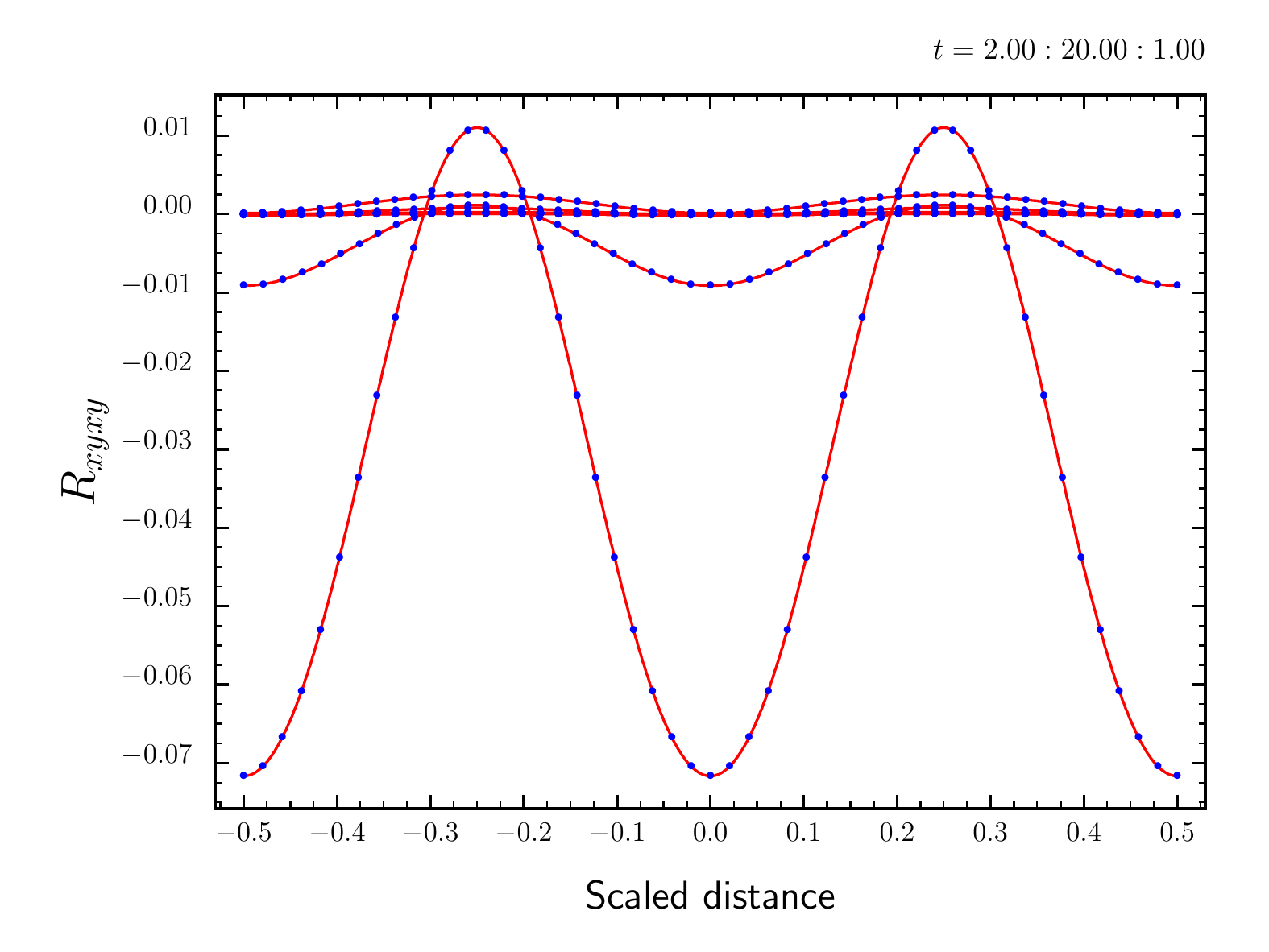}{\mywd}%
\caption{This figure is similar to the previous figure but this time for the
1+log slicing. The Cactus data (blue points) is based on $N_z=400$ with only
every fourth point shown. The lattice data (red lines) is based on $N_z=1024$.
Each figure contains 20 curves for $t=2$ to $t=20$ in steps of 1.}
\label{fig:GowdyC}
\end{figure}

\newpage

\begin{figure}[!ht]
\FigPair%
{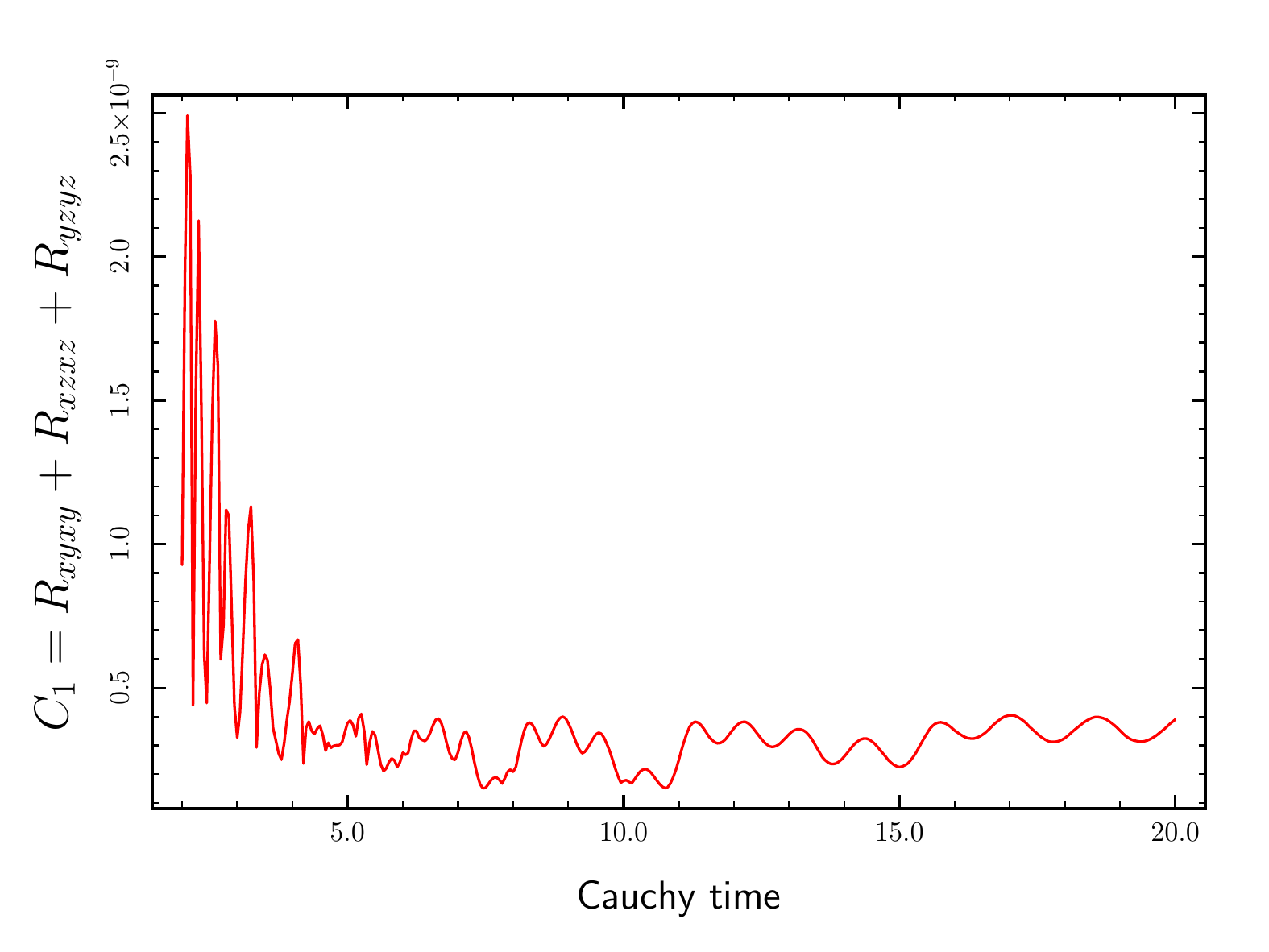}{\mywd}%
{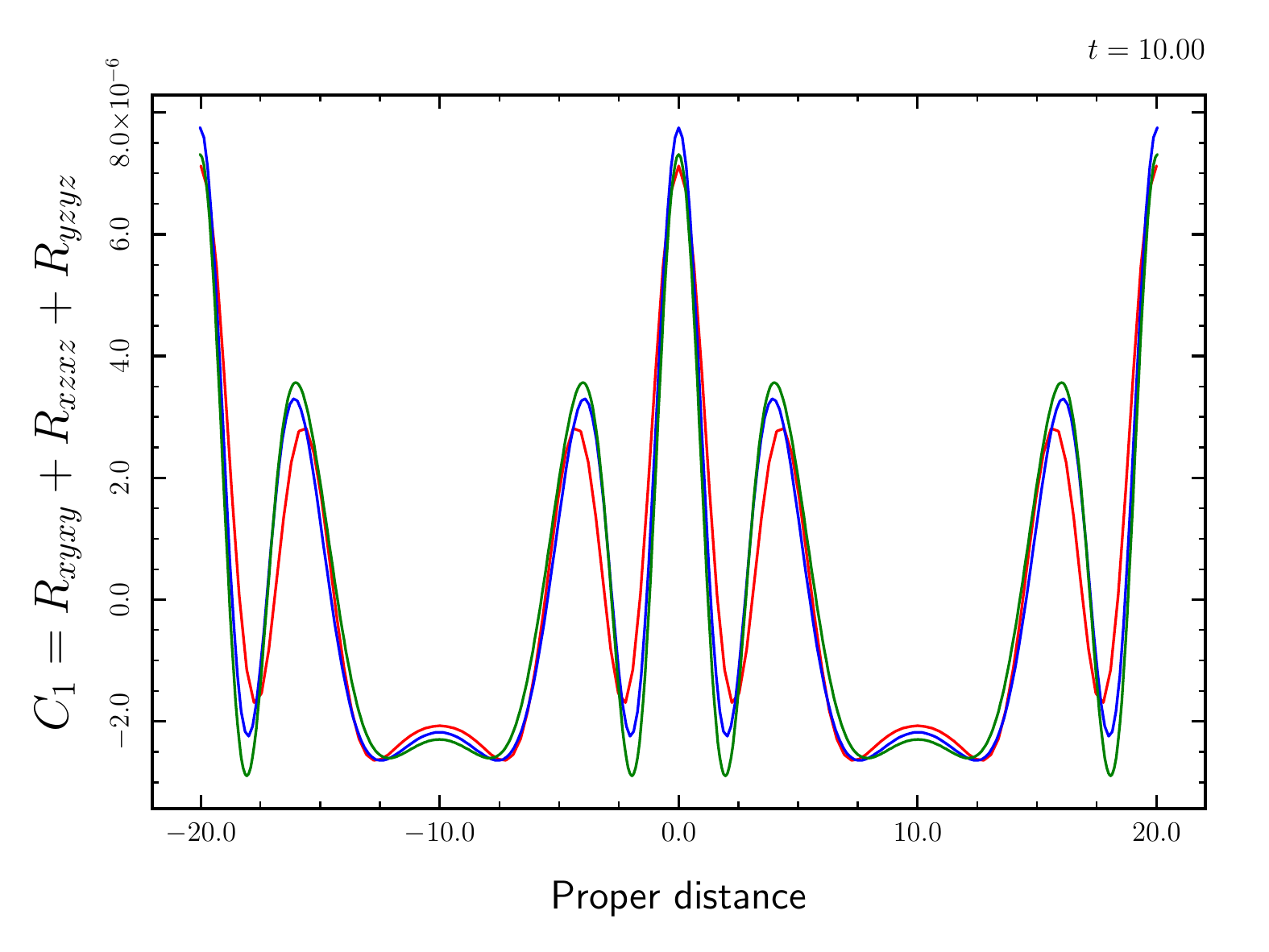}{\mywd}%
\caption{This figure shows the behaviour, in the 1+log slicing, of the $C_1$
constraint (\ref{eqn:GowdyC1}) over time (left panel) and across the grid at a
fixed time (right panel). The data in the left panel are for the case
$N_z=1024$ and show the maximum values of $C_1$ across the grid. The right
hand panel shows three curves, $N_z=256$ (red), $N_z=512$ (blue) and
$N_z=1024$ (green) with $y$ values, at $t=5$, scaled by $1$, $32$ and $1024$
respectively. The close agreement in the curves suggests that the constraints
converge to zero as $\BigO{N_z^{-5}}$. Similar behaviour was observed for the
remaining two constraints (\ref{eqn:GowdyC2},\ref{eqn:GowdyC3}). The somewhat
erratic behaviour in the left panel most likely arises by the fact that the
grid point on which the maximum occurs need not be a continuous function of
time.}
\label{fig:GowdyD}
\end{figure}


\begin{figure}[!ht]
\FigPair%
{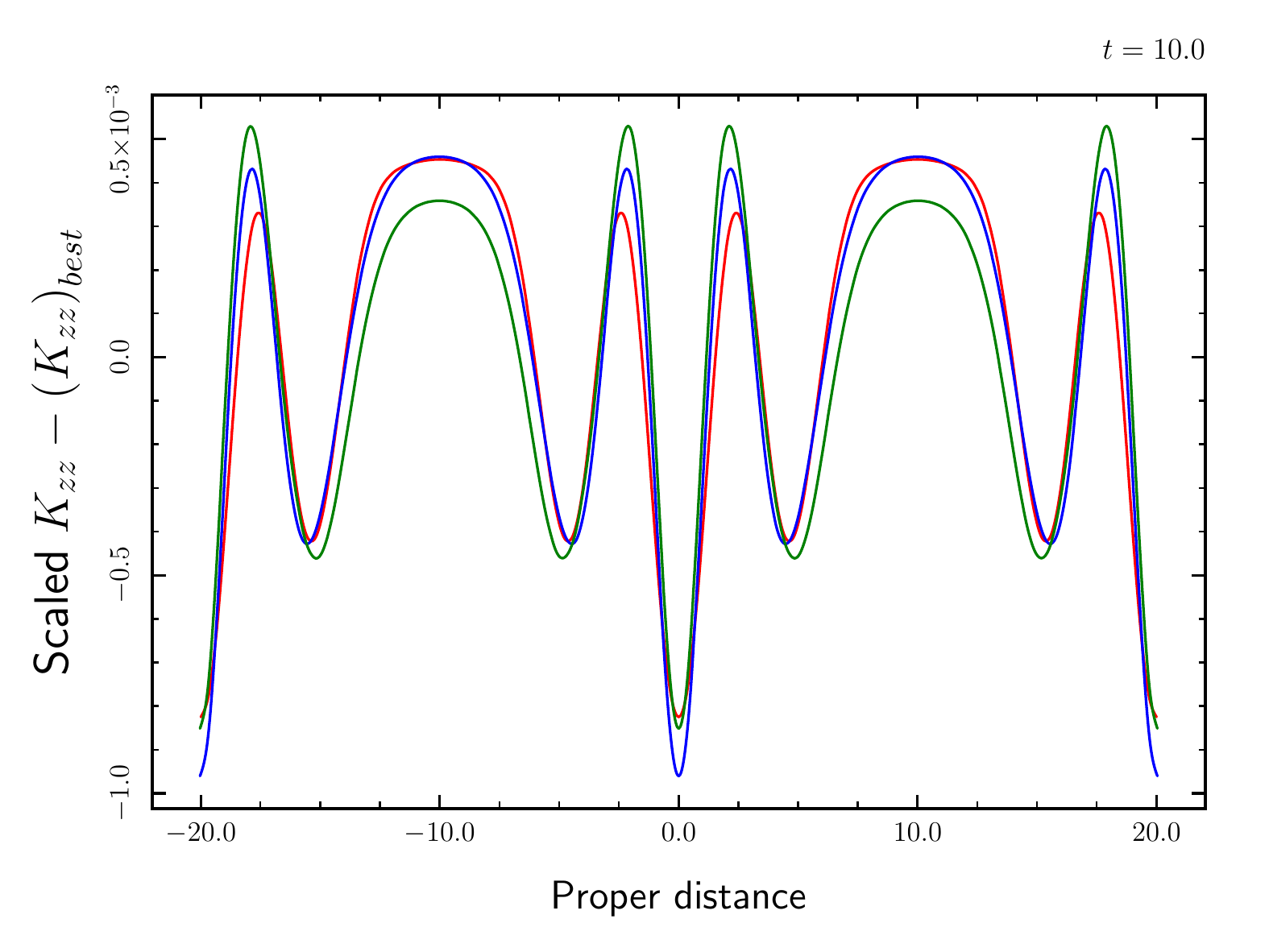}{\mywd}%
{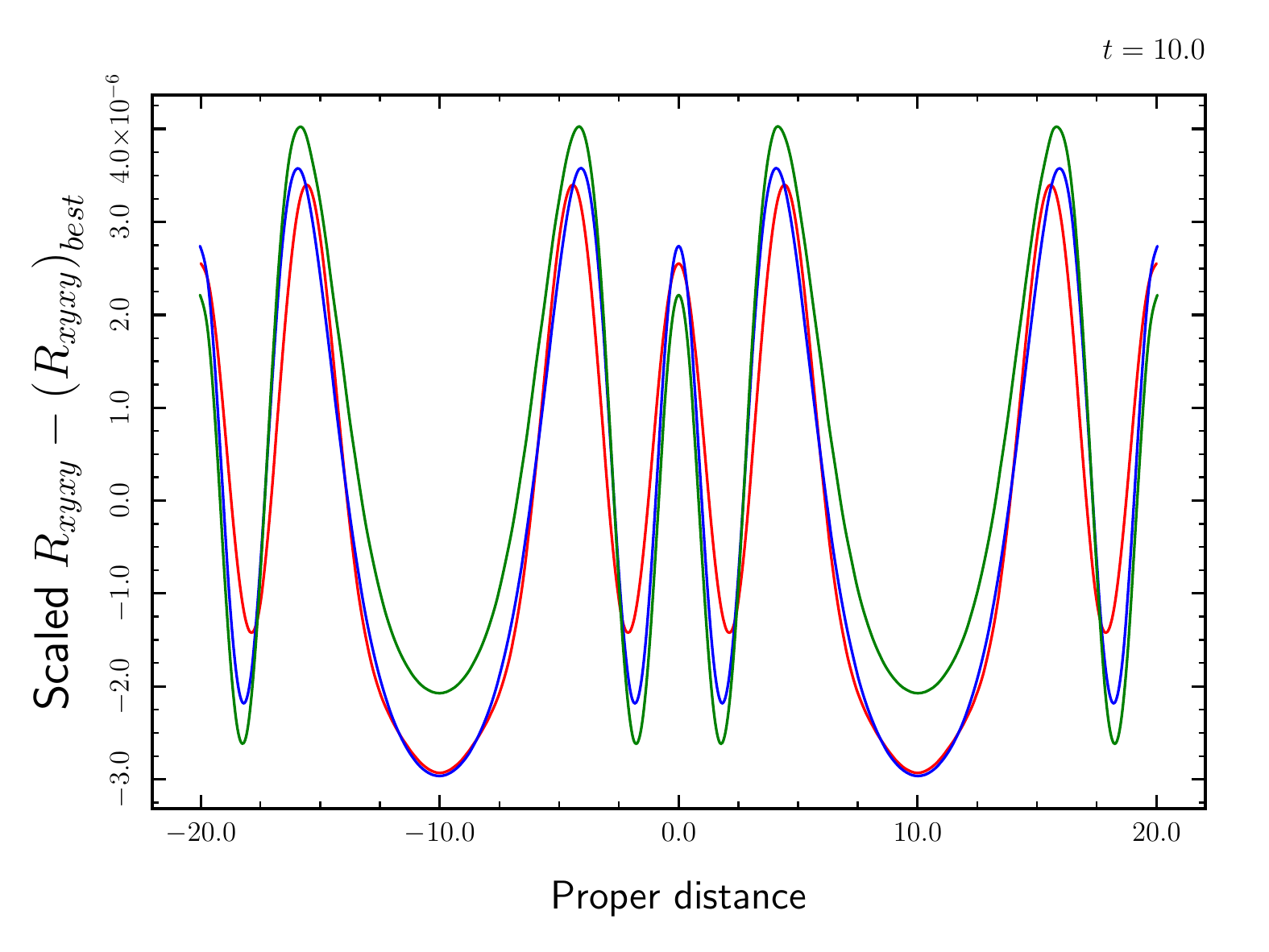}{\mywd}%
\caption{This figure show the convergence of two metric functions, $\Kzz$ and
$\Rxyxy$, as a function of $N_z$ in the 1+log slicing. The three curves
correspond to $N_z=128$ (red), $N_z=256$ (blue) and $N_z=512$ (green) and have
their $y$ values scaled by $1$, $32$ and $1024$ respectively. For the 1+log
slicing there is no exact solution available so the \emph{best} available data
(i.e., $N_z=1024$) was taken as a best estimate of the exact solution. This
suggests that the lattice data is converging to the exact solution as
$\BigO{N_z^{-5}}$.}
\label{fig:GowdyE}
\end{figure}

\newpage

\begin{figure}[!ht]
\FigPair%
{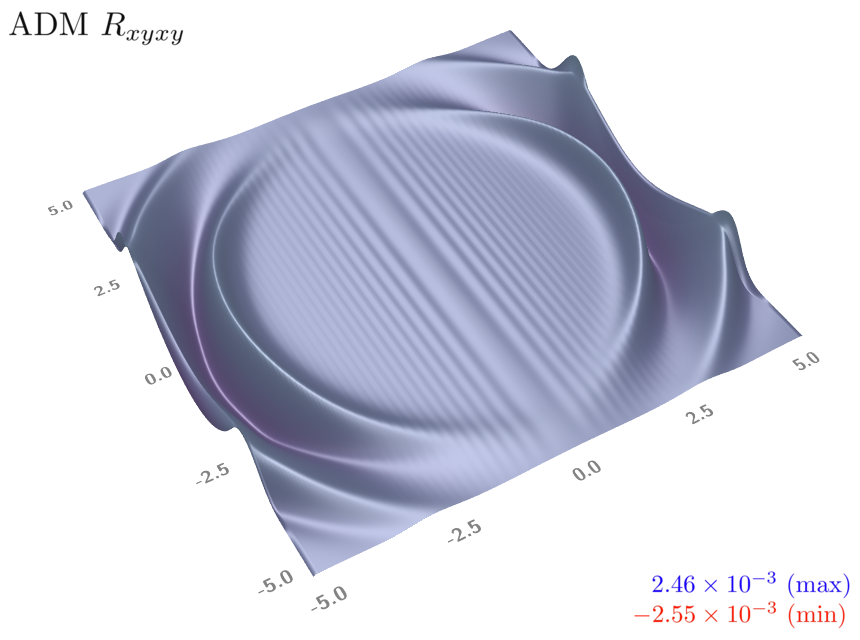}{\mywd}%
{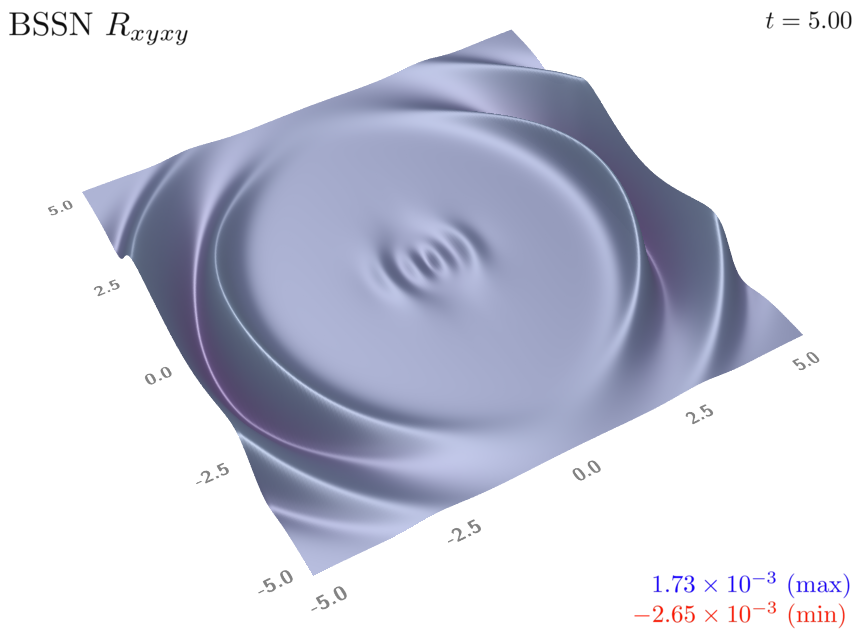}{\mywd}%
\FigPair%
{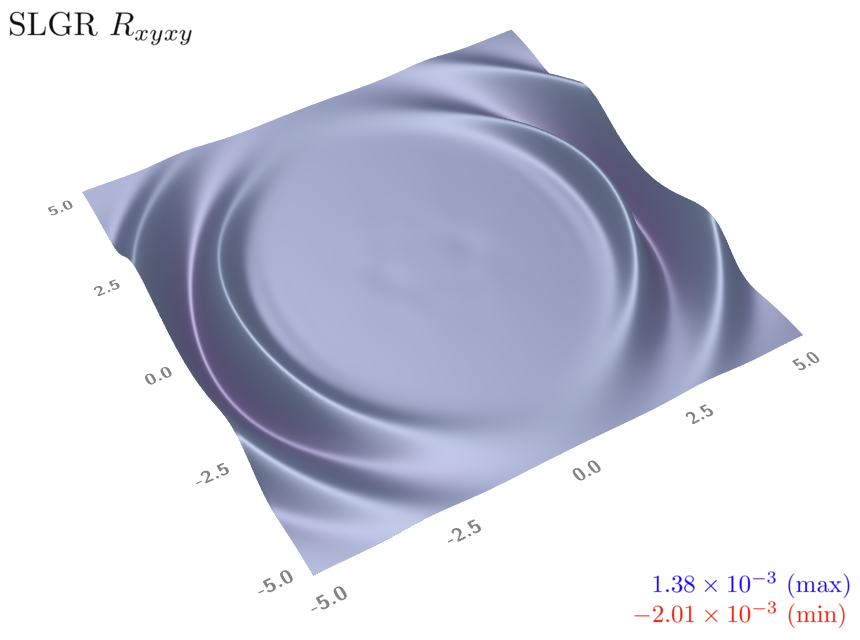}{\mywd}%
{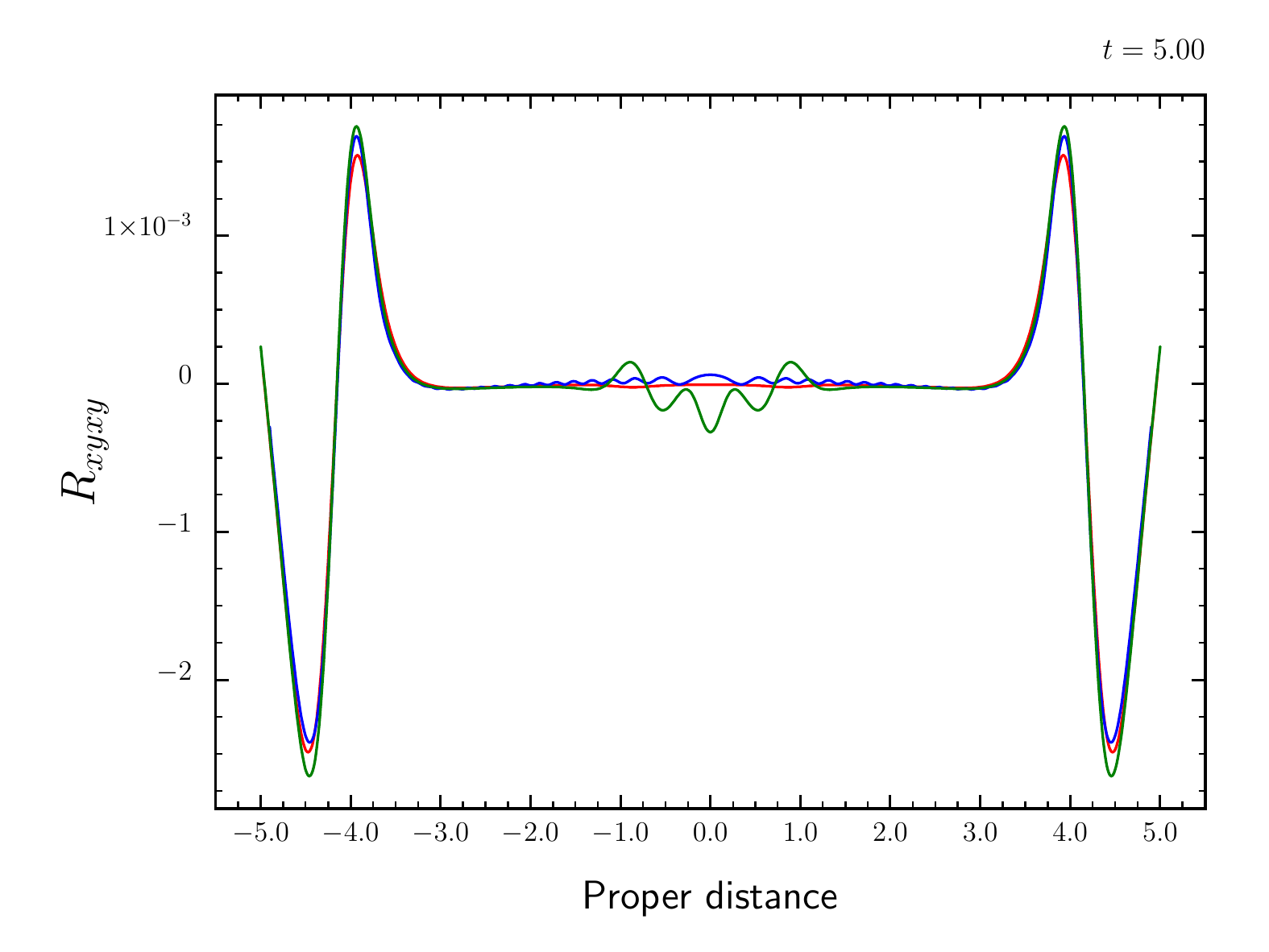}{\mywd}%
\caption{This figure shows a comparison between the lattice, ADM and BSSN
evolutions of $\Rxyxy$ for the Brill initial data at $t=5$. All three methods
agree well though the ADM and BSSN results show small waves near the symmetry
axis. The figure in the lower right shows the data for all three methods
(red, lattice), (blue, ADM) and (green, BSSN) along the $\Tx$ axis.}
\label{fig:BrillProfile05}
\end{figure}

\begin{figure}[!ht]
\FigPair%
{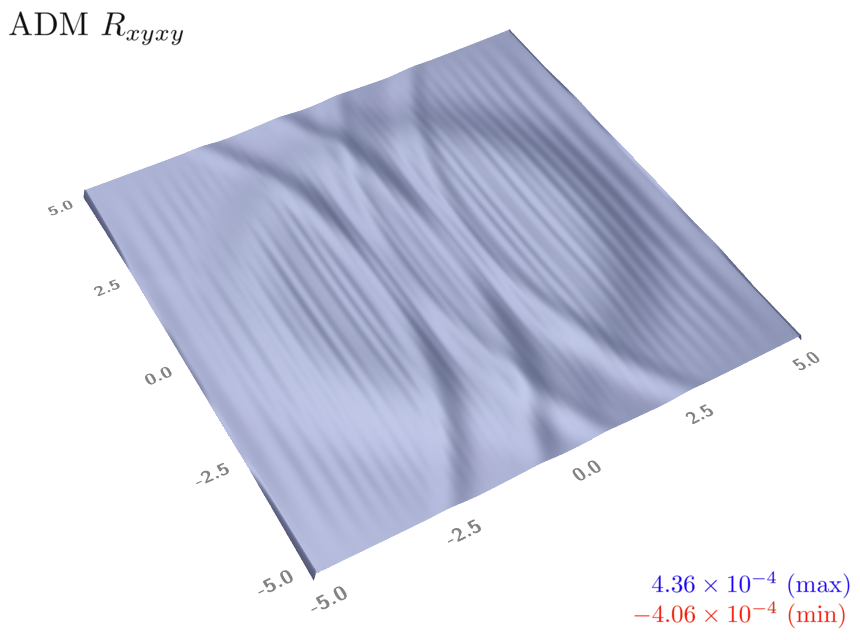}{\mywd}%
{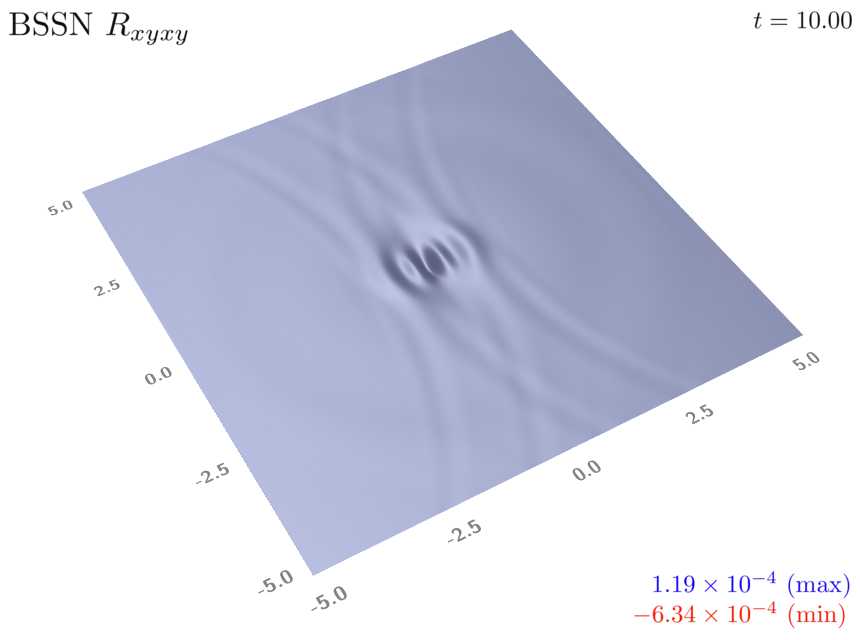}{\mywd}%
\FigPair%
{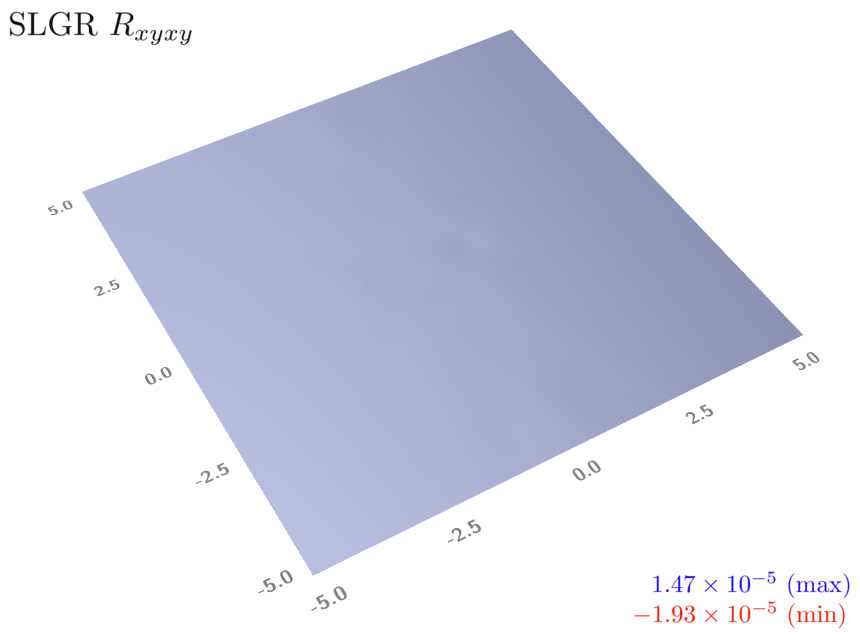}{\mywd}%
{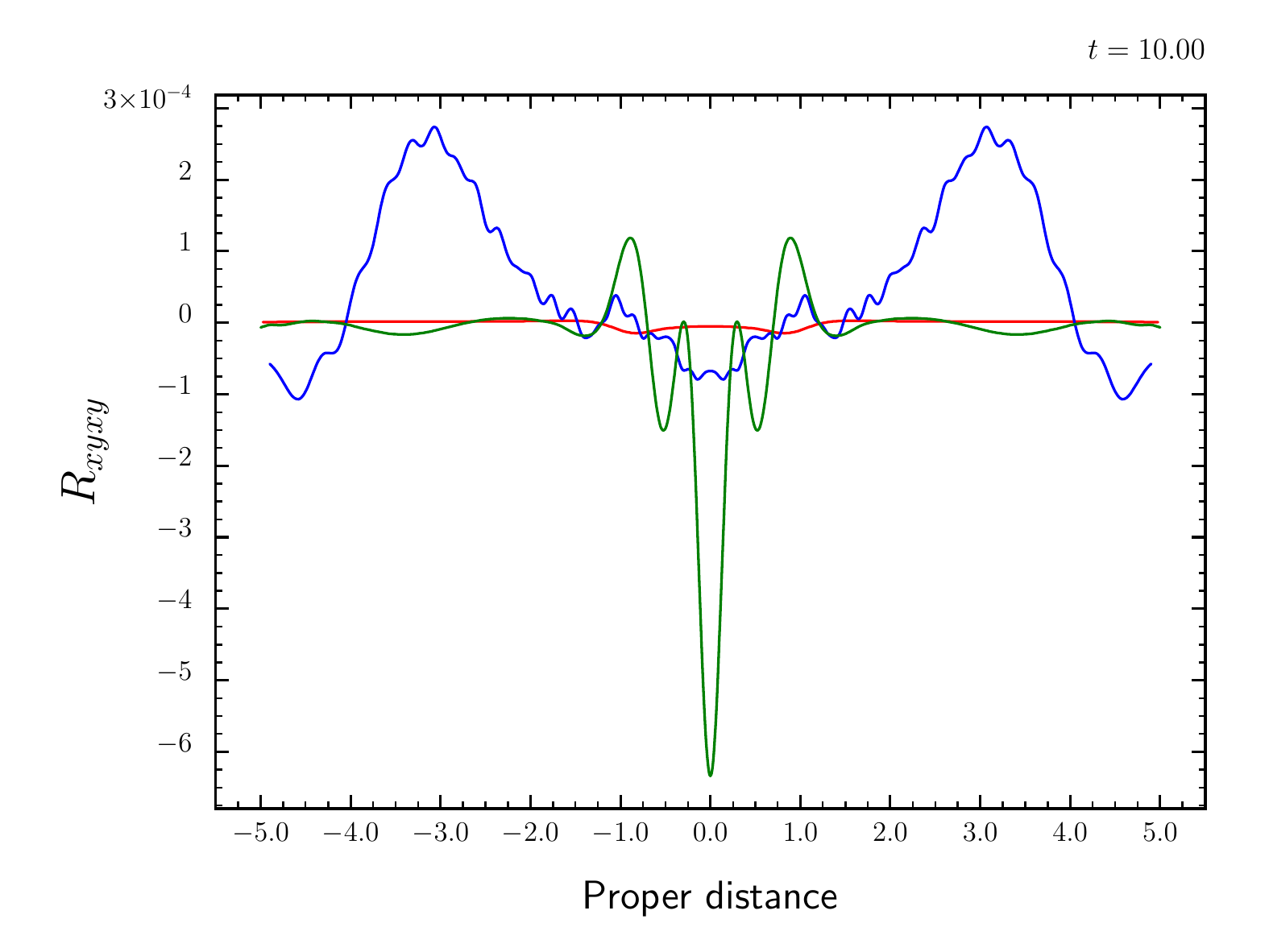}{\mywd}%
\caption{This is similar to figure (\ref{fig:BrillProfile05}) but for the
case $t=10$. It shows clears signs of reflected waves in the both ADM and
BSSN data while the lattice data is mostly flat apart from two small bumps
aligned to the wings of the BSSN bump.}
\label{fig:BrillProfile10}
\end{figure}

\begin{figure}[!ht]
\FigPair%
{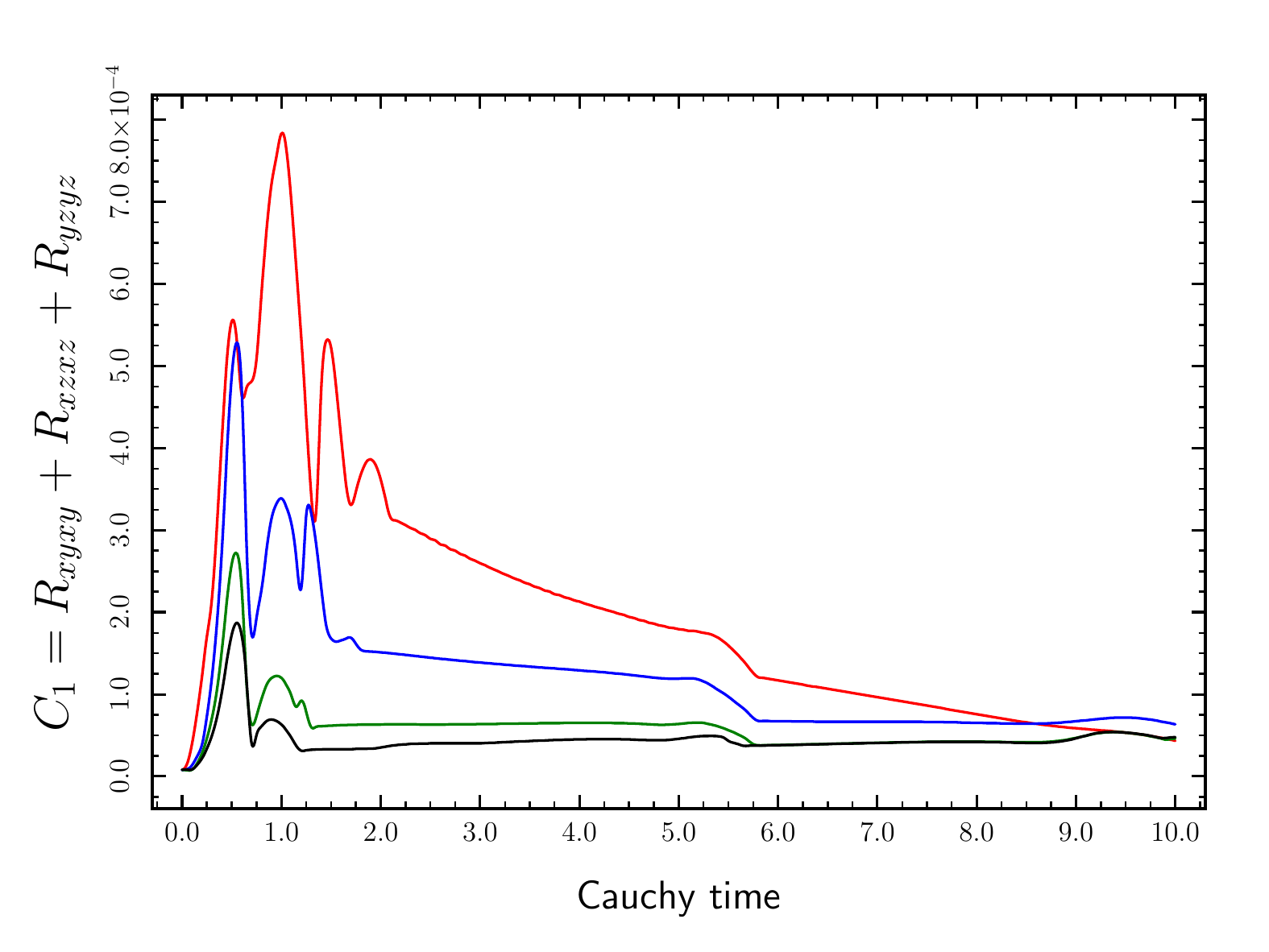}{\mywd}%
{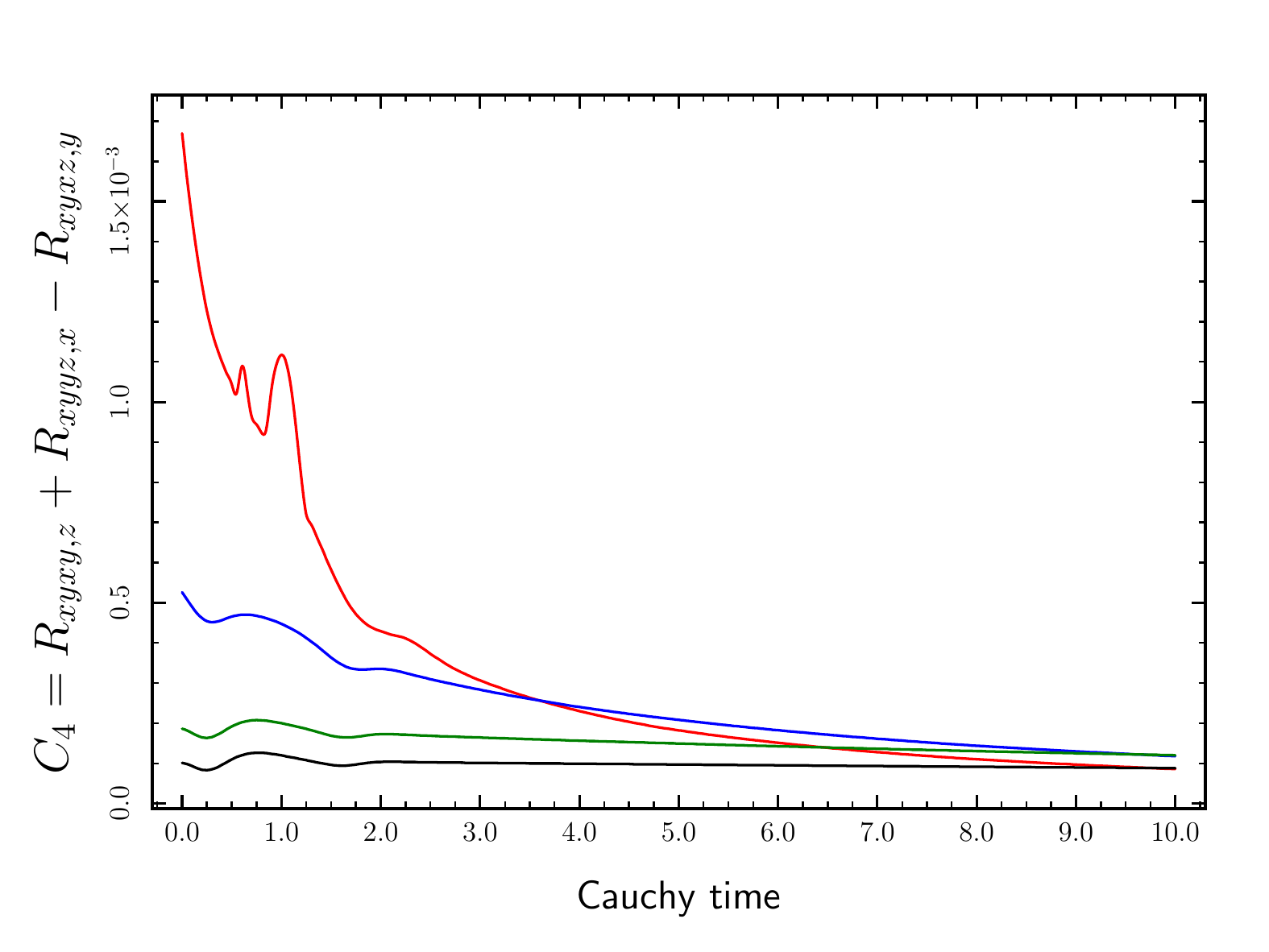}{\mywd}%
\caption{This pair of figures record the maximum value of the Brill
constraints $C_1$ and $C_4$ across the lattice for $0<t<10$. Note that the
constraints remain bounded and appear to decay towards a constant but
non-zero value during the evolution. The non-zero value is probably tied to
the truncation error in solving the Hamiltonian constraint
(\ref{eqn:BrillH}). The small bumps at approximately $t=5$ and $t=10$ in the
left hand figure are probably due to reflections from the outer boundary
(though this was not tested). The remaining constraints $C_2,C_3$ and $C_5$
are not included here as they show much the same behaviour as shown above.}
\label{fig:BrillConstC1C4}
\end{figure}

\begin{figure}[!ht]
\FigPair%
{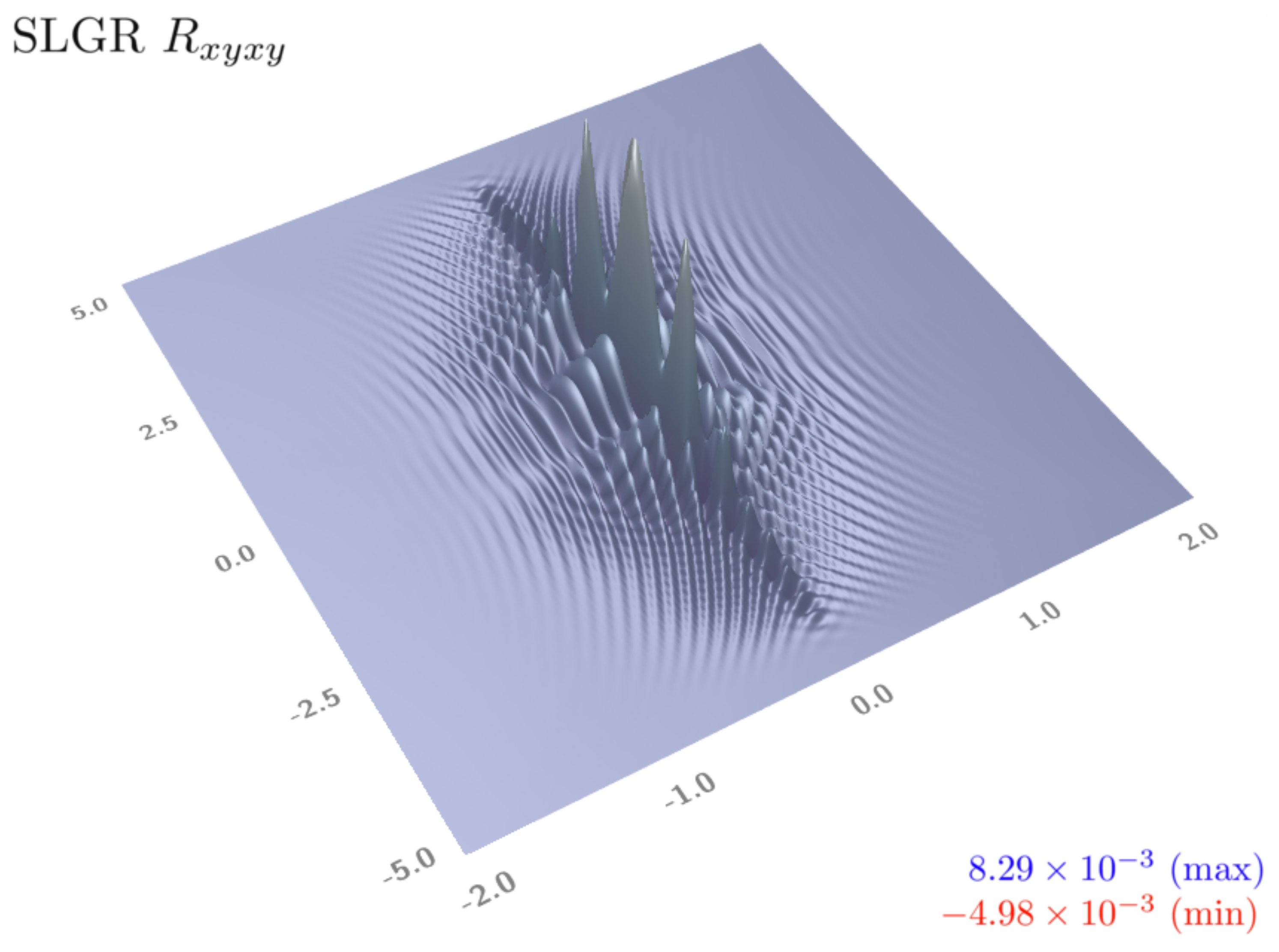}{\mywd}%
{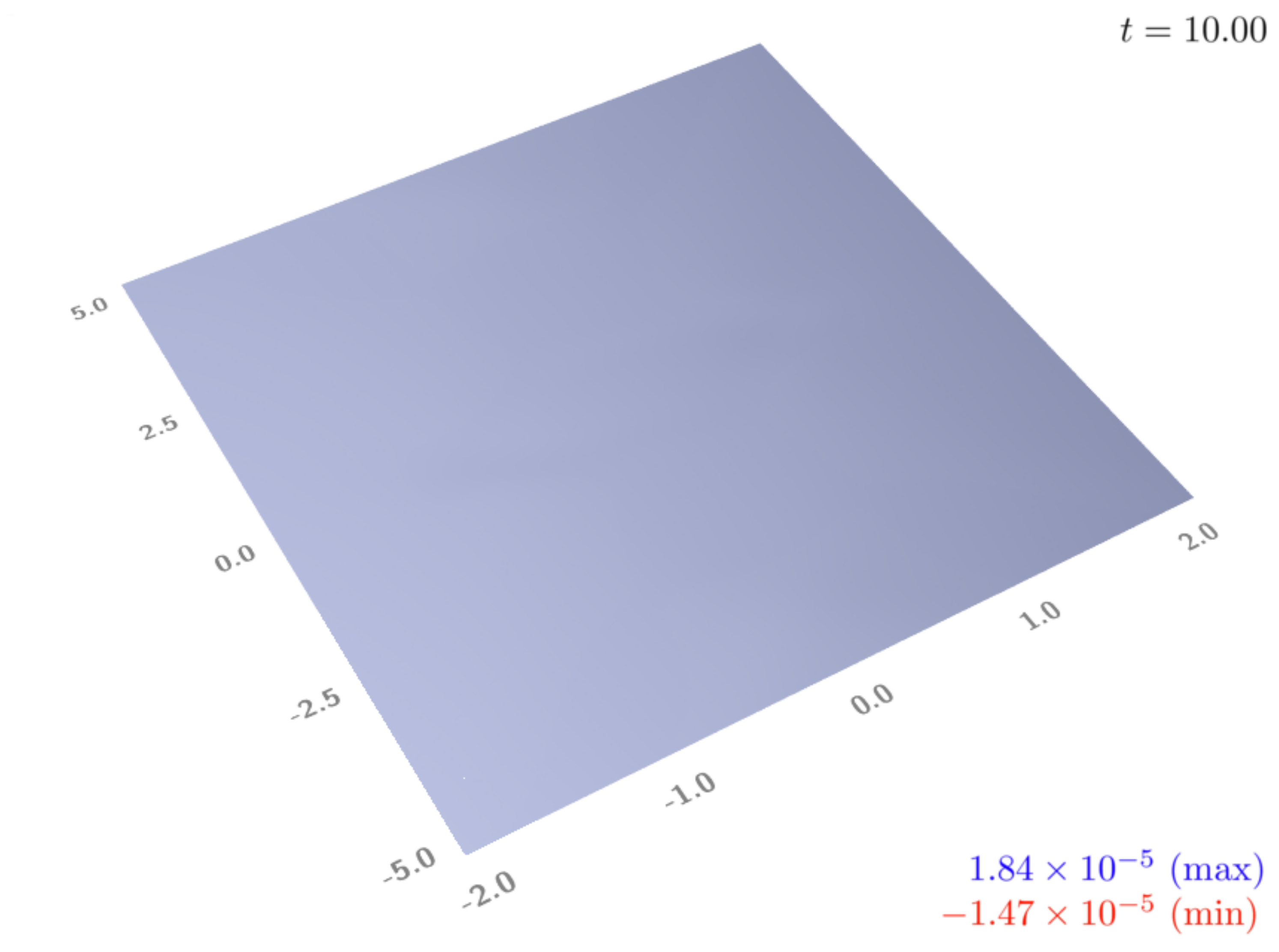}{\mywd}%
\FigPair%
{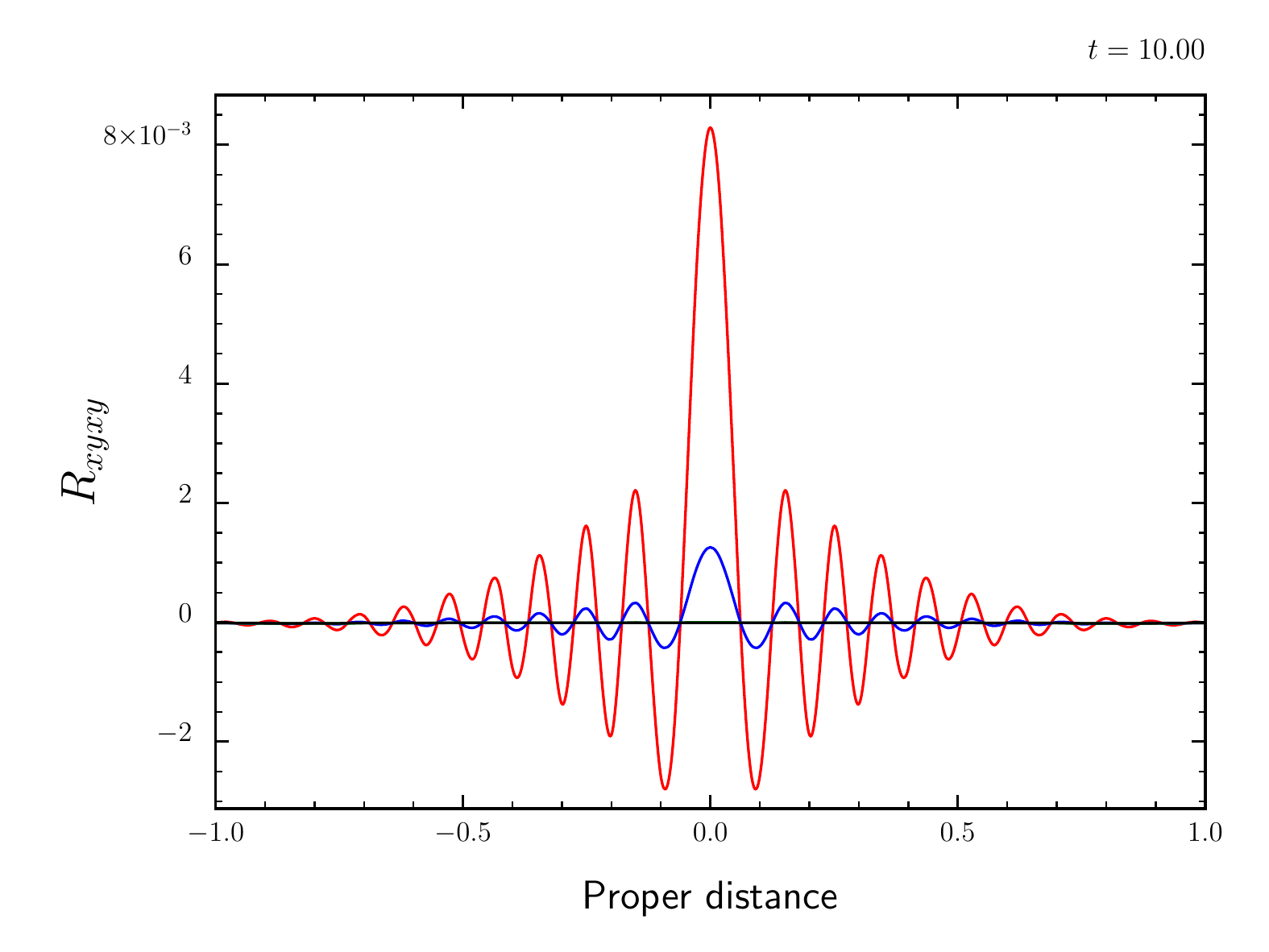}{\mywd}%
{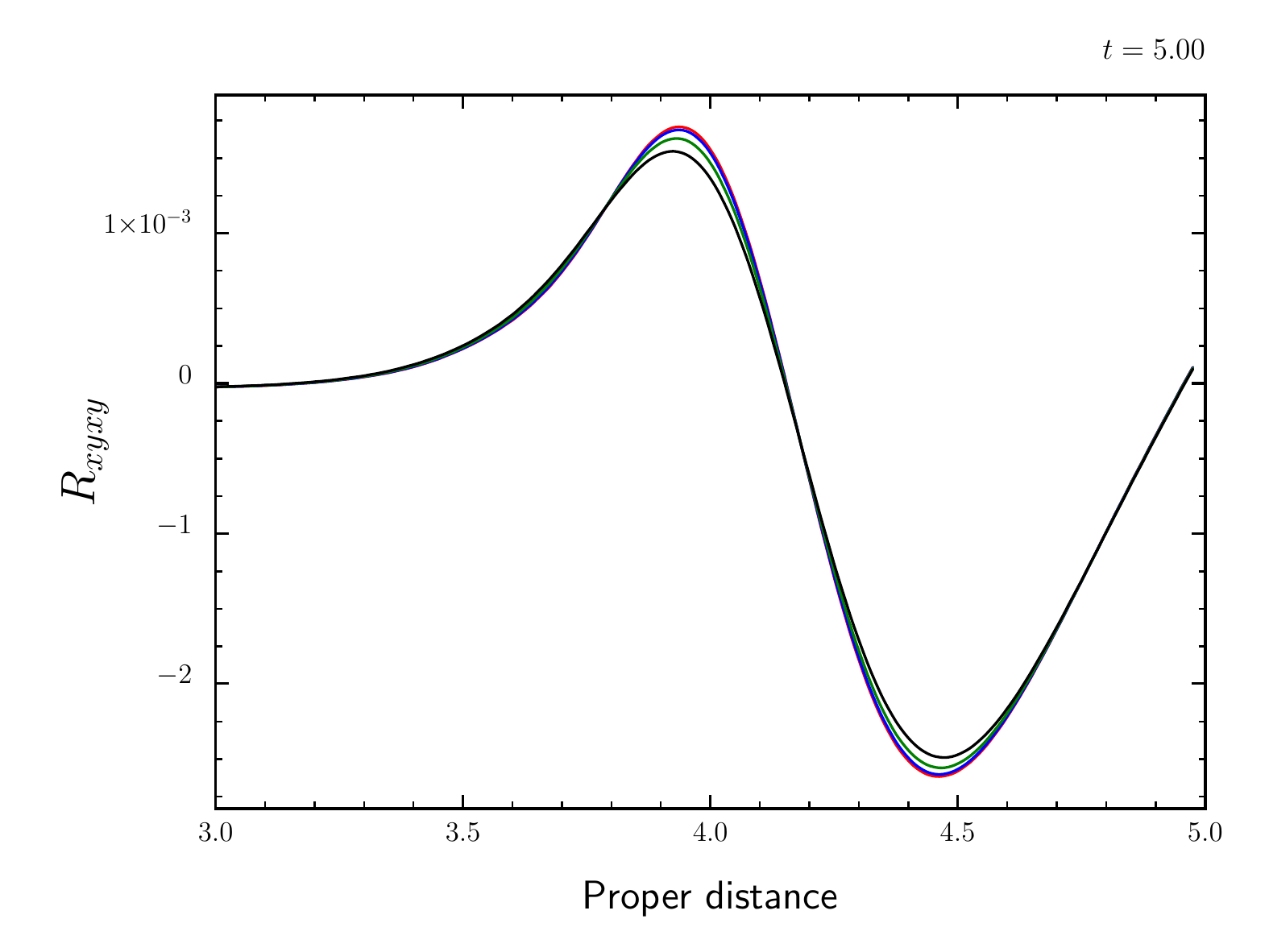}{\mywd}%
\caption{The top row of this figure shows how effective the numerical
dissipation can be in suppressing the axisymmetric instabilities. The data
differs only in the choice of the dissipation parameter, on the left
$\epsilon=0.1$ while on the right $\epsilon=1.0$. The bottom row shows
data along the $\Tx$ axis for four choices of the dissipation parameter,
$\epsilon=0.1$ (red), $\epsilon=0.2$ (blue), $\epsilon=0.5$ (green) and
$\epsilon=1.0$ (black). The lower right figure shows that the dissipation has
only a small effect on the peaks of the wave at $t=5$.}
\label{fig:BrillDissip}
\end{figure}

\newpage

\begin{figure}[!ht]
\FigPair%
{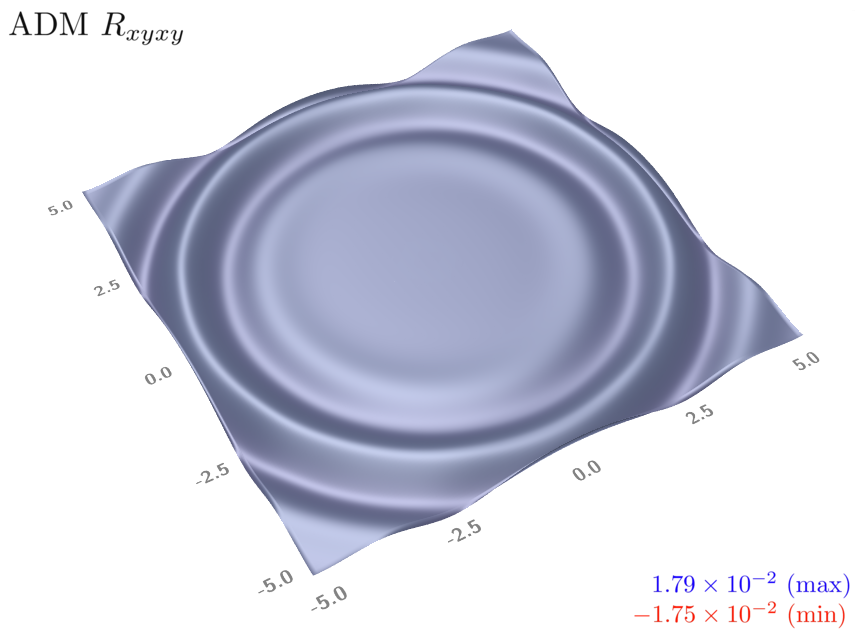}{\mywd}%
{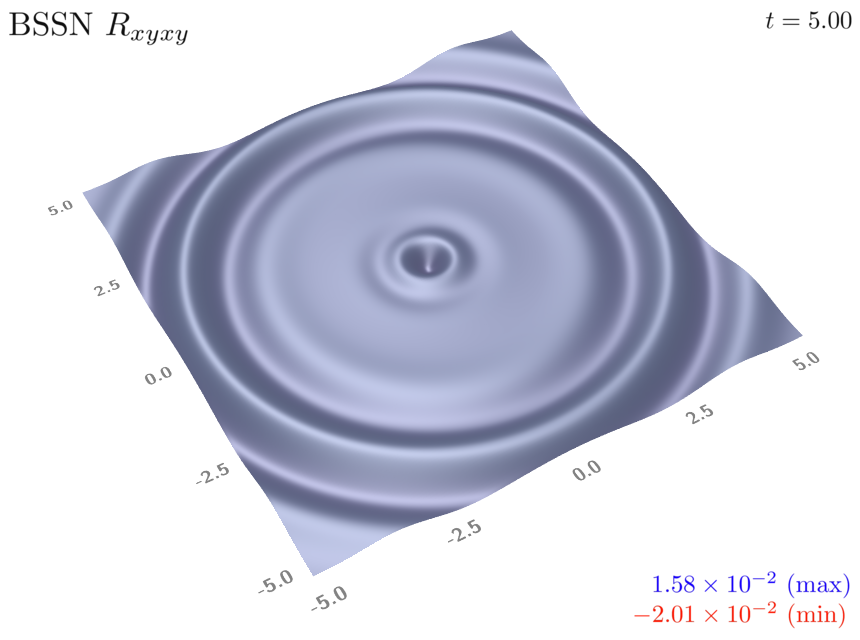}{\mywd}%
\FigPair%
{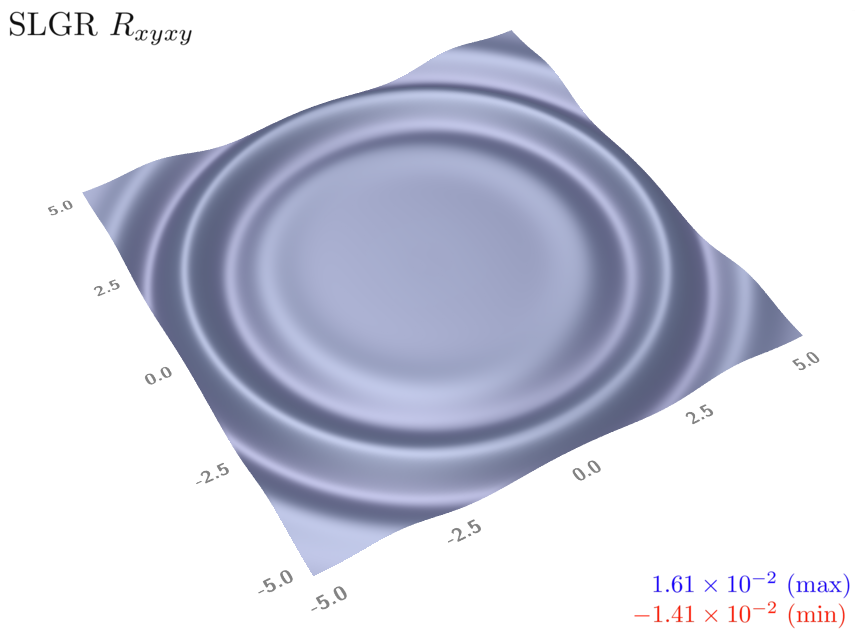}{\mywd}%
{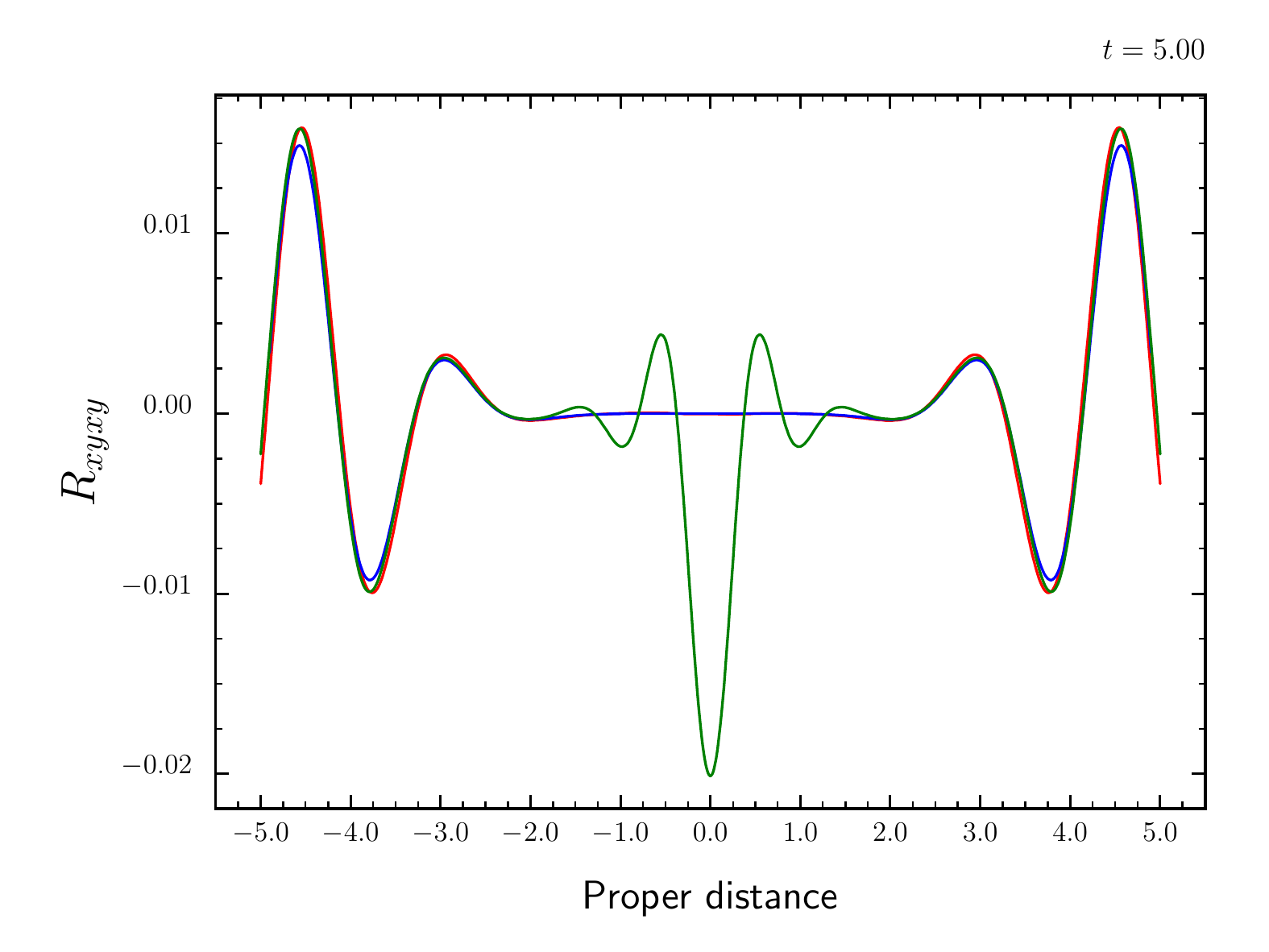}{\mywd}%
\caption{This figure is similar to figure (\ref{fig:BrillProfile05}) but in
this case showing the evolutions of the Teukolsky data. There are no obvious
boundary waves but the bump in the BSSN data remains. The lattice data again
looks smooth and flat behind the main wave.}
\label{fig:TeukProfile05}
\end{figure}

\begin{figure}[!ht]
\FigPair%
{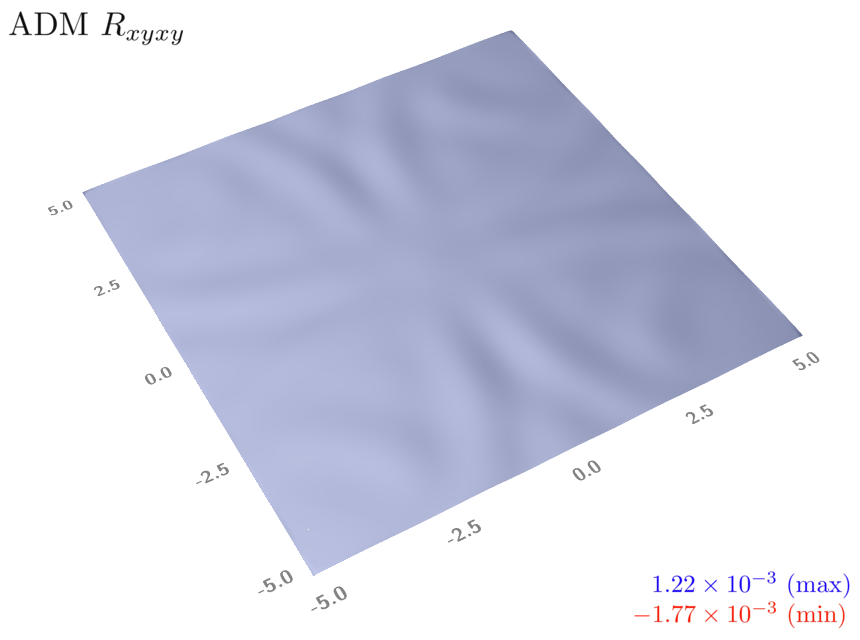}{\mywd}%
{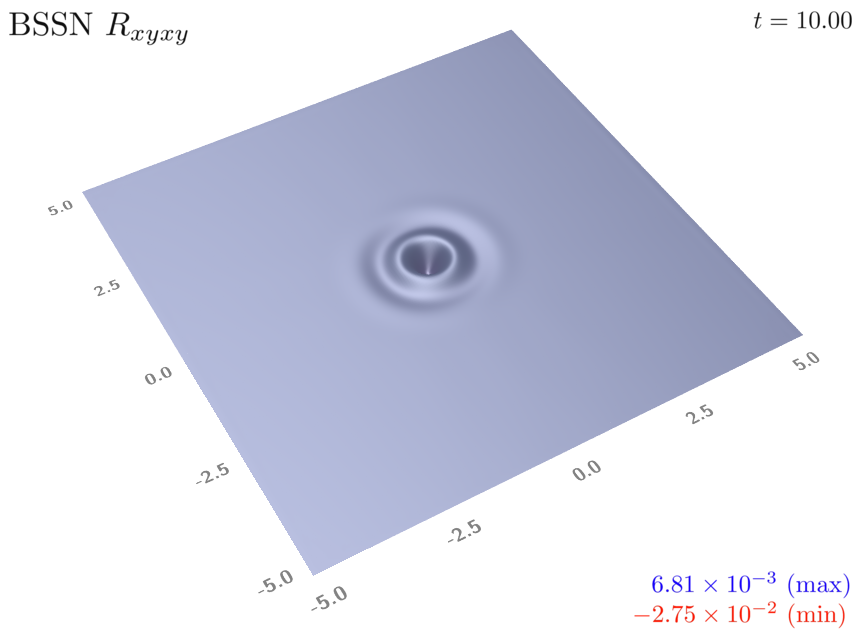}{\mywd}%
\FigPair%
{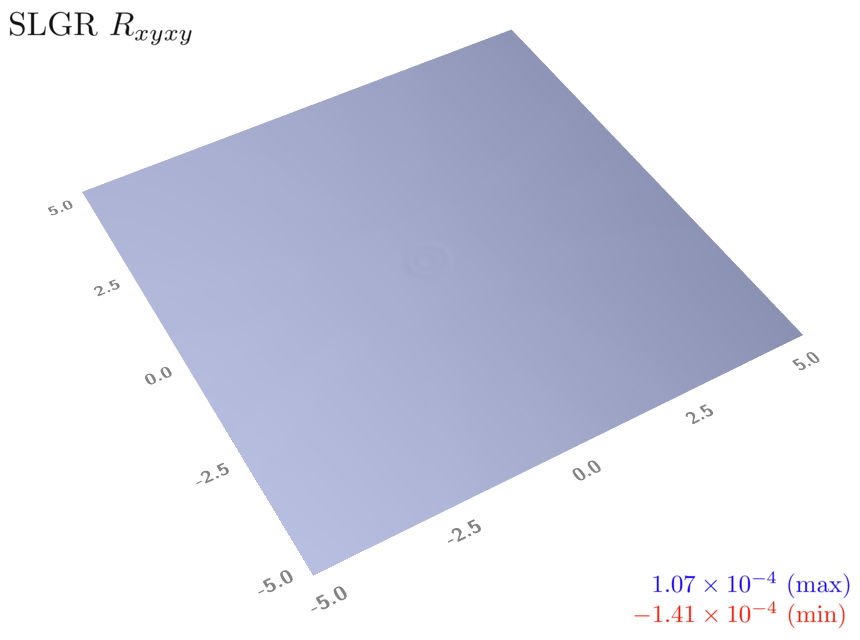}{\mywd}%
{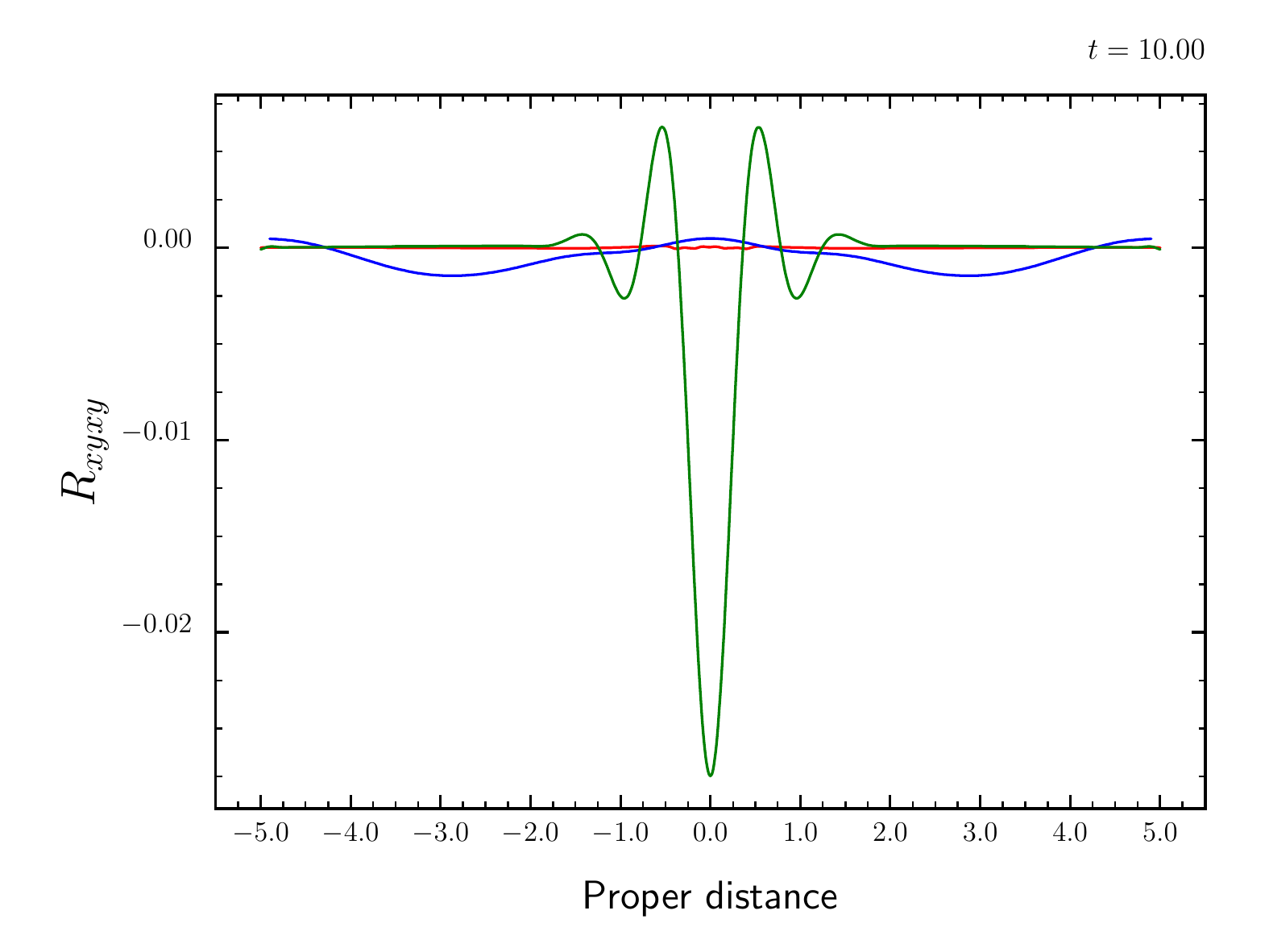}{\mywd}%
\caption{As per figure (\ref{fig:TeukProfile05}) but at $t=10$. The BSSN bump
has grown by a about 50\% over the period $t=0$ to $t=10$. There is also a
very small bump in the lattice data near the origin.}
\label{fig:TeukProfile10}
\end{figure}

\begin{figure}[!ht]
\FigQuad%
{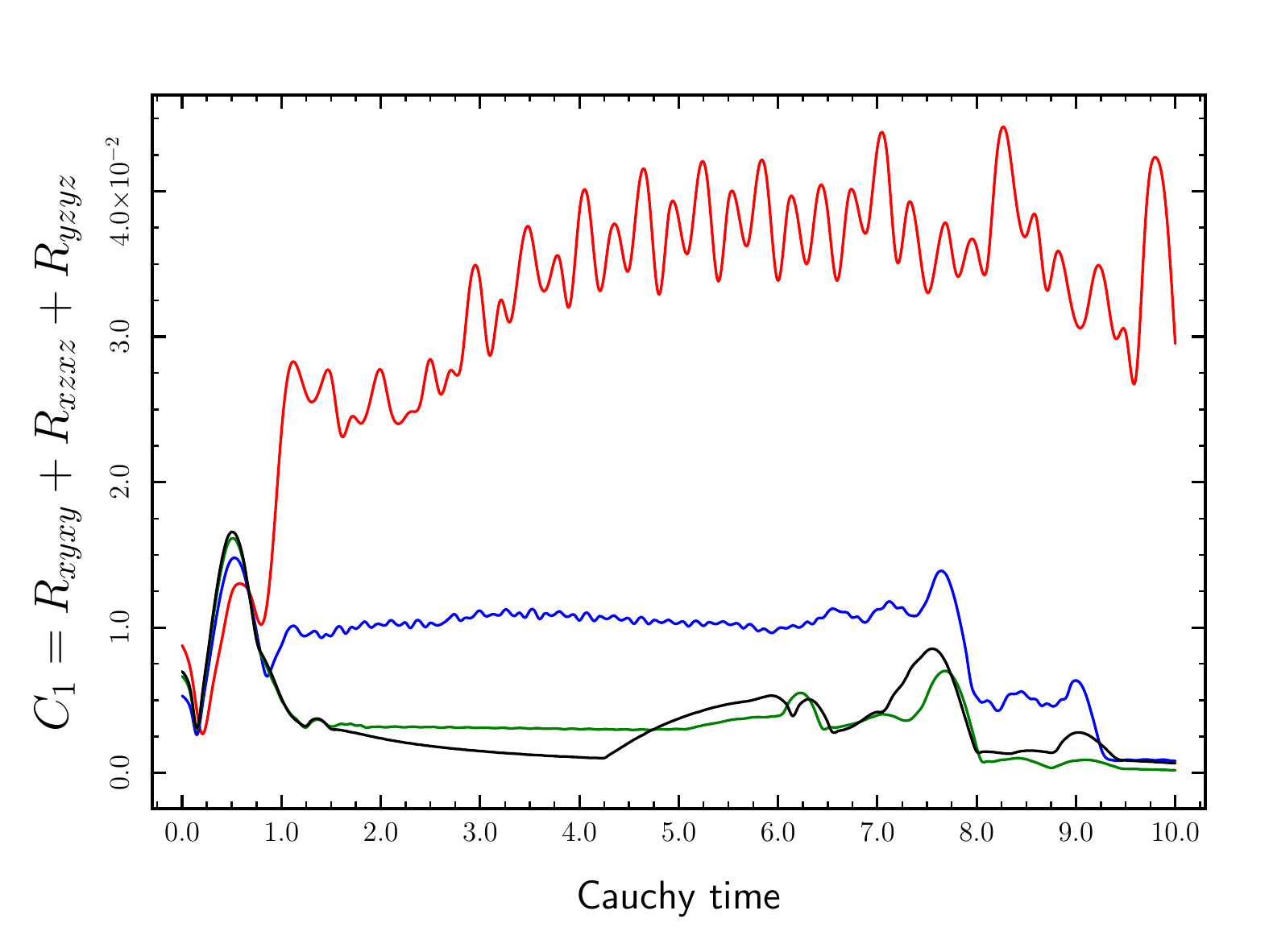}{\mywd}%
{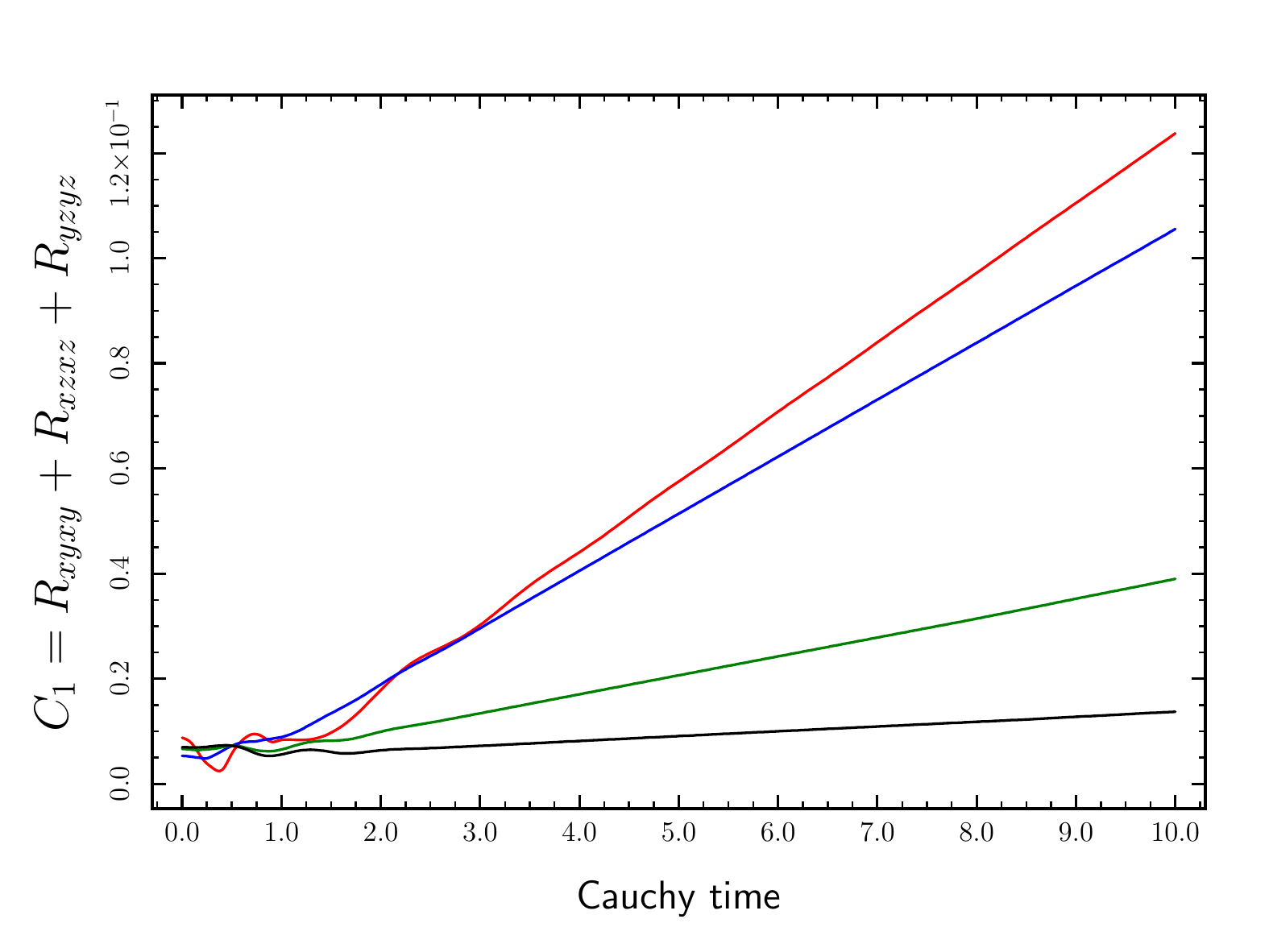}{\mywd}%
{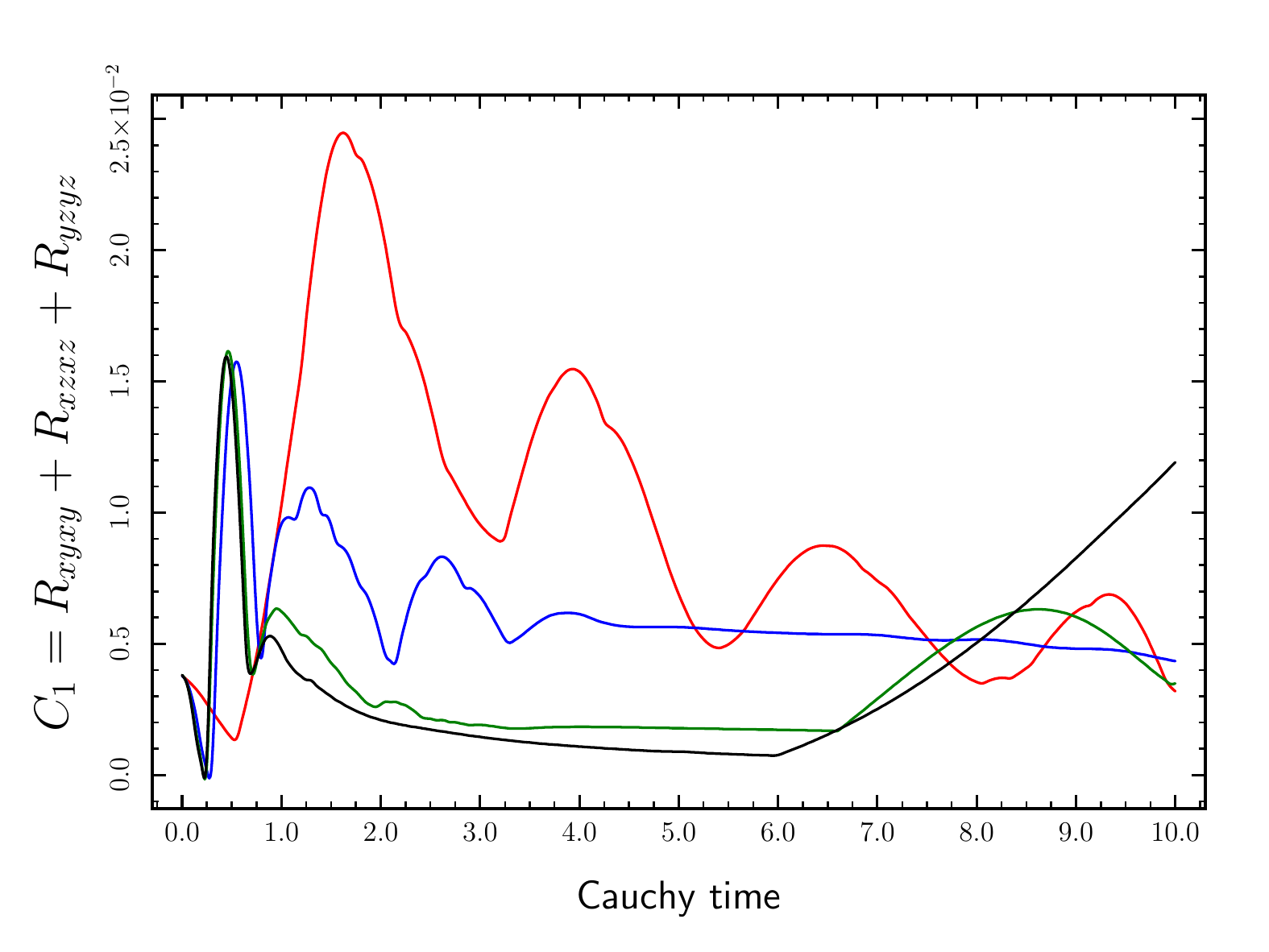}{\mywd}%
{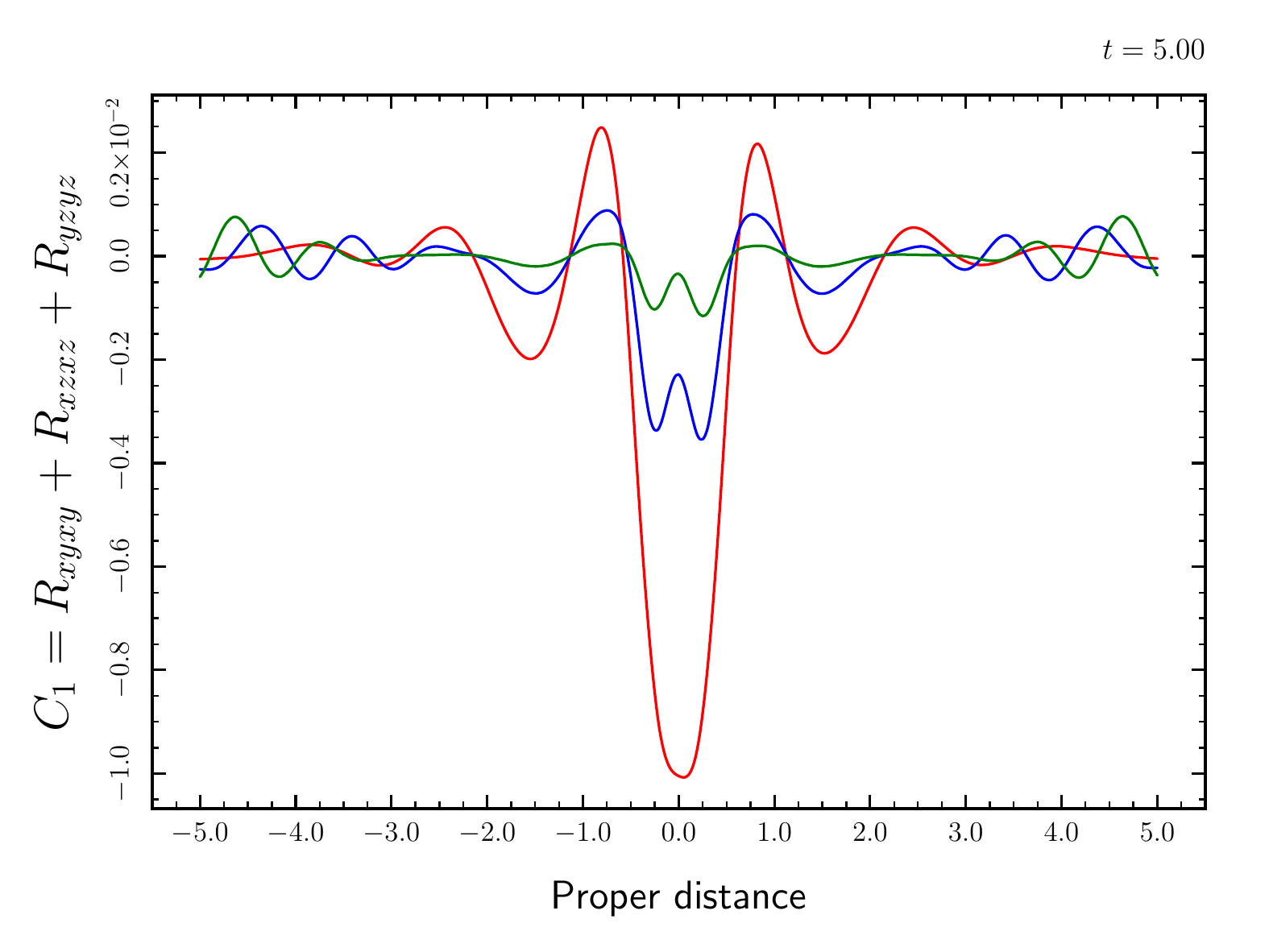}{\mywd}%
\caption{These plots show the behaviour of the $C_1$ constraint
(\ref{eqn:TeukConstC1}) for the evolution of the Teukolsky initial data. The
plots in the top left (ADM), top right (BSSN) and bottom left (SLGR) show the
evolution of the maximum of $C_1$ across the $xy$-plane. The colours in the
ADM and BSSN plots correspond to $N_x=N_y=N_z=26$ (red), 50 (blue), 100
(green) and 200 (black) while for the lattice the corresponding numbers are
25, 51,101 and 201. The plot in bottom right shows the values of $C_1$ along
the $\Tx$-axis for the lattice data at $t=5$ for three lattices,
$N_x=N_y=N_z=51$ (red), 101 (blue) and 201 (green).}
\label{fig:TeukConstC1}
\end{figure}

\begin{figure}[!ht]
\FigPair%
{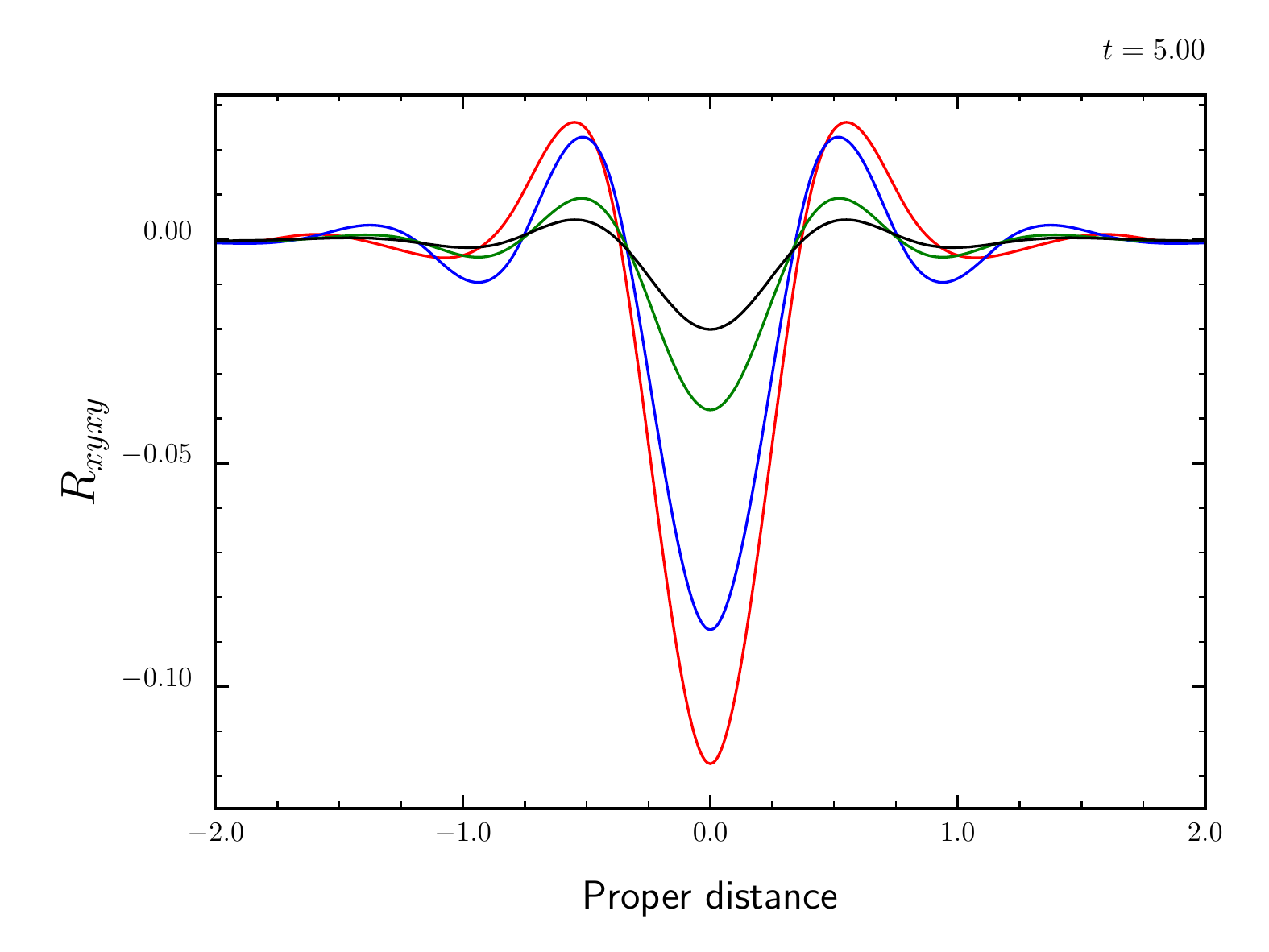}{\mywd}%
{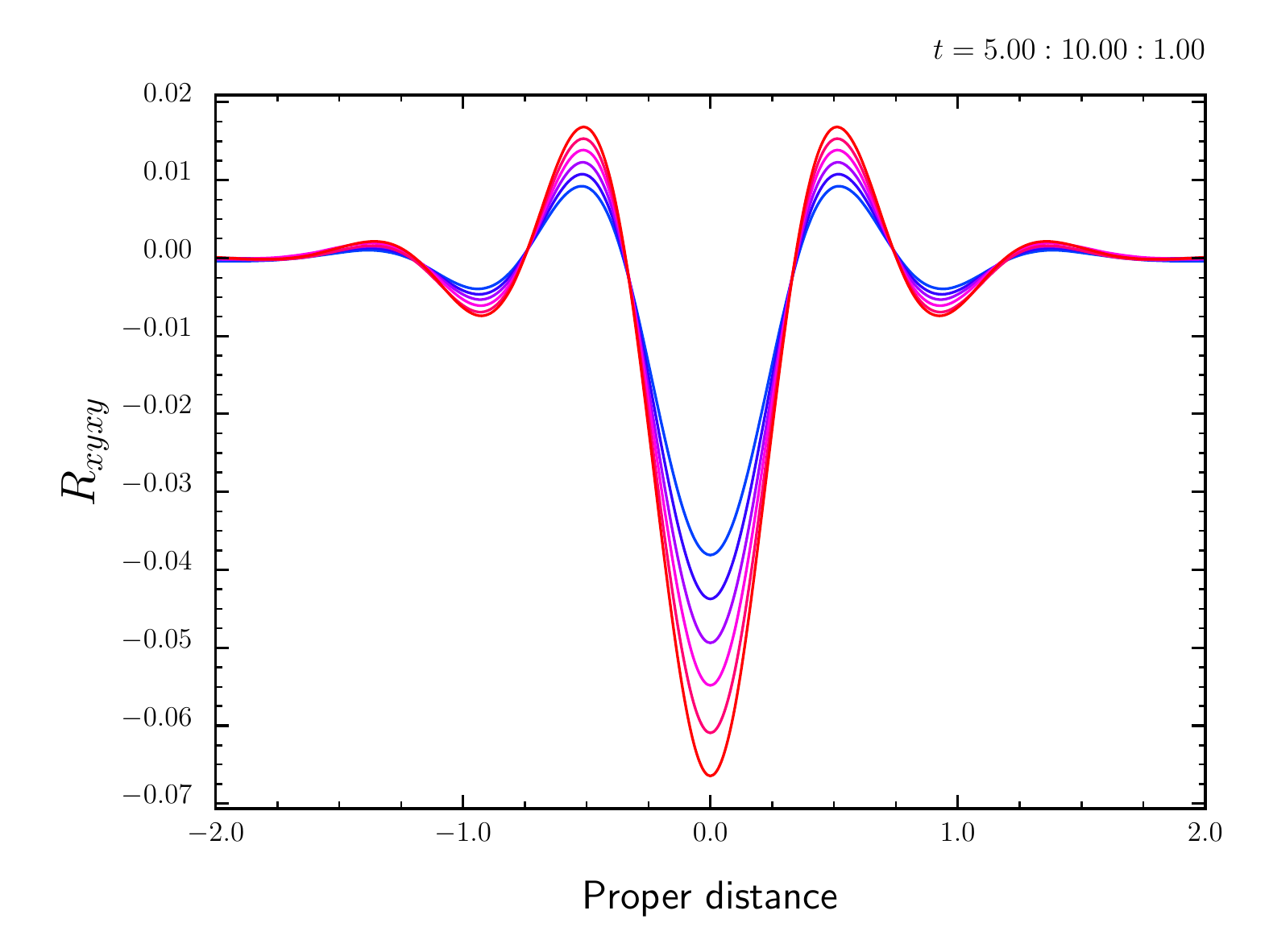}{\mywd}%
\caption{This pair of plots shows the behaviour the BSSN bump as a function of
the number of grid points (left plot with $N_z=26$ (red), $N_z=50$ (blue),
$N_z=100$ (green) and $N_z=200$ (black)) and as a function of time (right plot
for $t=5$ to $t=10$ in steps of 1). The left plot shows that as the number of
grid points is increased the size of the bump decreases while the right plot
shows that the bump increases linearly with time. This bump is the source of
the linear growth in the constraint seen in figure (\ref{fig:TeukConstC1}).}
\label{fig:TeukBump}
\end{figure}

\begin{figure}[!ht]
\Figure{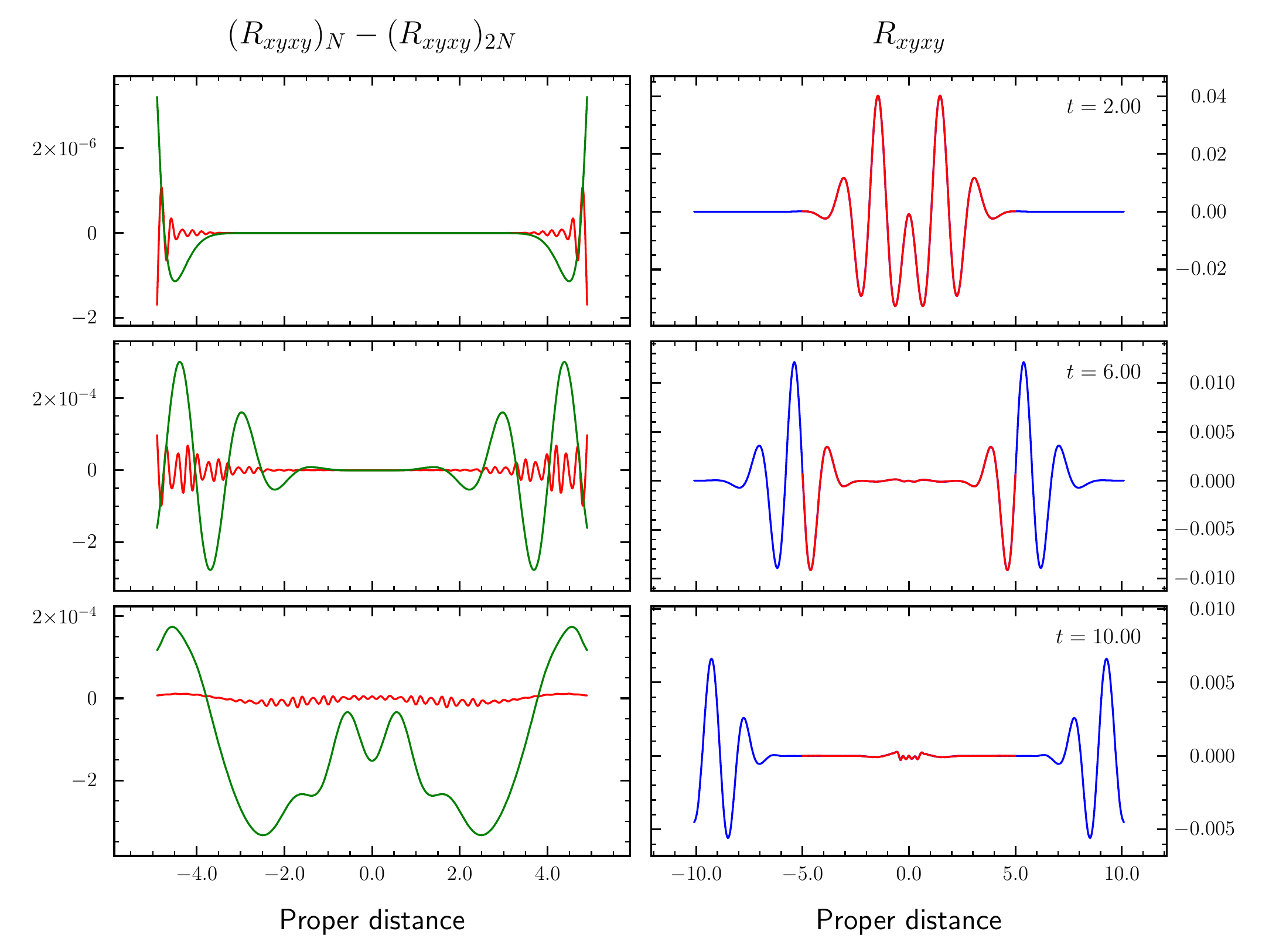}{1.00}%
\caption{
These plots were created by evolving two sets of initial data, one with
$N_x=N_y=N_z=101$, the other with $N_x=N_y=N_z=201$. Both initial data sets
used $\Delta x=\Delta y=\Delta z=0.1$. There are two curves in the right plot,
both for $\Rxyxy$, one on the small grid (red) and the other on the larger
grid (blue). Note how the red curve lies directly on top of the blue curve.
The plots on the left show the difference in $\Rxyxy$ between the two
evolutions on $\vert x\vert < 5$. The green curve is for the BSSN data while
the red curve is for the lattice data.}
\label{fig:TeukBCTest}
\end{figure}

\end{document}